\documentclass[aps,prx,10pt,twocolumn,superscriptaddress,nofootinbib,notitlepage,showpacs,floatfix,longbibliography]{revtex4-2}

\usepackage[english]{babel}

\usepackage{graphicx,graphics,epsfig,subfigure,times,bm,bbm,amssymb,amsmath,amsfonts,mathrsfs}
\usepackage[scr=boondoxo,scrscaled=1.05]{mathalfa}
\usepackage{color}
\usepackage{extarrows}
\usepackage{tikz}
\usepackage{pgfplotstable}
\usepackage{booktabs}
\usepackage{siunitx}
\usepackage{placeins}
\usetikzlibrary{external}
\usetikzlibrary{decorations.markings, arrows.meta}
\usetikzlibrary{fpu} 
\pgfkeys{/pgf/fpu=true}
\pgfmathparse{2*pi/(960e-9)}
\pgfmathfloattofixed{\pgfmathresult} 

\pgfkeys{/pgf/fpu=false}

\usepackage{mathtools}
\usepackage{pgfplots}
\usepackage{xr-hyper}
\usepackage[pdfstartview=FitH,colorlinks=true, allcolors=blue]{hyperref}
\hypersetup{bookmarksdepth=2}
\usepackage{cancel}
\usetikzlibrary{arrows.meta}
\usepackage[OT2,T1]{fontenc}
\usepackage{csvsimple}
\usepackage{xcolor}
\usepgfplotslibrary{groupplots} 

\newcommand{\secref}[1]{Sec.~\ref{#1}}
\newcommand{\appref}[1]{Appendix~\ref{#1}}

\DeclareSymbolFont{cyrletters}{OT2}{wncyr}{m}{n}
\DeclareMathSymbol{\Sha}{\mathalpha}{cyrletters}{"58}

\newtheorem{theorem}{Theorem}

\newtheorem{algorithm}[theorem]{Algorithm}

\long\def\ignore#1{}

\newcommand{\beq}{\begin{equation}}
\newcommand{\eneq}{\end{equation}}
\newcommand{\beqnn}{\begin{equation*}}
\newcommand{\eneqnn}{\end{equation*}}
\newcommand{\beqy}{\begin{eqnarray}}
\newcommand{\eneqy}{\end{eqnarray}}
\newcommand{\beqynn}{\begin{eqnarray*}}
\newcommand{\eneqynn}{\end{eqnarray*}}

\newcommand{\Tr}{\mathrm{Tr}}

\addto\captionsenglish{}

\pgfplotsset{compat=1.18}
\definecolor{clr1}{HTML}{4E79A7} 
\definecolor{clr2}{HTML}{F28E2C} 
\definecolor{clr3}{HTML}{e15759} 
\definecolor{clr4}{HTML}{76b7b2} 
\definecolor{clr5}{HTML}{59a14f} 
\definecolor{clr6}{HTML}{edc949} 
\definecolor{clr7}{HTML}{af7aa1} 
\definecolor{clr8}{HTML}{9c755f} 
\definecolor{clr9}{HTML}{bab0ab} 
\definecolor{clr10}{HTML}{111113} 

\newcommand{\dobib}{%
    \bibliographystyle{apsrev4-2}%
    \ifSubfilesClassLoaded{%
        \bibliography{../gauss_bibtex}
    }{}%
}

\usepackage{subfiles}

\newif\ifpaperExternalReferencesLoaded
\newcommand{\paperExternalReferences}{%
    \ifSubfilesClassLoaded{%
        \ifpaperExternalReferencesLoaded
        \else
            \global\paperExternalReferencesLoadedtrue
            \IfFileExists{../build/main.aux}{%
                \externaldocument[][nocite]{../build/main}[../build/main.pdf]%
            }{%
                \IfFileExists{../main.aux}{%
                    \externaldocument[][nocite]{../main}[../main.pdf]%
                }{%
                    \IfFileExists{build/main.aux}{%
                        \externaldocument[][nocite]{build/main}[build/main.pdf]%
                    }{%
                        \IfFileExists{main.aux}{%
                            \externaldocument[][nocite]{main}[main.pdf]%
                        }{%
                            \PackageWarning{CCspect}{Full-paper references
                              unavailable. Compile main.tex once before
                              compiling a standalone section}%
                        }%
                    }%
                }%
            }%
        \fi
    }{}%
}

\newcommand{\paperStandaloneSection}[4]{%
    \ifSubfilesClassLoaded{%
        \setcounter{section}{\numexpr#1-1\relax}%
        \setcounter{equation}{#2}%
        \setcounter{figure}{#3}%
        \setcounter{table}{#4}%
    }{}%
}

\makeatletter
\newcommand{\paperRecordCounter}[2]{%
    \begingroup
        \edef\@currentlabel{\number\value{#2}}%
        \label{#1}%
    \endgroup
}
\makeatother

\newcommand{\paperSetCounterFromReference}[2]{%
    \ifcsname r@#2\endcsname
        \setcounter{#1}{\getrefnumber{#2}}%
    \else
        \PackageWarning{CCspect}{Appendix #1 counter unavailable. Compile
          main.tex before compiling the standalone appendix}%
    \fi
}

\newcommand{\paperStandaloneAppendix}{%
    \ifSubfilesClassLoaded{%
        \paperSetCounterFromReference{figure}{paper:appendix-figure-offset}%
        \paperSetCounterFromReference{table}{paper:appendix-table-offset}%
        \paperSetCounterFromReference{theorem}{paper:appendix-theorem-offset}%
    }{%
        \paperRecordCounter{paper:appendix-figure-offset}{figure}%
        \paperRecordCounter{paper:appendix-table-offset}{table}%
        \paperRecordCounter{paper:appendix-theorem-offset}{theorem}%
    }%
}

\begin{document}
\renewcommand{\dobib}{} 

\title{
Control-centric quantum noise spectroscopy of time-ordered polyspectra
}

\newcommand{\griffithaffil}{Centre for Quantum Computation and Communication Technology (Australian Research Council), Centre for Quantum Dynamics, Griffith University, Brisbane, Queensland 4111, Australia}

\author{Kaiah Steven}
\thanks{These authors contributed equally.}
\author{Elliot Coupe}
\thanks{These authors contributed equally.}

\author{Qi Yu}
\affiliation{\griffithaffil}

\author{Gerardo A. Paz-Silva}
\email{g.pazsilva@griffith.edu.au} 
\affiliation{\griffithaffil}
\affiliation{Diraq Pty Ltd, Sydney, NSW, Australia}

\begin{abstract}
Precise environmental-noise characterisation in open quantum systems is a key step toward high-fidelity quantum control and targeted decoherence suppression in computing and sensing applications. Non-parametric quantum noise spectroscopy (QNS) provides a general-purpose, model-agnostic framework for estimating the spectral properties of an environment. The ability to perform such protocols under realistic constraints is key to their practical applicability. Notably, it is important to account for control constraints and understand how they limit the ability to learn about noise correlations as experiment-agnostic objects. We show how adopting a control-centric point of view allows one to recast the noise spectroscopy problem in such a way that (i) the central objects are now the time-ordered polyspectra, (ii) control filter functions are no longer encumbered by time ordering. In particular, we show that this approach enables the generalisation of frequency-comb QNS protocols to arbitrary control scenarios without introducing additional control symmetries that remove time ordering from filter functions, improving estimation in typically pathological scenarios.  We demonstrate the targeted reconstruction of the time-ordered polyspectra across classical Gaussian and quantum non-Gaussian environments via simulations.

\end{abstract}

\maketitle

\paperExternalReferences
\paperStandaloneSection{1}{0}{0}{0}

\section{Introduction}

Understanding how quantum systems interact with their environment is foundational to quantum technologies. In quantum computing, one is usually interested in the environment as a source of noise to be suppressed. Characterising it allows the effective and efficient use of control techniques, ultimately achieving the high-fidelity operations needed for fault-tolerant quantum computation~\cite{lidarQuantumErrorCorrection2013, khodjastehDynamicallyErrorCorrectedGates2009}. Conversely, in quantum sensing, the environment plays a dual role: it contains information of interest, such as magnetic field strength or temperature, while simultaneously acting as a source of noise that can obscure the target signal. A thorough understanding of the environment is therefore indispensable for success in both domains.

Environmental characterisation can take many forms, depending on prior information or assumptions. But, broadly speaking, it falls into two main approaches: The {\it parametric} approach assumes a specific model for the environment, such as a bath of two-level systems and a particular functional form for the correlation functions, e.g., $1/f^\alpha$~\cite{duttaLowfrequencyFluctuationsSolids1981, paladinoNoise2014, dialChargeNoiseSpectroscopy2013, yoshiharaFluxQubitNoise2014a, faoroDynamicalSuppression12004,wudarskiCharacterizingLowFrequencyQubit2023}, and seeks to determine the model parameters. While computationally efficient, this strategy can be overly restrictive when prior information is unavailable, e.g., when the model is a priori unknown. The alternatives are {\it non-parametric} approaches. While generally more resource-intensive, these minimise the model assumptions and thus provide less biased estimation. This is important when accurate characterisation is critical. However, zero-prior noise learning is generally an ill-posed problem: given a finite number of probes, there is no unique noise model that reproduces the data. Thus, even in the non-parametric scenario, assumptions must be made to render the problem tractable; otherwise, it is unlikely that a finite number of probing sequences will allow one to learn the full noise. These assumptions include defining the dimensions of the Hilbert space, the statistics of the induced noise~\cite{liConceptsQuantumNonMarkovianity2018a,caroleaddisDynamicalDecouplingEfficiency2015}, and the smoothness of the correlations. Ultimately, noise learning is most meaningful when driven by a specific objective; for example, environmental characterisation may require minimal assumptions, whereas an effective parametric model may suffice for control.

 Characterisation methods that use functional transforms (Fourier and frame decompositions) of bath correlations can be broadly categorised as quantum noise spectroscopy (QNS) techniques. Early research into dynamical decoupling (DD)~\cite{violaDYNAMICALDECOUPLINGOPEN1999,gotzs.uhrigExactResultsDynamical2008,gotzs.uhrigEfficientCoherentControl2010,kavehkhodjastehArbitrarilyAccurateDynamical2010,wenyangPreservingQubitCoherence2010} sequences revealed that the decay rate of a qubit undergoing DD can be expressed in the frequency domain as the overlap integral between the bath spectral density and the filter function generated by the DD control sequence~\cite{ajoyOptimalPulseSpacing2011a,biercukDynamicalDecouplingSequence2010,lukaszcywinskiHowEnhanceDephasing2008,kofmanAccelerationQuantumDecay2000,greenHighOrderNoiseFiltering2012, soareExperimentalNoiseFiltering2014a, paz-silvaGeneralTransferfunctionApproach2014a,stefanopasiniOptimizedDynamicalDecoupling2010,harrisonballWalshsynthesizedNoisefilteringQuantum2014}. This was exploited to perform environment characterisation by Alvarez and Suter~\cite{alvarezMeasuringSpectrumColored2011a}, utilising the fact that $M$-fold periodic repetition of a pulse sequence with cycle time $\tau_c$, generates a control function that approximates a Dirac comb with peaks localised at the $k\in\mathbb{Z}$ harmonic frequencies $\omega = k\omega_h$, where $\omega_h=2\pi/\tau_c$ denotes the fundamental comb frequency.
The narrowband comb `teeth' selectively probe the bath polyspectra at discrete frequency modes, forming a system of linear equations that relate the spectrum harmonics to measured decoherence. Inverting this system returns a discretised estimate of the spectrum. Conceptually similar approaches have also been used to characterise noisy environments in a range of settings~\cite{almogDirectMeasurementSystemEnvironment2011,yugeMeasurementNoiseSpectrum2011,szankowskiAccuracyDynamicaldecouplingbasedSpectroscopy2018,arianvezvaeeFourierTransformNoise2022}. 

Beyond conventional comb-based protocols, QNS can be further specialised based on the primary objective. For control-focused applications, frame representation of control elements $y_{a,b}(t)$ is constructed under explicit control constraints $\mathcal{C}_{\mathrm{ctrl}}$. These constraints define a restricted subspace of learnable noise correlators $\langle  B(t_1) \cdots B(t_k) \rangle \vert_{\mathcal{C}_{\mathrm{ctrl}}}$ that are \emph{adapted} to the limits imposed by the practical considerations such as control bandwidth, sampling frequency or gate speed. The resulting \emph{control-adapted} framework~\cite{chalermpusitarakFrameBasedFilterFunctionFormalism2020} establishes that $\mathcal{C}_{\mathrm{ctrl}}$ not only specifies the learnable noise correlators but also establishes their necessity and sufficiency for reconstructing the full interaction dynamics.

The rapid development of quantum technologies has precipitated a need for flexible QNS protocols capable of characterising arbitrary quantum environments across a diverse range of quantum devices. Recent refinements have extended QNS to higher-order non-Gaussian correlations~\cite{norrisQubitNoiseSpectroscopy2016c,paz-silvaDynamicalDecouplingSequences2016a,jana.krzywdaDynamicaldecouplingbasedSpatiotemporalNoise2019,dongResourceefficientDigitalCharacterization2023}, multi-axis noise~\cite{paz-silvaExtendingCombbasedSpectral2019a} and multi-qubit settings~\cite{paz-silvaMultiqubitSpectroscopyGaussian2017a}. Work has also been done on optimising resource use~\cite{huangRandomPulseSequences2025,tonekaboniGreedyMapbasedOptimized2023,dongResourceefficientDigitalCharacterization2023}. While these protocols rely on discretising the overlap integral through control sequences, complementary approaches exist, such as spin-locking protocols~\cite{Sung2021MultiLevelQNS,Capiglioni2022SpinLocking,khanMultiaxisQuantumNoise2024}, variational reconstruction schemes~\cite{Shitara2024VariationalQNS}, and single-shot Ramsey correlation spectroscopy~\cite{RojasArias2025NoiseCrossCorrelations}.

Despite these advances, contemporary QNS methods remain predicated on idealised assumptions that limit their applicability in experimental settings. Fundamentally, these assumptions can be understood as projecting the dynamics onto subspaces in which time ordering becomes irrelevant. This requirement is automatically satisfied for single-axis dephasing under instantaneous $\pi$-pulses, but fails generically with finite-duration controls that generate toggling-frame rotations through non-commuting axes. A number of strategies have been developed to extend QNS beyond the simplest realisations of this regime, though each ultimately retains some form of the underlying restriction: symmetrised control classes that ensure only factorisable filters contribute~\cite{paz-silvaExtendingCombbasedSpectral2019a}, weak-noise Magnus truncations suited to fidelity prediction~\cite{greenArbitraryQuantumControl2013}, adiabatic approximations valid only for slowly varying transverse noise~\cite{barnesFilterFunctionFormalism2016}, continuous-drive protocols limited to second-order characterisation under secular assumptions~\cite{khanMultiaxisQuantumNoise2024}, or ad hoc estimation corrections~\cite{sungNonGaussianNoiseSpectroscopy2019a}. While each extends QNS's reach in specific regimes, none provides a systematic framework for spectral reconstruction under general controls and arbitrary cumulant orders.

Nevertheless, the core machinery of comb-based QNS, repeated pulse sequences generating selective spectral sampling, has proven flexible, scalable, and experimentally validated across multiple solid-state platforms~\cite{bylanderNoiseSpectroscopyDynamical2011,sungNonGaussianNoiseSpectroscopy2019a,kylewillickEfficientContinuouswaveNoise2018,santiagohernandezgomezNoiseSpectroscopyQuantumclassical2018,guyramonTrispectrumReconstructionNonGaussian2019}. What is lacking is a unified formalism that retains the comb framework's scalability while lifting the self-commutativity constraint and accommodating the richer spectral structure that arises from temporal ordering.


In this work, we approach these limitations by reconsidering the convention of treating bath cumulants as symmetric functions of their time arguments. We find that embedding the causal structure of the integration simplex directly into the bath cumulants, via a product of Heaviside step functions, yields a useful alternative; the time-ordered polyspectra. Because the temporal ordering is handled by the spectrum rather than the control constraints, the filter function does not rely on traditional symmetry requirements to manage time-ordered contributions. Consequently, the multivariate convolution theorem can be naturally applied to a broader range of control sequences. This perspective provides a foundation for a control-centric framework in quantum noise spectroscopy, mapping time-ordered polyspectra directly to the system's evolution projected onto an operator basis, thereby building on the cumulant expansion of~\cite{norrisQubitNoiseSpectroscopy2016c}. For our purposes, this framework is particularly well-suited to accommodating realistic, finite-duration controls and non-commuting dynamics, simplifying the need for highly specialised pulse sequences. Importantly, the required number of distinct base control sequences matches the scaling of established comb-based methods.

Using the control-centric formalism, we provide a systematic method for performing comb spectroscopic tomography of environmental noise. The accompanying protocol can be understood as a form of quantum interferometry, in which the choice of preparation state and measurement axis projects onto a specific signal channel of the system evolution, thereby making it sensitive to particular multi-time bath correlators. With appropriate control-sequence engineering, it is possible to isolate distinct, symmetry-resolved noise components, including their classical or quantum nature, statistical order, and Gaussianity.

We demonstrate the utility of the control-centric approach through numerical simulations across diverse, experimentally relevant scenarios. Our results validate the framework's capacity to characterise environments with both Gaussian and non-Gaussian statistics, while naturally accommodating the non-commuting dynamics that arise from finite-width pulses, multi-axis interactions, and genuinely quantum baths. Additionally, we show that our formalism provides access to new diagnostic information encoded within the complex-valued polyspectra. Specifically, the real and imaginary components of the time-ordered polyspectra, which the Hilbert transform relates to one another, and their comparison serve as a powerful diagnostic tool. Discrepancies between components can be used in an adaptive, open-loop protocol to identify spectral regions that require higher sampling resolution. Our treatment of temporal ordering distinguishes this framework from existing approaches; in particular, a detailed comparison with the Keldysh-ordered cumulant formalism of Ref.~\cite{wangSpectralCharacterizationNonGaussian2020} appears in \secref{subsec:spectral_domain}.

The remainder of the paper follows the reconstruction pipeline. \secref{sec:theory} first projects the measured dynamics onto the $(\alpha,\gamma)$ tomographic channels, where the control functions $y_{a,b}$ build the control matrices $\mathbf{Y}^{(k)}$ and the bath cumulants $\mathcal{C}^{(k)}$ supply the environmental statistics. It then embeds the causal kernel into those cumulants to define the time-ordered polyspectra $\tilde{S}^{(k)}$, constructs the symmetry-dependent sampling domain, and reduces the comb overlap to the linear system $\vec{\mathcal{I}}=\mathbf{G}\vec{\tilde{S}}$. \secref{sec:results} applies this pipeline to Gaussian and non-Gaussian examples, including adaptive Kramers--Kronig refinement, while \secref{sec:discussion} interprets the reconstructed spectral sectors and discusses their broader implications.

\dobib

\paperExternalReferences
\paperStandaloneSection{2}{0}{0}{0}

\section{Theoretical framework}
\label{sec:theory}
We consider a generalised QNS setting, in which an open quantum system with a Hilbert space $\mathcal{H}_S$ of dimension $d_s$ is coupled to an unknown environment (bath) $\mathcal{H}_B$. The evolution of the joint Hilbert space $\mathcal{H}_S\otimes\mathcal{H}_B$ in the laboratory frame is governed by the total Hamiltonian $H^{(\mathrm{lab})}(t) = H_S + H_B + H_{SB}^{(\mathrm{lab})}(t) + H_{\mathrm{ctrl}}^{(\mathrm{lab})}(t)$, where $H_{\mathrm{ctrl}}^{(\mathrm{lab})}(t)$ describes the control Hamiltonian acting non-trivially on $\mathcal{H}_S$. We express the bipartite interaction as
$$H_{SB}^{(\mathrm{lab})}(t)= \sum_{b\in I_b} \Lambda_b \otimes \beta_b(t),$$
where $\{ \Lambda_b | b \in I_b\} \subset \mathcal{B}(\mathcal{H}_S)$ is a set of orthonormal system operators and $\{\beta_b |b\in I_{b}\}\subset\mathcal{B}(\mathcal{H}_B)$ are the associated time-dependent bath operators encoding noise processes which may be classical, quantum or both. The index set $I_b$ labels the bath-coupled system operators, and the choice of these operators designates the interpreted nature of the interaction. For a qubit system where $\left\{\Lambda_b\right\}$ corresponds to the Pauli basis, the choice of coupling axis determines the decoherence mechanism. Bath coupling only along the eigenbasis $\Lambda_b= \sigma_z$ induces a pure dephasing interaction, provided that $[H_S, \sigma_z ] =0$. This special case preserves energy eigenstates while destroying phase coherence. In contrast, transverse coupling $\Lambda_b \in \{\sigma_x,\sigma_y\}$ leads to both relaxation and dephasing due to energy exchange with the bath, as the system operators no longer commute with the free Hamiltonian. 

To isolate the effect of environmental noise, we transform $H^{(\mathrm{lab})}(t)$ into the interaction picture with respect to the free Hamiltonians $(H_S+H_B)$, whereupon bath operators $B_b(t)=U_B^\dagger(t)\beta_b(t)U_B(t)$ now evolve under the free bath dynamics $U_B(t)\equiv\mathcal{T}_+ e^{-\mathrm{i}\int_0^tdsH_B(s)}$. Similarly, the control propagator takes the form, $U_{\mathrm{ctrl}}(t)\equiv\mathcal{T}_+\exp{\!\left(-\mathrm{i}\int_0^t ds\,U_S^\dagger(s)\,H_{\mathrm{ctrl}}^{(\mathrm{lab})}(s)\,U_S(s)\right)}$. Expanding the resulting system operators in a complete orthonormal basis of invertible operators $\{\Lambda_a \,|\, a \in I_a \}$ satisfying $\Tr[\Lambda_a^\dagger \Lambda_{a'}] = d_s\, \delta_{a,a'}$, the effective Hamiltonian takes the form
\begin{equation}
H_I(t) = \sum_{a\in I_a,\,b\in I_b} y_{a,b}(t) \Lambda_a \otimes B_b(t).
\label{eq:H_I}
\end{equation}
Since control generically mixes the bath-coupled operators $\{\Lambda_b |b \in I_b \}$ across all basis directions, the index set $I_a\supseteq I_b$ spans the full operator space. The control-matrix elements
$$y_{a,b}(t)=\frac{1}{d_s}\Tr\left[\Lambda_a^{\dagger}\,U_{\mathrm{ctrl}}^\dagger(t)\, U^{\dagger}_{S}(t)\,  \Lambda_b\, U_{S}(t)\, U_{\mathrm{ctrl}}(t)\right],$$
encode how control redistributes the interaction. We refer to these elements as control functions and collect them in the control matrix $\mathbf{Y}(t)$.

Let $O\in\mathcal{B}(\mathcal{H}_S)$ denote an invertible observable of interest for the system. Provided that the initial system--bath state is uncorrelated, $\rho(0)=\rho_S\otimes\rho_B$, the expectation value of $O$ evolving under the complete open dynamics at time $T$ is given by
\begin{align}
\langle O(T) \rangle = \sum_{\alpha} o_{\alpha}(T)\, \Tr \left[ V_{\Lambda_\alpha}(T)\, \rho_S\, \Lambda_\alpha \right],
\end{align}
where the Hilbert-Schmidt decomposition, applied to the toggling-frame observable
\begin{align*}
\bar{O}(T)
&= U^\dagger_{\text{ctrl}}(T)\, U^{\dagger}_{S}(T)\, O\, U_S(T)\, U_{\text{ctrl}}(T) \\
&= \sum_\alpha o_\alpha(T)\,\Lambda_\alpha,
\end{align*}
is expanded in the orthonormal basis $\{\Lambda_{\alpha}\}$ with $o_\alpha(T) =\Tr[ \Lambda_\alpha^{\dagger} \bar{O}(T)]/d_s$. The operator $
V_{\Lambda_\alpha}(T) = \langle\Tr_B [ \Lambda_\alpha^{-1} \, U_I^\dagger(T)\, \Lambda_\alpha\, U_I(T)\, \rho_B ]\rangle_c$, defined in terms of the propagator $U_I(t)$ generated by $H_I(t)$, encapsulates all bath-mediated decoherence effects accumulated over the evolution. 

The corresponding effective propagator may be expressed as
\begin{equation}
\label{eq:unexpanded V}
V_{\Lambda_{\alpha}}(T)=\left\langle\mathcal{T}_+ e^{-\mathrm{i} \int_{-T}^{T} H_{\Lambda_{\alpha}}(t) d t}\right\rangle_{c, q},
\end{equation}
of the piecewise effective Hamiltonian
\begin{align*}
H_{\Lambda_{\alpha}}(t) = \begin{cases} 
H_1(t), & t \in (0,T] \\
H_0(t), & t \in [-T,0)
\end{cases}
\end{align*} 
where the binary subscript distinguishes the constituent positive and negative time Hamiltonians, respectively defined as
\begin{align*}
H_1(t) &= -\Lambda_{\alpha}^{-1} H_I(T-t) \Lambda_{\alpha}, \\
H_0(t) &= H_I(T+t).
\end{align*}
In the QNS context, this piecewise construction was originally introduced in Ref.~\cite{paz-silvaMultiqubitSpectroscopyGaussian2017a}. In~\appref{app:reduced_dynamics} we demonstrate that this representation naturally arises from the Schwinger--Keldysh closed-time-path formalism~\cite{keldyshDiagramTechniqueNonequilibrium1965,schwingerBrownianMotionQuantum1961} of non-equilibrium field theory. As this approach admits both classical and quantum noise, we denote the average $\langle \cdot \rangle_{c,q} = \left\langle\Tr_{B}[\cdot \rho_B]\right\rangle_c$ as the trace over the environment and, if applicable, an ensemble average over any classical noise realisations on the system. We make no assumptions on the form of $\rho_B$, so long as it is initially uncorrelated with the system; the ensuing formalism accommodates arbitrarily separated initial bath states.

The central task of QNS is summarised in the evaluation of Eq.~\eqref{eq:unexpanded V} as $V_{\Lambda_\alpha}(T)$ governs the decay and phase evolution of the probe coherence in response to an applied control. This problem of determining the dynamics of this operator is typically addressed perturbatively with either Dyson or cumulant (Kubo) expansions~\cite{kuboGeneralizedCumulantExpansion1962, szankowskiEnvironmentalNoiseSpectroscopy2017}. The former expresses $V_{\Lambda_\alpha}(T)$ explicitly as a series of nested time-ordered integrals of multi-time bath correlation functions, while the cumulant expansion reorganises the series into exponentials of connected correlation functions. 

The remainder of this section develops the spectral estimation problem in two stages. \secref{subsec:cumulant_expansion} expands the effective propagator in bath cumulants and projects each order onto the $(\alpha,\gamma)$ tomographic channels, exposing how the control matrix $\mathbf{Y}^{(k)}$ filters the bath statistics $\mathcal{C}^{(k)}$ and how the toggling operator $\lambda_\alpha$ routes distinct symmetry sectors to distinct signal channels. \secref{subsec:spectral_domain} then moves to the frequency domain, where the time-ordering constraint on the integration simplex is the central obstacle to spectral reconstruction under general controls. We resolve it by embedding the causal structure into the bath cumulants through a Heaviside-kernel convolution, defining the time-ordered polyspectra $\tilde{S}^{(k)}$ whose real and imaginary parts satisfy generalised Kramers--Kronig relations. With this embedding in place, $M$-fold periodic repetition of a base control sequence generates a frequency comb that discretises the spectral overlap on the symmetry-dependent effective principal domain, and the reconstruction reduces to the linear system $\vec{\mathcal{I}}=\mathbf{G}\vec{\tilde{S}}$. A summary of the principal symbols is provided in Table~\ref{tab:notation} of \appref{app:notation_derivation}.

\subsection{Cumulant expansion}
\label{subsec:cumulant_expansion}
In this work, we adopt the cumulant expansion for its rapid convergence in weak-coupling regimes and for its compact structure, which naturally reflects the presence of (non-) Gaussian statistics with truncation order ($k\geq3$) $k\leq2$. This convergence is not universal, however: the expansion is systematically extensible in $k$ rather than guaranteed convergent, so for strong coupling or long-lived correlations a finite truncation is adequate only when successive connected contributions decrease over the observation window~\cite{kuboGeneralizedCumulantExpansion1962,norrisQubitNoiseSpectroscopy2016c}, a qualification we return to in \secref{subsec:reconstruction_limitations}. We write the cumulant expansion of $V_{\Lambda_{\alpha}}(T)$ as the exponentiated series
\begin{align*}
V_{\Lambda_{\alpha}}(T) = \exp\left(\sum_{k=1}^\infty (-\mathrm{i})^k \hat{\mathcal{I}}_\alpha^{(k)}\right),
\end{align*}
with $k$th-order contributions given by the integrals 
\begin{align}
    \hat{\mathcal{I}}^{(k)}_\alpha(T) = \int_0^T\!\! d_{>} \vec{t}_{[k]} \sum_{\substack{\vec{j} \in \mathcal{J}_k,\\\vec{l} \in \mathcal{L}_{\vec{j}}}} C^{(k)}\left(H_{\vec{j}}(t_{\vec{l}})\right)\label{eq:cumulant_overlap},
\end{align}
over the $k$th-order cumulants of the effective, piecewise Hamiltonian. We have employed the shorthand notation
$\int_{0}^{T} d_>\vec{t}_{[k]} = \int_{0}^{T} \int^{t_1}_{0} \cdots \int^{t_{k-1}}_{0} dt_1 \cdots dt_k$~\cite{paz-silvaExtendingCombbasedSpectral2019a} to denote the positive-time integration simplex $T\geq t_1\geq\dots\geq t_k\geq0$. The summation variable $\vec{j}\in\mathcal{J}_k\subset\{0,1\}^k$ is a binary vector that indexes admissible sequences of products of toggled Hamiltonians, which have the order $k$, with the constraint that all reverse-time $H_{j_m}$ with $j_m=1$ must be ordered to the left of all forward-time terms with $j_m =0$, preserving the time ordering from the integral in Eq.~\eqref{eq:unexpanded V}. This forms two contiguous blocks of operators within the product $(H_1(t_{l_1})\cdots H_1(t_{l_p})H_0(t_{l_{p+1}})\cdots H_0(t_{l_k}))$, which originate from this reverse--forward time ordering of the integration branches.
For a fixed branch vector $\vec{j}$, each $\vec{l} \in \mathcal{L}_{\vec{j}}$ denotes an admissible sequence of time indices with anti-time ordering for $j_m=1$ (reverse) and normal time ordering for $j_m=0$ (forward). Notably, a degree of `ordering freedom' between the two blocks of operators allows for any ordering between operators associated with the paired arguments $t_{l_p}$ and $t_{l_{p+1}}$. Equivalently, the full order-$k$ set $\mathcal{L}_k = \bigcup_{\vec{j}\in\mathcal{J}_k}\mathcal{L}_{\vec{j}}$ can be systematically generated by enumerating all bipartitions of $\left\{1,2,\dots,k\right\}$ into subsets of cardinality $p$ and $k-p$, then respectively reordering each subset as anti-chronological and chronological.

To systematically analyse the reduced dynamics, we examine each $k$-th order contribution, $\hat{\mathcal{I}}_{\alpha}^{(k)}$, by projecting it onto a complete operator basis $\{\Lambda_\gamma\}$. The phase-free, order-resolved overlaps are
\begin{equation}
    \mathcal{I}^{(k)}_{\alpha,\gamma} =\frac{1}{d_s} \Tr\left[ \hat{\mathcal{I}}^{(k)}_{\alpha} \Lambda_\gamma^{\dagger} \right].
\label{eq:I_projection}
\end{equation}
For a single-qubit Pauli basis, define the selector
$s(\gamma,\alpha)=\delta_{\gamma,\alpha}+\delta_{\gamma,\mathbbm{1}}$, which is
unity for the identity and parallel components and zero for the transverse
components. The real coefficients without an order superscript are then
\begin{align}
\mathcal{I}_{\alpha,\gamma}(T)
&=\sum_{k=1}^{\infty}
(-\mathrm{i})^{k-1+s(\gamma,\alpha)}
\mathcal{I}^{(k)}_{\alpha,\gamma}(T).
\label{eq:I_order_sum}
\end{align}
With $T$ implicit and $\gamma\ne\alpha$ restricted to Pauli axes, the
propagator is
\begin{equation}
V_{\Lambda_{\alpha}}(T)
=\exp\!\Big[
\mathcal{I}_{\alpha,\mathbbm{1}}\Lambda_{\mathbbm{1}}
+\mathcal{I}_{\alpha,\alpha}\Lambda_\alpha
-\mathrm{i}\sum_{\gamma\ne\alpha}
\mathcal{I}_{\alpha,\gamma}\Lambda_\gamma
\Big].
\label{eq:V_lambda}
\end{equation}
Eq.~\eqref{eq:cumulant_overlap} expresses the complete $k$th-order interaction through cumulants of toggling Hamiltonians. To expose how individual control functions couple to the environment, we pass to the moment expansion and refactor the toggling-Hamiltonian cumulants $C^{(k)}(H_{\vec{j}}(t_{\vec{l}}))$ into bath-operator cumulants
\begin{align}
\mathcal{I}^{(k)}_{\alpha,\gamma}(T) = \int_0^T\!\!d_> \vec{t}_{[k]} \sum_{\substack{\mathbf{a},\,\mathbf{b}, \\ \pi,\varpi,\\ \vec{l} \in \mathcal{L}_k}}\bar{\lambda}_{\alpha, \gamma}(\mathbf{a},\pi,\vec{l})\mathbf{Y}^{(k)}_{\mathbf{a},\mathbf{b}}(t_{\vec{l}})\, \mathcal{C}^{(k)}_{\mathbf{b};\vec{l};\pi; \varpi}(\vec{t}_k).\label{eq:I_a_gamma_k}
\end{align}
Here the bath moments are factored back into cumulant form $\mathcal{C}^{(k)}_{\mathbf{b};\vec{l};\pi;\varpi}$ via moment-cumulant expansion
\begin{align*}
    \mathcal{B}^{(k)}_{\mathbf{b};\vec{l};\pi}(\vec{t}_k) = \sum_{\varpi} \mathcal{C}^{(k)}_{\mathbf{b};\vec{l};\pi; \varpi}(\vec{t}_k),
\end{align*}
where, for an ordered partition $\pi=(p_1,\ldots,p_r)$, we have consolidated the $k$th-order product of moments into
\begin{align*}
\mathcal{B}^{(k)}_{\mathbf{b};\vec{l};\pi}
\equiv \prod_{v=1}^{|\pi|}\left\langle \prod_{q\in p_v} B_{b^{(q)}}(t_{l_q})\right\rangle_{c,q}.
\end{align*}
The indices $\mathbf{b}$ specify the respective bath operator axes. The blocks $p_v$ are multiplied in their displayed order, while the indices within each block retain their original order. Thus $\pi\in\mathrm{OP}(\{1,\ldots,k\})$ is an ordered set partition, with non-commuting moment-expansion weight $\mu(\pi)=(-1)^{r-1}/r$ (for a complete derivation see \appref{app:notation_derivation}). The contributions of $\mathcal{C}^{(k)}_{\mathbf{b};\vec{l};\pi;\varpi}(\vec{t}_k)$ toward the reduced dynamics are filtered by the equivalent order control matrix
\begin{align}
\mathbf{Y}^{(k)}_{\mathbf{a},\mathbf{b}}(t_{\vec{l}})
    \;\triangleq\;
    \prod_{u=1}^{k} y_{a^{(u)},b^{(u)}}\!\bigl(t_{l_{u}}\bigr),
\label{eq:Y_k}
\end{align}
such that the control $y_{a^{(u)},b^{(u)}}(t_{l_u})$ is applied along the axis $a^{(u)}$, and modulates the coupling to the bath operator $b^{(u)}$. The effective system response axis is captured by the scalar coefficient, which is defined as the projection onto the basis $\Lambda_\gamma$:
$$
\bar{\lambda}_{\alpha,\gamma}(\mathbf{a},\pi,\vec{l})= \frac{1}{d_s}\Tr  \left[  \left( \sum_{\vec{j} \in\mathcal{J}^{\prime}_{\vec{l}}}\lambda_{\alpha}(\mathbf{a},\pi,\vec{j})\right) \Lambda_{\gamma}^\dagger\right].
$$
The toggling operator, $\lambda_{\alpha}(\mathbf{a},\pi,\vec{j}\,)$, implements an algebraic projection that determines the symmetry of the system--bath interaction. This projection occurs through the adjoint action of the measurement basis operator $\Lambda_\alpha$ on the system operators $\Lambda_\mathbf{a}$. For a qubit system, the conjugate action of the system operators on the measurement observable follows the rule $\Lambda_\alpha\Lambda_{a^{(q)}}\Lambda_\alpha^{-1}=(-1)^{1-s(a^{(q)},\alpha)}\Lambda_{a^{(q)}}$, with the selector introduced above, $s(a,\alpha)=\delta_{a,\alpha}+\delta_{a,\mathbbm{1}}$. This results in the toggling operator,
$$
\lambda_\alpha(\mathbf{a},\pi,\vec{j})  = \mu(\pi )\prod_{v=1}^{|\pi|}\left( \prod_{q \in p_v} (-1)^{j_{q}s(a^{(q)},\alpha)}\Lambda_{a^{(q)}} \right).
$$
For a single-qubit probe, this formalism can be understood as a form of quantum interferometry, where the system evolves along a forward-time branch $(j_q=0)$ and a reverse-time branch $(j_q=1)$. An emergent mechanism for symmetry resolution is encoded in the toggling sign, which controls the interference between these two paths. The alignment of the measurement axis $\alpha$ relative to the control axes $\mathbf{a}$ determines the underlying symmetry of the interaction being probed. For instance, specific alignments can lead to constructive interference, isolating symmetric (anticommutator) correlations, whereas other alignments can produce destructive interference, isolating antisymmetric (commutator) correlations. This interference defines a specific, symmetry-filtered signal channel. The $(\gamma, \alpha)$ decompose the channel into tomographic projectors, allowing for the selective measurement of distinct symmetry-resolved $\mathcal{C}^{(k)}_{\mathbf{b};\vec{l};\pi}$ through appropriately engineered $\mathbf{Y}^{(k)}_{\mathbf{a},\mathbf{b}}(\vec{t}_k)$.
By initialising and measuring the probe in an orthonormal, complete basis, such as $\{\rho_\xi = \left(\Lambda_\xi+\mathbbm{1}\right) / \Tr [\Lambda_\xi+\mathbbm{1}] \}$, the observable response of the $k$th-order bath correlator with respect to some basis may be calculated through algebraic manipulation (for details see \appref{appendix:cbar_calcs}), yielding direct calculations from the right side of Eq.~\eqref{eq:I_a_gamma_k}.

\subsection{Spectral domain of cumulants}\label{subsec:spectral_domain}

The preceding section generalises the well-established formalism of~\cite{norrisQubitNoiseSpectroscopy2016c} by introducing a systematic organisation of the cumulant-control overlap integrals. Calculation reveals that the measured observable is associated with specific control axes, enabling the isolation and subsequent identification of specific bath statistics. This property is highlighted in \secref{subsec:results_gaussian_classical}, where, for a single-qubit pure-dephasing interaction, classical or quantum bath operators are isolated by the chosen control and readout scheme.

Notwithstanding, the central task remains to extract arbitrary-order bath correlators from the reduced dynamics by inverting Eq.~\eqref{eq:I_a_gamma_k}; however, a direct inversion in the time domain is impractical without discretising the associated integral. Non-parametric QNS protocols achieve this in the frequency domain, where comb-based pulse-sequence engineering~\cite{alvarezMeasuringSpectrumColored2011a, biercukDynamicalDecouplingSequence2010, norrisQubitNoiseSpectroscopy2016c, paz-silvaDynamicalDecouplingSequences2016a} recovers noise spectra at discrete harmonic intervals set by the control periodicity. The noise environments that these reconstructions access are, however, restricted to dephasing-preserving interactions, where the temporal symmetry of the integrand in Eq.~\eqref{eq:I_a_gamma_k} admits a direct multivariate convolution.


\subsubsection*{Time ordering in the frequency domain}

Fundamentally, the reduced dynamics that couple the control matrix $\mathbf{Y}^{(k)}_{\mathbf{a},\mathbf{b}}$ and the bath cumulant $\mathcal{C}^{(k)}_{\mathbf{b}}$ in Eq.~\eqref{eq:I_a_gamma_k} occur on the time-ordered integration simplex. Throughout, we take the $k$-variate Fourier transforms of both the control matrix and the bath cumulant to exist. We define the corresponding frequency domain representation of the control matrix over the finite time-ordered domain as the \emph{time-ordered} filter function (FF)


\begin{align*}
    \tilde{F}^{(k)}_{\mathbf{a},\mathbf{b}}(\vec{\omega}_k, T) \equiv \int_0^T\!\! d_{>}\vec{t}_{[k]}\, e^{\mathrm{i}\vec{\omega}_k \cdot \vec{t}_k}\, \mathbf{Y}^{(k)}_{\mathbf{a},\mathbf{b}}(\vec{t}_k),
\end{align*}

where the tilde distinguishes $\tilde{F}^{(k)}_{\mathbf{a},\mathbf{b}}$ from its unrestricted counterpart $F^{(k)}_{\mathbf{a},\mathbf{b}}$. The restricted FF $\tilde{F}^{(k)}_{\mathbf{a},\mathbf{b}}$ does \emph{not} factorise as a product of single-variable transforms, because the simplex constraint couples all frequency arguments. It is this mismatch between the simplex integral defining the overlap and the hypercube integral required by the multivariate convolution theorem that constitutes the primary barrier to frequency-domain QNS under general controls.

To circumvent this barrier, the goal of standard spectral formulations is to trade the time-ordered simplex for an unrestricted frequency integral of the form
\begin{align}
\int_{0}^{T}\!\!\! d_{>}\vec{t}_{[k]}\mathbf{Y}^{(k)}_{\mathbf{a},\mathbf{b}}(\vec{t}_k) \mathcal{C}^{(k)}_{\mathbf{b}}\!(\vec{t}_k)
\!\to\!\!\int_{\mathbb{R}^k}\!\!\frac{d\vec{\omega}_k}{(2\pi)^k}F^{(k)}_{\mathbf{a},\mathbf{b}}(\vec{\omega}_k, T) S^{(k)}_{\mathbf{b}}(\vec{\omega}_k),
\label{eq:time_to_freq}
\end{align}
Here the noise polyspectrum $S^{(k)}_{\mathbf{b}}(\vec{\omega}_k)\!\!\equiv\!\!\int_{\mathbb{R}^{k}}d\vec{t}_ke^{-i\vec\omega_k\cdot\vec{t}_k}\mathcal{C}^{(k)}_\mathbf{b}(\vec{t}_k),$ is the $k$-variate Fourier transform of the bath cumulant, paired with the unrestricted filter $F^{(k)}_{\mathbf{a},\mathbf{b}}(\vec{\omega}_k,T)$ introduced above. The bath spectra follow the negative-exponential Fourier convention and the control filters the dual, positive-exponential convention, so their product in the overlap is evaluated at a common frequency without relabelling. This replacement requires specific symmetry conditions: it holds when both the control matrix and bath cumulants belong to the class of functions symmetric under all permutations of their time arguments, $\mathbf{Y}^{(k)}_{\mathbf{a},\mathbf{b}},\,\mathcal{C}^{(k)}_{\mathbf{b}}\in\mathrm{Sym}(\vec{t}_k)$. Without loss of generality, the prerequisite reduces to self-commutativity of the interaction Hamiltonian, $[H_I(t_i), H_I(t_j)] = 0$ for all times $\{t_i, t_j\} \in [-T, T]$---as is the case for dephasing-preserving interactions (e.g., qubit systems with single-axis coupling to classical noise). Such regimes, however, fall short of those most relevant to contemporary experiments.

Because generic non-commuting dynamics do not satisfy these symmetries, existing QNS protocols operate by actively forcing the simplification $\tilde{F}^{(k)}_{\mathbf{a},\mathbf{b}} \to F^{(k)}_{\mathbf{a},\mathbf{b}}$. These strategies can be broadly divided into two categories. The first imposes conditions on the \emph{system--bath interaction}—either through ad hoc estimation corrections, weak-noise Magnus truncations, or adiabatic and secular approximations~\cite{greenArbitraryQuantumControl2013, sungNonGaussianNoiseSpectroscopy2019a, barnesFilterFunctionFormalism2016, khanMultiaxisQuantumNoise2024}—whereby the dynamics are projected onto an approximate dephasing-preserving regime in which time ordering becomes irrelevant. The second imposes conditions on the \emph{control}, such as comb-based protocols for multi-axis noise~\cite{paz-silvaExtendingCombbasedSpectral2019a}, which explicitly symmetrise the integration domain by restricting the class of pulse sequences. Both strategies are inherently limited: the former by perturbative validity, and the latter by restrictive constraints on admissible controls. Neither extends naturally to a general spectral reconstruction valid for arbitrary pulse sequences, non-commuting dynamics, and arbitrary cumulant orders.

A distinct perspective is offered by the \emph{control-adapted} approach of~\cite{chalermpusitarakFrameBasedFilterFunctionFormalism2020}, which inverts the conventional logic by taking the control's experimental limitations as the starting point for analysis. Rather than symmetrising the filter, this formulation shifts time ordering from the FFs to the bath spectra, extending spectral characterisation to a broader class of controls. This reframing is the principal motivation for the present work. However, the control-adapted formulation relies on a finite-frame condition, and it remains unclear how to achieve a complete spectral characterisation of the bath when this condition is not satisfied.

\subsubsection*{Control-centric resolution: time-ordered polyspectra}

Inspired by the control-adapted interpretation of~\cite{chalermpusitarakFrameBasedFilterFunctionFormalism2020} and the limitations identified above, we resolve the time-ordering problem by reassigning the causal structure of the integration simplex from the FF to the bath cumulants. In the spirit of Kubo's linear response formalism underlying the fluctuation-dissipation theorem~\cite{kuboStatisticalMechanicalTheoryIrreversible1957}, we embed the chronological constraint $T \geq t_1 \geq \dots \geq t_k \geq 0$ through the product of Heaviside step functions $\prod_{j=1}^{k-1} \theta(\tau_j)$, with intervals $\tau_j = t_j - t_{j+1}$, and define the \emph{time-ordered polyspectra} as
\begin{align}
    \tilde{S}^{(k)}_{\mathbf{b}}(\vec{\omega}_k) \equiv \mathcal{F}\left[\prod_{j=1}^{k-1} \theta(\tau_j)\, \mathcal{C}^{(k)}_{\mathbf{b}}(\vec{t}_k)\right].
    \label{eq:S_tilde_def}
\end{align}
The step functions ensure that all orders of response information encoded in the time-dependent stochastic correlators reflect only the influence of perturbations at prior times $t_{j+1} < t_j$. From Titchmarsh's theorem~\cite{titchmarshIntroductionTheoryFourier1937}, for square-integrable functions satisfying $\mathcal{C}^{(k)}_{\mathbf{b}}\!\!\in\!L^2(\mathbb{R}^k)$ with $\operatorname{supp}(\mathcal{C}^{(k)}_{\mathbf{b}}) \subseteq [0, \infty)^k$, imposing forward-time causality via $\theta(\tau)$ guarantees analyticity of $\tilde{S}^{(k)}_{\mathbf{b}}$ in the upper half-plane with at most $(k-1)$ singularities, each located at the origin $\nu_j = 0$. This is reflected in the Fourier transform of the $j$-th kernel
\begin{align}
    \Theta(\nu_j) \equiv \mathcal{F}[\theta(\tau_j)] = \pi\delta(\nu_j) - \mathrm{i}\mathcal{P}(1/\nu_j),
\label{eq:theta_kernel}
\end{align}
where the forward-time structure encoded in the Cauchy principal value integral $\mathcal{P}(1/\nu_j)$ is resolved by the Sokhotski--Plemelj theorem. The time-ordered polyspectra are derived from $(k-1)$ convolutions of the $\Theta(\nu)$ kernel with the unordered spectrum
\begin{align} 
\tilde{S}^{(k)}_{\mathbf{b}}(\vec{\omega}_{k})
&=\int_{\mathbb{R}^{k-1}}
\frac{d\vec{\nu}_{k-1}}{(2\pi)^{k-1}}
\prod_{j=1}^{k-1}\Theta(\nu_j)\,
S^{(k)}_{\mathbf{b}}\!\left(
\vec{\omega}_k+\mathbf{T}\vec{\nu}_{k-1}
\right)
\label{eq:S_ordered}
\end{align}
where $\mathbf{T}$ is a $(k \times (k-1))$ lower bidiagonal transformation matrix with non-zero elements $T_{i,i} = -1$ and $T_{i+1,i} = 1$. Each $\Theta(\nu_j)$ encodes the causal constraint $\theta(\tau_j)$ through auxiliary frequencies $\nu_j$, which sequentially mediate spectral correlations across $S^{(k)}_{\mathbf{b}}$ with each convolution conditioned on the outcomes of the preceding $(j-1)$ integrations. In analogy to the fluctuation-dissipation theorem, the kernel structure partitions the frequency response for each $\tau_j$ interval: the coherent component $\pi\delta(\nu_j)$ preserves the equilibrium fluctuations of the unordered spectrum $S^{(k)}_{\mathbf{b}}$. In contrast, the dispersive component $-\mathrm{i}\mathcal{P}(1/\nu_j)$ generates the causal response through the Hilbert transform. Accordingly, the real and imaginary parts of $\tilde{S}^{(k)}_{\mathbf{b}}$ are necessarily related~\cite{kuboStatisticalMechanicalTheoryIrreversible1957} and together satisfy a $(k-1)$-dimensional generalisation of the Kramers--Kronig relations
\begin{align}
\mathrm{Im}[
\tilde{S}^{(k)}_{\mathbf{b}}(\vec{\omega}_{k})
]\!=\!-\mathcal{P}\!\!
\int_{\mathbb{R}^{k-1}}\!\!\!\frac{d\vec{\nu}_{k-1}}{\pi^{k-1}}\!
\prod_{j=1}^{k-1}\!\mathrm{Re}[
\tilde{S}^{(k)}_{\mathbf{b}}
\!(\vec{\omega}_k\!+\!\mathbf{T}\vec{\nu}_{k-1})
]/\nu_j,
\label{eq:kramers_kronig}
\end{align}
where $\mathrm{Re}[\tilde{S}^{(k)}_{\mathbf{b}}]$ follows from interchanging $\mathrm{Re} \leftrightarrow \mathrm{Im}$ and negating the result. Full knowledge of either the real or imaginary part of the time-ordered spectra thus allows direct calculation of the other. When only partial information is extracted, however, estimating both components provides a more reliable characterisation, as further illustrated in \secref{sec:results}.

At second order ($k=2$), the unordered spectra of the Hermitian bath operators considered here are both real but have opposite parity. The anticommutator correlator is real and even in the time lag, yielding a real, even $S^{(+)}(\omega)$; the commutator correlator is imaginary and odd, yielding a real, odd $S^{(-)}(\omega)$. Consequently, $\mathrm{Im}[\tilde{S}^{(\pm)}(\omega)]$ arises entirely from the causal embedding and represents the Hilbert-transform shadow of $\mathrm{Re}[\tilde{S}^{(\pm)}(\omega)]$. Although the Hilbert transform yields no independent information about the bath when the real part is fully known, under discrete and finite sampling, it encodes novel temporal correlation structures that are useful for diagnosing sub-harmonic spectral features~\cite{Skingle1974}.

As shown in \secref{sec:results}, this causal-shadow structure extends to higher cumulant orders through the multidimensional Kramers--Kronig relations~\eqref{eq:kramers_kronig}, where $\mathrm{Im}[\tilde{S}^{(k)}_{\mathbf{b}}]$ arises from the reconstruction procedure itself rather than from the underlying dynamics.

Equipped with the time-ordered polyspectra $\tilde{S}^{(k)}_{\mathbf{b}}$ defined in~\eqref{eq:S_tilde_def}, non-commuting contributions to the reduced dynamics $\mathcal{I}^{(k)}_{\alpha,\gamma}(T)$ are confined to the temporally asymmetric integration region generated by the dispersive component $\mathrm{Im}[\tilde{S}^{(k)}_{\mathbf{b}}]$. Consequently, the FF $F^{(k)}_{\mathbf{a},\mathbf{b}}$ is liberated from ordering restrictions, and the $k$th-order convolution theorem thus holds for arbitrary controls (cf.\ \appref{app:ft_comb}). This yields the $\mathcal{I}^{(k)}_{\alpha,\gamma}(T)$ dual
\begin{align}
\mathcal{I}^{(k)}_{\alpha,\gamma}(T)&=\sum_{\mathbf{a},\mathbf{b}} \frac{\bar{\lambda}_{\alpha,\gamma}(\mathbf{a})}{(2\pi)^k}\int_{\mathbb{R}^k}\!d\vec{\omega}_k\,F^{(k)}_{\mathbf{a},\mathbf{b}}(\vec{\omega}_k,T)\tilde{S}^{(k)}_{\mathbf{b}}(\vec{\omega}_k).
\label{eq:I_a_k_spectra}
\end{align}
Here and in the comb-sampling equations that follow, $\sum_{\mathbf{a},\mathbf{b}}$ abbreviates the full summation over $\mathbf{a}$, $\mathbf{b}$, $\pi$, $\varpi$, and $\vec{l}\in\mathcal{L}_k$, while $\bar{\lambda}_{\alpha,\gamma}(\mathbf{a})$ and the spectra $S^{(k)}_{\mathbf{b}}$, $\tilde{S}^{(k)}_{\mathbf{b}}$ suppress the associated $(\pi,\varpi,\vec{l})$ sector labels carried by each summand.

For stationary noise environments, the $k$th-order cumulants are translationally invariant, $\mathcal{C}^{(k)}_{\mathbf{b}}(\vec{t}_k) = \mathcal{C}^{(k)}_{\mathbf{b}}(\bar{\vec{\tau}}_{k-1})$, depending only on the vector of $(k-1)$ time separations $\bar{\tau}_i \equiv t_i - t_k$ for $i \in \{1, \dots, k-1\}$ rather than absolute times. The overbar distinguishes these final-coordinate lags from the successive differences $\tau_j=t_j-t_{j+1}$ used in Eq.~\eqref{eq:S_tilde_def}; either coordinate system is valid provided it is maintained consistently. Setting $|\vec{\omega}_k| \equiv \sum_{i=1}^k \omega_i$, time-translation invariance implies that in a distributional sense
\begin{align}
    \tilde{S}^{(k)}_{\mathbf{b}}(\vec{\omega}_k) = 2\pi\delta(|\vec{\omega}_k|)\, \tilde{S}^{(k)}_{\mathbf{b}}(\vec{\omega}_{k-1}),
\label{eq:stationary_cond}
\end{align}
and~\eqref{eq:I_a_k_spectra} reduces accordingly to an integral over $(k-1)$ independent frequencies, with the delta constraint fixing $\omega_k=-|\vec{\omega}_{k-1}|$.

The control-centric formalism recasts causality and non-commuting interactions within the reduced dynamics as intrinsic properties of the environment, formally characterised by the real and imaginary Hilbert-conjugate components of the time-ordered polyspectra $\tilde{S}^{(k)}_{\mathbf{b}}(\vec{\omega}_{k})$. The observed system response under arbitrary control isolates symmetry-resolved projections of $\tilde{S}^{(k)}_{\mathbf{b}}(\vec{\omega}_{k})$ that selectively couple to specific channels within the $(\alpha,\gamma)$ tomographic decomposition. This approach offers three key advantages: (i) the FF $F^{(k)}_{\mathbf{a},\mathbf{b}}$ is now an unrestricted dual Fourier transform, adapted to practical control capabilities without restriction; (ii) the formalism naturally accommodates quantum environments with non-commuting Hamiltonians; and (iii) the imaginary component $\mathrm{Im}[\tilde{S}^{(k)}_{\mathbf{b}}(\vec{\omega}_k)]$ captures temporal signatures analogous to impulse-response in the reduced dynamics, providing a diagnostic to validate the sampling protocol and enable the detection of instantaneous frequency shifts and sharp sub-resolution dispersive features, as demonstrated in \secref{subsec:results_gaussian_classical}.

The present framework is related to the Keldysh-ordered quantum polyspectra introduced by Wang and Clerk~\cite{wangSpectralCharacterizationNonGaussian2020}, who defined operationally meaningful polyspectra via the forward--backward evolution structure of qubit decoherence. Both approaches recover $\tilde{S}^{(\pm)}(\omega)$ at second order. At higher orders, however, the frameworks differ in their treatment of temporal structure: the Keldysh-ordered cumulant kernels remain permutation-symmetric in their time arguments, whereas the time-ordered polyspectra introduced here break this symmetry by construction through the $\theta(\tau_j)$ embedding. Related in spirit is the Keldysh self-energy approach of Huang \textit{et al.}~\cite{Huang2023KeldyshMap}, where $\Sigma(\tau)=\ln\Pi(\tau)$ plays the role of a superoperator-valued ordered-cumulant generator for the reduced map. Unlike that map-based construction, however, our framework remains polyspectral, with temporal ordering embedded directly into the bath spectra via the $\theta(\tau_j)$ kernels. Together, these distinctions provide a complementary route to complex-valued spectra that integrates naturally with the comb-based reconstruction developed below and the multi-axis filter design presented in \secref{sec:results}.

\subsubsection{Frequency-comb spectroscopy}
We have shown how the detailed framework represents these system dynamics spectrally using time-ordered polyspectra. We now establish its connection to the frequency-comb approach introduced by Alvarez and Suter~\cite{alvarezMeasuringSpectrumColored2011a}, extending their methodology to arbitrary controls while addressing key limitations that emerge in realistic experimental scenarios: non-commuting system--bath dynamics, and the preservation of temporal correlation information encoded in $\tilde{S}^{(k)}$. Since our time ordering is embedded into the spectrum, the FF is free to take any form, so long as it yields an adequate discretisation of the integral. The key insight behind comb-based spectroscopy is that repeating any control sequence generates an FF with ``comb-like'' features, enabling selective sampling of the noise spectrum. For stationary noise processes, when a base control sequence of duration $\tau_c$ is repeated $M\gg1$ times, the spectral overlap integrand may be written as the product~\cite{norrisQubitNoiseSpectroscopy2016c}
\begin{align}
\mathcal{I}^{(k)}_{\alpha,\gamma}(M\tau_c)
&=\sum_{\mathbf{a},\mathbf{b}} \frac{\bar{\lambda}_{\alpha,\gamma}(\mathbf{a})}{(2\pi)^{k-1}}\!\! \int_{\mathbb{R}^{k-1}}\!\!\!d\vec{\omega}_{k-1}\,
\Sha(\vec{\omega}_{k-1},M\tau_c)
\nonumber\\[-2pt]
&\quad\quad\quad\quad\quad\quad\quad\quad\times
F^{(k)}_{\mathbf{a},\mathbf{b}}(\vec{\omega}_{k-1})
\tilde{S}^{(k)}_{\mathbf{b}}(\vec{\omega}_{k-1}),
\label{eq:comb_overlap}
\end{align}
consisting of the $M=1$ \textit{fundamental} FF $F^{(k)}_{\mathbf{a},\mathbf{b}}(\vec{\omega}_{k-1})\equiv F^{(k)}_{\mathbf{a},\mathbf{b}}(\vec{\omega}_{k-1},\tau_c)$ and the universal ``comb function''
$$\Sha(\vec{\omega}_{n}, M\tau_c) = \frac{\sin(|\vec{\omega}_{n}|M\tau_c/2)}{\sin(|\vec{\omega}_{n}|\tau_c/2)}\prod_{j=1}^{n}\frac{\sin(\omega_jM\tau_c/2)}{\sin(\omega_j\tau_c/2)}.$$

For a smooth test function $H(\vec{\omega}_{k-1})$ and sufficiently large $M$, the $\Sha(\vec{\omega}_{k-1}, M\tau_c )$ behaves asymptotically like a multidimensional Dirac comb
\begin{align*}
    \int_{\mathbb{R}^{n}}d\omega_{n} \Sha(\vec{\omega}_n, M\tau_c )H(\vec{\omega}_{n}) \approx M\omega_h^k\!\sum_{\vec{s}_{n}\in\mathcal{R}_{n}} H(\,\vec{s}_{n}\,),
\end{align*}
where $\mathcal{R}_{n}\equiv \{\vec{\nu} \cdot \omega_h \,|\, \vec{\nu} \in \mathbb{Z}^{n}\}$ defines a uniform sampling lattice in $\mathbb{R}^{n}$ with spacing set by the fundamental comb frequency $\omega_h = 2\pi/\tau_c$. This comb structure was first observed in the $k=2$ case in~\cite{ajoyOptimalPulseSpacing2011a,alvarezMeasuringSpectrumColored2011a}, later extended to arbitrary $k$ in~\cite{lukaszcywinskiDynamicaldecouplingNoiseSpectroscopy2014}, and subsequently exploited in~\cite{norrisQubitNoiseSpectroscopy2016c} to deal with models of increasing generality, including non-Gaussian and quantum bosonic bath models in multi-axis and multi-qubit settings~\cite{paz-silvaMultiqubitSpectroscopyGaussian2017a, paz-silvaExtendingCombbasedSpectral2019a, yuanlongwangBroadbandSpectroscopyQuantum2024}.

All real noise processes exhibit structural symmetries that spectroscopy protocols exploit to avoid redundant sampling. The minimal representative region, known as the principal domain $\mathcal{D}_{k-1}\subset\mathbb{R}^{k-1}$, is constructed by identifying the symmetry group $\mathrm{Sym}_{\mathbf{b}}$, under which $S_\mathbf{b}^{(k)}$ is invariant, partitioning frequency space into the orbits this group generates, and retaining a single representative from each orbit~\cite{chandranGeneralProcedureDerivation1994,NikiasPetropulu1993HigherOrderSpectra}. Symmetries under conjugation, $S^{(k)}_{\mathbf{b}}(\vec{\omega}_{k-1}) = S^{(k)}_{\mathbf{b}}(-\vec{\omega}_{k-1})^{*}$, and stationarity, which from~\eqref{eq:stationary_cond} fixes $\omega_k = -|\vec{\omega}_{k-1}|$, are two generic transformations that respectively emerge from properties intrinsic to the Fourier transform and the underlying noise model. For wide-sense stationary classical noise processes, $\mathrm{Sym}_{\mathbf{b}}$ additionally includes symmetry under permutations $S_\mathbf{b}^{(k)}(\ldots,\omega_i,\ldots,\omega_j,\ldots)=S_\mathbf{b}^{(k)}(\ldots,\omega_j,\ldots,\omega_i,\ldots)$ as a tautological consequence of commuting random variables~\cite{NikiasPetropulu1993HigherOrderSpectra}. Existing spectroscopy protocols have predominantly operated within dephasing-preserving regimes, where the full $\mathrm{Sym}_{\mathbf{b}}$ is retained and reconstruction on $\mathcal{D}_{k-1}$ is sufficient. More generally, however, the bath is only ever accessed through its coupled interaction with the control, so $\mathcal{D}_{k-1}$ is set by the symmetries the bath and control jointly retain, and any reconstruction protocol must account for both. The control-centric framework does so by construction, representing the reduced dynamics through symmetry-dependent projections of the time-ordered bath correlators under arbitrary control.

\begin{figure}[!htbp]
\includegraphics[width=\columnwidth]{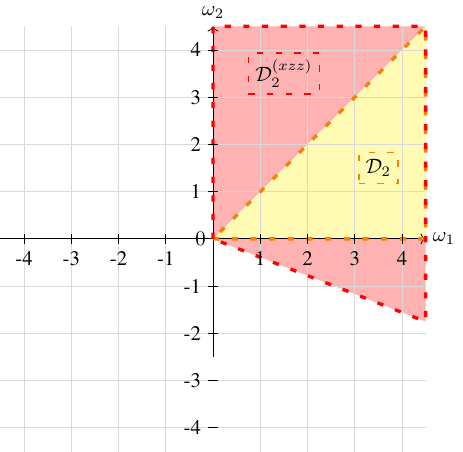}
    \caption{%
        Two-dimensional frequency-space representation of the time-ordered
        bispectrum.
        The plotted domain illustrates how the differing symmetries of the FF alter the principal domain of the sampled space.
        Distinct symmetry-breaking features highlight the importance of carefully considering  non-instantaneous controls and the embedding of time-ordered spectral
        structure in $\tilde{S}^{(2)}(\omega_1,\omega_2)$.
    }
    \label{fig:principal_domain}
\end{figure}

We formalise this through the \textit{effective principal domain} $\mathcal{D}^{(\mathbf{a},\mathbf{b})}_{k-1}\supseteq\mathcal{D}_{k-1}$, the minimal representative region of the reduced dynamics rather than of the bath alone. The observable response couples the bath cumulant to the control through the filter $F^{(k)}_{\mathbf{a},\mathbf{b}}$, and only those elements of $\mathrm{Sym}_{\mathbf{b}}$ that the filter also respects survive to constrain the reconstruction. Conjugation and stationarity are inherited unchanged, whereas the permutation content is retained only insofar as the control axes are left invariant under argument exchange. The surviving transformations form a subgroup $\mathrm{Sym}_{\mathbf{a},\mathbf{b}}\subseteq\mathrm{Sym}_{\mathbf{b}}$, and $\mathcal{D}^{(\mathbf{a},\mathbf{b})}_{k-1}$ is the corresponding representative region, obtained by the same orbit construction applied to $\mathrm{Sym}_{\mathbf{a},\mathbf{b}}$ in place of the full group (\appref{app:effective_domain}). Fewer transformations remain to fold frequency space, so the effective region is no smaller than the standard one, with equality only when the control preserves the whole of $\mathrm{Sym}_{\mathbf{b}}$. The effective principal domain thus delineates the boundary of \textit{observability} under a given protocol, not the boundary of what exists in the environment; spectral content that the bath symmetries alone would render redundant becomes, once the control breaks those symmetries, independent data to be reconstructed on the enlarged region. This aligns with the control-adapted formalism of Ref.~\cite{chalermpusitarakFrameBasedFilterFunctionFormalism2020}, where the learnable noise projections are fixed by the control frame rather than by the full spectral content. The group $\mathrm{Sym}_{\mathbf{a},\mathbf{b}}$ depends on how the control axes are assigned to the ordered arguments, so the domain does too; permuting that assignment reorders the orbits and yields a distinct region, $\mathcal{D}^{(xzz)}_2\neq\mathcal{D}^{(zxz)}_2$. As illustrated in Fig.~\ref{fig:principal_domain}, single-axis control $\mathbf{a}=zzz$ preserves the full permutation symmetry of the bath and reproduces the standard domain, $\mathcal{D}_2^{(zzz)}\equiv\mathcal{D}_2$, whereas multi-axis control $\mathbf{a}=xzz$ breaks it and enlarges the effective domain, $\mathcal{D}^{(xzz)}_2\supset\mathcal{D}^{(zzz)}_2$.

Provided a justified truncation to the finite set, $\Omega_{k-1}^{(\mathbf{a},\mathbf{b})}\subset\mathcal{D}_{k-1}^{(\mathbf{a},\mathbf{b})}$ and $\tilde{S}^{(k)}_{\mathbf{b}}$ is sufficiently smooth, the frequency-comb approximation yields the finite polynomial (see \appref{app:ft_comb} for details)
\begin{align}
\lim_{M \gg 1} \mathcal{I}^{(k)}_{\alpha, \gamma}(M\tau_c)
\approx \frac{M}{(\tau_c)^{k-1}}\!\!\!\!&\sum_{\substack{\mathbf{a},\mathbf{b}\\ \vec{s}_{k-1} \in\, \Omega_{k-1}^{(\mathbf{a},\mathbf{b})}}}\!\!\!\bar{\lambda}_{\alpha,\gamma}(\mathbf{a})m_{\mathbf{a},\mathbf{b}}(\vec{s}_{k-1})\nonumber\\
&\quad\quad\quad\,\times F^{(k)}_{\mathbf{a},\mathbf{b}}(\vec{s}_{k-1})
\tilde{S}^{(k)}_{\mathbf{b}}(\vec{s}_{k-1}),
\label{eq:spectrum_comb_estimate}
\end{align}
where the multiplicity factor $m_{\mathbf{a},\mathbf{b}}(\vec{s}_{k-1})\equiv \text{card}\{\vec{r}_{k-1} \in \mathcal{R}_{k-1} \mid \exists\, g \in \text{Sym}_{\mathbf{a},\mathbf{b}}: g(\vec{r}_{k-1}) = \vec{s}_{k-1}\}$ denotes the cardinality of the equivalence class of points in the full domain $\mathcal{R}_{k-1}$ that, under a transformation in the symmetry group $\text{Sym}_{\mathbf{a},\mathbf{b}}$, map to points in the non-redundant region $\vec{s}_{k-1}\in\mathcal{D}_{k-1}^{(\mathbf{a},\mathbf{b})}$. In many practical settings, noise correlations have finite correlation time (are absolutely integrable), in which case the corresponding polyspectra decay at large frequencies: $\lim_{\|\vec{s}\| \to \infty} \tilde{S}^{(k)}_{\mathbf{b}}(\vec{s}) = 0$. This property justifies imposing a high-frequency cut-off, $\omega_{\mathrm{max}}$, which bounds the span of relevant reconstruction frequencies as $\text{span}(\vec{s}) = \{\vec{s}_{k-1} \in \Omega_{k-1}^{(\mathbf{a},\mathbf{b})} \mid |s_i| \le \omega_{\mathrm{max}}, \forall\, i \in \{1, \dots, k-1\}\}$, where the total number of harmonic coordinates that need to be estimated within this bounded sampling domain is defined as $N_h$.
\subsubsection{Spectral reconstruction}
By applying $M\gg1$ repetitions of $N_s \geq N_h$ distinct ``base'' control sequences, Eq.~\eqref{eq:spectrum_comb_estimate} gives the phase-free matrix equation
\begin{align*}
\vec{\mathcal{I}}^{(k)}_{\alpha,\gamma}
&=\sum_{\mathbf{a},\mathbf{b}}\!\bar{\lambda}_{\alpha,\gamma}(\mathbf{a})\,\mathbf{G}_{\mathbf{a},\mathbf{b}}\,
\vec{\tilde{S}}^{(k)}_{\mathbf{b}}.
\end{align*}
Tomography directly yields the real coefficient vector
$\vec{\mathcal{I}}_{\alpha,\gamma}$.  When the protocol isolates a single
cumulant order $k$, the two vectors are related by the known phase
$\vec{\mathcal{I}}_{\alpha,\gamma}
=(-\mathrm{i})^{k-1+s(\gamma,\alpha)}
\vec{\mathcal{I}}^{(k)}_{\alpha,\gamma}$; when several orders contribute,
their phase-weighted sum is taken according to Eq.~\eqref{eq:I_order_sum}.
The vector $\vec{\tilde{S}}^{(k)}_{\mathbf{b}}$ contains the polyspectrum values at
the $N_h$ harmonics sampled from the effective principal domain,
$\Omega_{k-1}^{(\mathbf{a},\mathbf{b})}
=\{\vec{s}_1,\dots,\vec{s}_{N_h}\}$. The $(N_s\times N_h)$ matrix
$\mathbf{G}_{\mathbf{a},\mathbf{b}}$ collects the control-filter row vectors,
with elements
\begin{align*}
[\mathbf{G}_{\mathbf{a},\mathbf{b}}]_{j,i} = \frac{M}{\tau_{c,j}^{k-1}} m_{\mathbf{a},\mathbf{b}}(\vec{s}_i) F^{(k)[j]}_{\mathbf{a},\mathbf{b}}(\vec{s}_i).
\end{align*}
Provided a set of $N_s$ control FFs with sufficiently disjoint spectral support on $\Omega_{k-1}^{(\mathbf{a},\mathbf{b})}$ that together formulate a well-conditioned $\mathbf{G}_{\mathbf{a},\mathbf{b}}$, the noise spectra may be reconstructed from standard numerical matrix inversion methods.

Obtaining precise polyspectra characterisation is a non-trivial problem that requires tailored design of FFs through an appropriate control optimisation protocol. While the FF matrix must exhibit a low condition number, $\kappa(\mathbf{G}_{\mathbf{a},\mathbf{b}})= \mathcal{O}(1)$ for accurate computation of its inverse, formulating an FF optimisation objective as $\min\kappa(\mathbf{G}_{\mathbf{a},\mathbf{b}})$ is generally insufficient for a low error reconstruction of $\tilde{S}^{(k)}_{\mathbf{b}}$. Instead, optimal FF design requires a multi-criteria approach that simultaneously addresses several competing objectives, collectively enforcing a diagonally dominant FF matrix $\mathbf{G}_{\mathbf{a},\mathbf{b}}\propto\mathbbm{1}$, from which control-FF linear independence follows by construction. In particular, to reliably reproduce $\tilde{S}^{(k)}_{\mathbf{b}}$ an optimal control protocol must realise FFs with: (i) band-limited spectral support within $\text{span}(\vec{s})$ to minimise leakage beyond $\omega_{\mathrm{max}}$; (ii) diagonal dominance such that each FF achieves maximal normalised response at a unique harmonic coordinate, with normalised matrix elements satisfying $|[\mathbf{G}_{\mathbf{a},\mathbf{b}}]_{j,i}|/\max_q |[\mathbf{G}_{\mathbf{a},\mathbf{b}}]_{j,q}| \approx \delta_{j,i}$ (after reordering); and (iii) when the zero-frequency harmonic $\vec{0}_{k-1}$ is included in $\Omega_{k-1}^{(\mathbf{a},\mathbf{b})}$, appropriate attenuation of the corresponding FF amplitude such that $|[\mathbf{G}_{\mathbf{a},\mathbf{b}}]_{\cdot,\vec{0}_{k-1}}|  \lesssim\,\max_{q \neq 0} |[\mathbf{G}_{\mathbf{a},\mathbf{b}}]_{\cdot,q}|$ prevents biasing the reconstruction toward the DC component and coupling to higher-order cumulants. Since these criteria constrain the geometry of $\mathbf{G}_{\mathbf{a},\mathbf{b}}$ rather than the spectrum being reconstructed, they are agnostic to the underlying control family; any protocol whose objective enforces them provides a sufficient, though not unique, route to accurate polyspectra reconstruction.

\dobib

\paperExternalReferences
\paperStandaloneSection{3}{18}{1}{0}

\section{Numerical demonstrations}
\label{sec:results}

We validate the proposed control-centric framework by characterising the leading-order time-ordered polyspectra from numerical simulations of a single-qubit probe interacting with a dephasing environment. The recovered polyspectra reflect experimentally realistic system--bath dynamics drawn from the physical constants of Ref.~\cite{sungNonGaussianNoiseSpectroscopy2019a}. By employing arbitrary control sequences matched to physically meaningful sampling rates, the simulations explicitly generate the non-dephasing dynamics induced by finite-duration pulses. Our approach resolves these effects ab initio, recovering the dispersive spectral features that arise from the underlying causal ordering.

\subsubsection*{System configuration}
The probe qubit evolves under the system Hamiltonian $H_S = (\omega_s/2)\Lambda_z \otimes \mathbbm{1}$ with resonant control applied in the $xy$-plane
\begin{equation}
H_{\mathrm{ctrl}}^{(\mathrm{lab})}(t) = \frac{f(t)}{2}\left[\Lambda_x\sin(\omega_s t \!+\! \phi(t))\! +\! \Lambda_y\cos(\omega_s t\! +\! \phi(t))\right],
\end{equation}
where $f(t)$ and $\phi(t)$ denote the envelope and phase functions, respectively. The probe couples to the environment via the bath operator $B(t)$ through a generalised dephasing interaction $H_{SB}^{(\mathrm{lab})}(t) = \Lambda_z \otimes B(t)$ satisfying $\left[H_S,H_{SB}^{(\mathrm{lab})}(t)\right]=0$. The control protocol may also apply a global rotation $R_y(\phi_R)$ immediately before and after the controlled evolution, with sequence-dependent values specified in \appref{app:control_sequences}.

Working in the rotating frame with $\phi(t) = 0$ for $y$-axis control, the control-toggling transformation $U_{\mathrm{ctrl}}(t) = e^{-\mathrm{i}\varphi(t)\Lambda_y/2}$ with $\varphi(t) = \int_0^t f(s)ds$ yields the effective Hamiltonian
\begin{equation}
H_I(t) = \sum_{a \in \{x,z\}} y_{a,z}(t) \Lambda_a\otimes B(t),
\end{equation}
where the control functions $y_{x,z}(t) = -\sin(\varphi(t))$ and $y_{z,z}(t) = \cos(\varphi(t))$ emerge from the interaction picture rotated with the control unitary. Idealised QNS protocols treat $f(t)$ as a sequence of instantaneous $\pi$-pulses, confining the interaction to the longitudinal axis with $y_{z,z}(t)=\pm 1$ and $y_{x,z}(t)=0$. However, finite-duration control pulses induce continuous evolution of $\varphi(t)$, generating transverse control leakage $y_{x,z}(t)\neq0$ that couples $B(t)$ to the non-commuting operator $\Lambda_x$. Within the control-centric framework, the longitudinal and transverse control axes together manifest causal correlations that the $\Theta$-kernel contributions encode in the time-ordered polyspectra.

We apply our framework to analyse this non-commuting scenario using the split Pauli-basis form of the effective propagator in Eq.~\eqref{eq:V_lambda}. The decomposition index $\gamma$ specifies the axis along which the effective noise generator acts. The unsuperscripted quantities $\mathcal{I}_{\alpha,\gamma}(T)$ are the real coefficients in the exponent. The phase-free, order-resolved overlaps
$\mathcal{I}^{(k)}_{\alpha,\gamma}(T)$ given in \appref{app:direct_calc_3} are converted to these coefficients by
Eq.~\eqref{eq:I_order_sum}.

The coefficients $\mathcal{I}_{\alpha,\gamma}(T)$ are extracted through state tomography. For each measurement axis $\alpha$, the qubit is initialised in states $\rho_\xi = (\Lambda_\xi + \mathbbm{1})/\Tr[\Lambda_\xi + \mathbbm{1}]$ for $\xi \in \{\mathbbm{1}, x, y, z\}$, yielding expectation values $\langle \Lambda_\alpha \rangle_\xi$. The four preparation--measurement configurations per $\alpha$ form a linear system (detailed in \appref{appendix:cbar_calcs}) that, upon inversion, yields the four coefficients $\{\mathcal{I}_{\alpha,\mathbbm{1}}, \mathcal{I}_{\alpha,x},\mathcal{I}_{\alpha,y}, \mathcal{I}_{\alpha,z}\}$. Exhaustive characterisation across all three measurement axes $\alpha \in \{x,y,z\}$ therefore requires at least twelve independent preparation--measurement configurations, comprising a minimum of four for each measurement axis. Additional configurations may be included to form a redundant, overcomplete tomographic dataset. In practice, a minimal reconstruction task targeting a specific $\alpha$ uses four configurations; for instance, the classical Gaussian spectrum reconstruction of \secref{subsec:results_gaussian_classical} uses $\alpha = x$ alone, while \secref{subsec:results_quantum} requires all $\alpha \in \{x,y,z\}$ to characterise the non-Gaussian quantum noise polyspectra. The phase-free overlap expressions up to order $k=3$, provided in \appref{app:direct_calc_3}, collectively govern the dynamics in the following subsections. The bath correlations enter the reduced dynamics as anticommutator and commutator bracket cumulants: i.e.\ at second order, $\{B(t_1),B(t_2)\}_C^\pm\equiv C(B(t_1), B(t_2))\pm C(B(t_2), B(t_1))$, when Fourier transformed and convolved with $\theta(\tau_1)$, yield time-ordered spectra $\tilde{S}^{(\pm)}(\omega)$, while higher orders involve nested combinations of these brackets with corresponding polyspectra $\tilde{S}^{(\pm\pm\dots)}(\omega_1,\omega_2,\dots)$. For notational simplicity, the cumulant order $k-1$ has been suppressed in favour of the number of $\pm$-superscripts. Here, the $+$ and $-$ are read left to right and denote the bracket type from inner to outer commutator brackets, with brackets always left-nested.

We apply the control-centric framework to reconstruct the time-ordered spectra in two complementary scenarios. In both cases, finite-duration pulses induce multi-axis control leakage, generating time-ordered interactions through non-commuting operators in the system Hilbert space. The second scenario accounts for additional time-ordered dynamics arising from non-commuting, multi-axis operators within the bath Hilbert space.
\secref{subsec:results_gaussian_classical} analyses a classical Gaussian process with sharp spectral features. Reconstruction of both real and imaginary components of the time-ordered spectrum $\tilde{S}^{(+)}(\omega)$ demonstrates how the Hilbert transform relationship provides diagnostic information for adaptive sampling strategies.
\secref{subsec:results_quantum} addresses a quantum bath exhibiting both non-Gaussian statistics and non-commuting dynamics. We reconstruct the classical and quantum spectra $\tilde{S}^{(\pm)}(\omega)$ and introduce a systematic protocol to characterise symmetry-dependent principal domains of complex polyspectra, exemplified through reconstruction of $\tilde{S}^{(++)}(\omega_1,\omega_2)$.

\subsection{Gaussian, classical noise}\label{subsec:results_gaussian_classical}
We consider the evolution of a qubit probe over the time interval $t\in[0,T]$ subject to a dephasing interaction with a stationary, classical Gaussian process $B(t)=\beta(t)\otimes\mathbbm{1}$ with zero-mean, $\langle B(t)\rangle=0$. Our numerical simulations sample $\beta(t)$ from the spectral profile shown in Fig.~\ref{fig:gauss_ordered_spectrum}(a). We embed a sharp Gaussian feature within the interval $(3\omega_h,4\omega_h)$ with spectral width narrower than the comb spacing $\omega_h$. Such features can arise in qubit systems due to unwanted couplings, resonant frequencies, or crosstalk between qubits~\cite{romachSpectroscopySurfaceinducedNoise2015,zhaoMitigationQuantumCrosstalk2023}. We introduce this scenario to demonstrate the framework's ability to resolve sub-harmonic features below the frequency-comb sampling resolution.

The zero-mean, stationary, Gaussian statistics of $\beta(t)$ are fully characterised by the $(k=2)$-order spectrum $\tilde{S}^{(+)}(\omega)$ shown in Fig.~\ref{fig:gauss_ordered_spectrum}(a). All cumulants other than $k=2$ vanish, so the sum in Eq.~\eqref{eq:I_order_sum} reduces exactly to its phase-adjusted second-order contribution; no higher-order terms enter the unsuperscripted coefficients below. These coefficients $\mathcal{I}_{\alpha,\gamma}(T)$ relate a measurement axis $\alpha$ and a noise-component axis $\gamma$ to symmetry-resolved components of $\tilde{S}^{(+)}(\omega)$, determined by the symmetry of the filter function. We focus on measurements along $\alpha=x$, as dephasing noise primarily degrades transverse coherence where the longitudinal control component $y_{z,z}(t) \approx \pm 1$ dominates.

From the cumulant expansion~\eqref{eq:cumulant_overlap}, introducing the time-ordering kernel $\theta(\tau)$ and applying the convolution theorem, the non-zero coefficients in the $\gamma=\{\mathbbm{1},y\}$ sectors are
\begin{align}
\mathcal{I}_{x,\mathbbm{1}}(T) &= -\frac{1}{2\pi}\int_{\mathbb{R}}  d\omega F_{zz}(\omega,T) \tilde{S}^{(+)}(\omega) \\
\mathcal{I}_{x,y}(T) &= \frac{1}{2\pi}\int_{\mathbb{R}}  d\omega F_{zx}(\omega,T)\tilde{S}^{(+)}(\omega),\label{eq:I_xy_freq}
\end{align}
where the order- and sector-dependent phases of
Eq.~\eqref{eq:I_order_sum} have been applied, and the time-ordered bracket
spectrum is defined as
\begin{align}
\tilde{S}^{(+)}(\omega ) &\equiv\mathcal{F}\left[\theta(t_1-t_2)\{B(t_1),B(t_2)\}^+_C\right]\nonumber\\
& = \tfrac{1}{2} S^{(+)}(\omega)  - \frac{\mathrm{i}}{2\pi}\mathcal{P} \int_{\mathbb{R}} d\nu\,\frac{S^{(+)}(\omega - \nu )}{\nu}.
\end{align}
The noise-component axis $\gamma$ decomposes the symmetry-mediated system response to the filter interacting with the noise. Using the conjugation symmetry $\tilde{S}^{(+)}(\omega)^* = \tilde{S}^{(+)}(-\omega)$ of the Fourier transform, the spectral content partitions into components of definite parity. The even (real) part corresponds to the traditional power spectrum $\mathrm{Re}[\tilde{S}^{(+)}(\omega)] = S^{(+)}(\omega)/2$ governing the dissipative decay of coherence. The odd (imaginary) part is its Hilbert transform, encoding global information about the spectral shape through weighted averaging over all frequencies. It quantifies the reactive response and induces a coherent, noise-induced rotation of the Bloch vector. The pseudo-Hermiticity argument of \appref{app:real_structure}, together with the explicit phase convention in Eq.~\eqref{eq:I_order_sum}, guarantees that the coefficients $\mathcal{I}_{\alpha,\gamma}(T)$ are real-valued. Each spectral component therefore couples exclusively to filter function components of matching parity---symmetric filters extract real power spectra, antisymmetric filters extract imaginary dispersive components.

At fixed order, the component $\gamma = \mathbbm{1}$ extracts the trace of the phase-free noise generator, $\mathcal{I}^{(k)}_{\alpha,\mathbbm{1}}=\frac{1}{2}\Tr[\hat{\mathcal{I}}^{(k)}_\alpha]$. The measured coefficient is $\mathcal{I}_{\alpha,\mathbbm{1}}=\sum_k(-\mathrm{i})^k\mathcal{I}^{(k)}_{\alpha,\mathbbm{1}}$. Since this channel carries no directional information, it couples exclusively to the symmetric bath correlator $\{B(t_1), B(t_2)\}^+_C$. This element can be interpreted as the signal's decay. Because the bath couples only along $z$, we abbreviate the single-factor filter $F_{a,z}(\omega,T)$ as $F_a(\omega,T)$; both letters in the second-order shorthand $F_{ab}$ therefore denote control axes. For $\alpha=x$, the resulting filter function $F_{zz}(\omega,T) = |F_z(\omega,T)|^2$ is manifestly even, projecting onto the real component of the time-ordered spectrum
\begin{equation}
\mathcal{I}_{x,\mathbbm{1}}(T) = -\frac{1}{2\pi}\int_{\mathbb{R}} d\omega\, F_{zz}(\omega,T)\,\mathrm{Re}[\tilde{S}^{(+)}(\omega)].
\end{equation}
The components $\gamma \in \{x,y,z\}$ encode directional contributions to the reduced dynamics that reflect noise-induced rotations about specific axes. The orientation of these rotations is intrinsically linked to the temporal order of control--bath interactions, as sequential operations along non-commuting axes yield distinct geometric outcomes. Temporally asymmetric interactions are characteristic of a non-zero imaginary (dispersive) component of the noise spectra that identically couple to the imaginary component of the filter functions. For $\gamma = y$, the cross-filter $F_{zx}(\omega,T) = F_z(\omega,T)F_x^*(\omega,T)$ decomposes into even $\mathrm{Re}[F_{zx}]$ and odd $\mathrm{Im}[F_{zx}]$ components that respectively couple to $\mathrm{Re}[\tilde{S}^{(+)}]$ and $\mathrm{Im}[\tilde{S}^{(+)}]$, such that the imaginary spectrum may then be isolated from the real as
\begin{align}
\mathcal{I}_{x,y}(T)=\frac{1}{2\pi}\int_{\mathbb{R}} d\omega\, \mathrm{Re}[F_{zx}(\omega,T)]\,\mathrm{Re}[\tilde{S}^{(+)}(\omega)] \nonumber\\ -\frac{1}{2\pi}\int_{\mathbb{R}} d\omega\, \mathrm{Im}[F_{zx}(\omega,T)]\,\mathrm{Im}[\tilde{S}^{(+)}(\omega)].
\end{align}

\subsubsection{Spectral reconstruction and adaptive resampling}
By constructing a set of $N_s$ distinct filter functions $\{F^{[j]}\}_{j=1}^{N_s}$, where $N_s \geq N_h$ exceeds the number of harmonics to estimate (up to some finite cut-off $\omega_{\mathrm{max}}$), we construct a linear system to determine the relevant sampling points of the principal domain $\mathcal{D}$. The linear equation matrices formed by the system of equations are labelled $\mathbf{G}_{zz}$ and $\mathbf{G}_{zx}$, respectively. To estimate the real part of the time-ordered spectrum, we solve the following linear system
\begin{equation}
\vec{\mathcal{I}}_{x,\mathbbm{1}} = -\mathbf{G}_{zz}\, \mathrm{Re}[\vec{\tilde{S}}_z^{(+)}],
\end{equation}
where $\vec{\mathcal{I}}_{x,\mathbbm{1}}$ is calculated from basis measurements (see \appref{appendix:cbar_calcs} for details). We solve this system via regularised matrix inversion; more robust methods, such as maximum likelihood estimation, apply when sampling error is significant. Subtracting the reconstructed real contribution $\mathrm{Re}[\mathbf{G}_{zx}]\mathrm{Re}[\vec{\tilde{S}}_z^{(+)}]$ from the measured $\vec{\mathcal{I}}_{x,y}$ then isolates the imaginary part on the transverse sampling domain $\Omega_1$
\begin{equation}
    \vec{\mathcal{I}}\!_{x,y}-\mathrm{Re}[\mathbf{G}_{zx}]\mathrm{Re}[\vec{\tilde{S}}_z^{(+)}]=
-\mathrm{Im}[\mathbf{G}_{zx}]\mathrm{Im}[\vec{\tilde{S}}_z^{(+)}].
\end{equation}

\begin{figure}[!htb]
    \centering
    \tikzset{external/export next=false}
    \hspace*{-0.5cm}
    \includegraphics[width=\columnwidth]{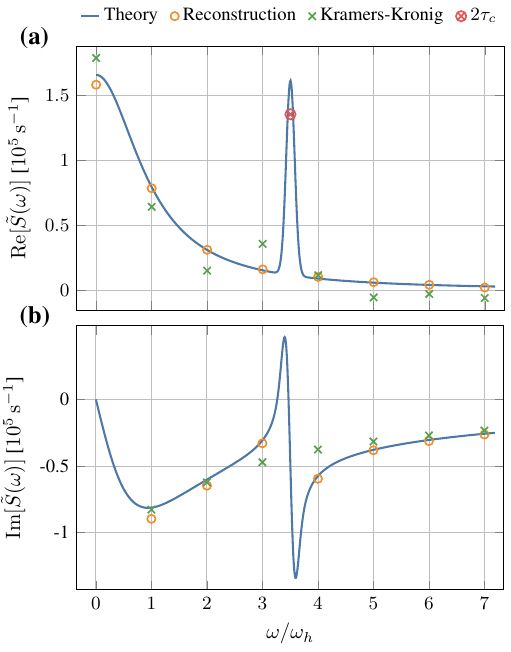}
    \vspace{-0.3cm}
    \caption{Comparison of theoretical and reconstructed time-ordered spectra $\tilde{S}^{(+)}(\omega)$ for classical Gaussian noise. \textbf{(a)} Real component $\mathrm{Re}[\tilde{S}^{(+)}(\omega)]$. The initial comb spacing misses a sharp sub-harmonic feature in the interval $(3\omega_h, 4\omega_h)$, which is recovered by a targeted sample at $\omega = 3.5\,\omega_h$ (red crossed circle) obtained by doubling $\tau_c$. \textbf{(b)} Imaginary component $\mathrm{Im}[\tilde{S}^{(+)}(\omega)]$. Divergent behaviour in the estimate between harmonics $(3\omega_h,4\omega_h)$ flags the location of a sharp local maximum undetected in (a). Green crosses indicate numerical integration of the Kramers--Kronig relations in Eq.~\eqref{eq:kramers_kronig} from complementary reconstruction spectra estimates.}
    \label{fig:gauss_ordered_spectrum}
\end{figure}

The resulting spectrum estimate is shown in Fig.~\ref{fig:gauss_ordered_spectrum}. Panels (a) and (b) compare the theoretical real and imaginary components of the time-ordered spectrum, respectively, to their recovered estimates. We observe good agreement between the theoretical curves and the reconstructed $\mathrm{Re}[\tilde{S}^{(+)}(\omega)]$ across the sampling domain $\Omega_1\equiv\{0, \dots, 7\omega_h\}$, and similarly for $\mathrm{Im}[\tilde{S}^{(+)}(\omega)]$, whose odd parity accounts for the vanishing DC component.

Despite the agreement at the sampled harmonics, an interpolant of the real spectral samples in Fig.~\ref{fig:gauss_ordered_spectrum}(a) would fail to capture the sharp Gaussian feature embedded within the $(3\omega_h, 4\omega_h)$ interval. However, the $\mathrm{Im}[\tilde{S}^{(+)}(\omega)]$ reconstruction in Fig.~\ref{fig:gauss_ordered_spectrum}(b) exhibits a pronounced inflection between the third and fourth harmonics, signalling the presence of additional spectral structure that the coarse comb spacing fails to resolve directly. We diagnose this behaviour by comparing the reconstructed $\mathrm{Im}[\tilde{S}^{(+)}(\omega)]$ with an independent estimate, obtained from a numerical evaluation of the Kramers--Kronig relations, Eq.~\eqref{eq:kramers_kronig}, applied to an interpolant of the reconstructed power spectrum $\mathrm{Re}[\tilde{S}^{(+)}(\omega)]$ on $\Omega_1$. Agreement between the two validates the interpolation, whereas a discrepancy—such as the inflection observed above—reveals frequency regions where the comb spacing is insufficient. This diagnostic forms the basis for the iterative refinement protocol below.

Increasing the pulse period $\tau_c$ allows a finer sampling grid to probe the deviated area further. It is not necessary to resample the entire domain; rather, we target the specific interval flagged by the inflection in the imaginary estimate. As shown in Fig.~\ref{fig:gauss_ordered_spectrum}(a), a single additional control sequence with doubled period $2\tau_c$ successfully recovers the amplitude of the sharp sub-harmonic peak. \appref{app:adaptive_sampling} formalises this procedure as a resource-efficient adaptive sampling method: the discrepancy between the Kramers--Kronig prediction and the directly reconstructed $\mathrm{Im}[\tilde{S}^{(+)}(\omega)]$ defines a residual on the sampled harmonics, a sign-flip score flags unresolved intervals, and Algorithm~\ref{alg:adaptive_refinement} iterates the targeted resampling until the residual falls below a set tolerance. In the validation of \appref{app:residual_demo}, the deterministic trigger flags only the interval $(3\omega_h, 4\omega_h)$, with interval score $\chi^{*} = 0.22$ against tolerance $\vartheta = 0.1$. The $2\tau_c$ resampling pins the peak amplitude, the residual sign pattern reverses, and a concluding width fit recovers the feature's full width as $0.20\,\omega_h$ against the true $0.18\,\omega_h$. The loop terminates at $\chi^{*} = 0.001$, two orders of magnitude below tolerance, with the recovered spectral weight within $5\%$ of its true value (Table~\ref{tab:residual_demo}).

A finer-grained control sequence with cycle time $\tau_c' > \tau_c$ will have additional harmonics proportional to the ratio $\tau_c'/\tau_c$; however, the harmonics which fall in the well-approximated areas (where the splines are sufficiently close) can be estimated using a spline of the original reconstruction.

While the control sequences we employ here (for details see \appref{app:control_sequences}) recover low-error spectra estimates, additional care must be taken to design sufficiently band-limited filter functions for characterisation of $\mathrm{Im}[\tilde{S}^{(+)}(\omega)]$. The principal-value structure of the Hilbert transform causes truncation errors to accumulate logarithmically when spectral weight persists beyond $\omega_{\mathrm{max}}$, yielding biased estimates if filter functions lack adequate spectral localisation. For sub-optimal control sequences, the Kramers--Kronig relations correct by approximating the out-of-band contribution. Splicing an interpolant of the $\mathrm{Re}[\tilde{S}^{(+)}(\omega)]$ reconstruction with an exponential tail beyond the sampled domain, the leakage is obtained as the difference between Eq.~\eqref{eq:kramers_kronig} integrated over $[-\omega_{\mathrm{max}}, \omega_{\mathrm{max}}]$ and the corresponding integral extended to approximate the full real line.

\subsection{Non-Gaussian, quantum noise}\label{subsec:results_quantum}
We now demonstrate that our framework accommodates genuinely quantum noise, characterised by bath operators that fail to commute at different times, $[B(t_1), B(t_2)] \neq 0$. This non-commutativity generates additional quantum contributions to the reduced dynamics through commutator bracket cumulants $\{B(t_1), B(t_2)\}^{-}_C$, which vanish identically for classical noise processes.

We adopt a minimal model consisting of an auxiliary bath qubit with an interaction that couples to multiple system axes, 
\begin{equation}
B(t) = \beta^2(t) (\Lambda_x + \Lambda_z),\label{eq:quantum_bath_operator}
\end{equation}
where $\beta(t)$ is the stationary, zero-mean Gaussian process from \secref{subsec:results_gaussian_classical}, now excluding the sharp spectral feature. The squared dependence on $\beta(t)$ produces non-Gaussian statistics in $B(t)$, yielding non-zero third-order cumulants and the associated classical bispectrum $\tilde{S}^{(++)}(\vec{\omega}_2)$. Although the underlying stochastic process $\beta(t)$ remains Gaussian, the multi-axis coupling yields $[B(t_1), B(t_2)] \neq 0$, generating quantum signatures in the commutator spectrum $\tilde{S}^{(-)}(\omega)$ and mixed commutator-anticommutator bispectra $\tilde{S}^{(-+)}(\vec{\omega}_2)$ and $\tilde{S}^{(--)}(\vec{\omega}_2)$.

\secref{subsec:results_gaussian_classical} demonstrated the role of the noise-component index $\gamma$ in selecting real or imaginary spectral components through filter function parity; here, the reconstruction of $\tilde{S}^{(+)}(\omega)$ proceeds identically. We now show that the \textit{relative orientation} of $\alpha$ and $\gamma$ axes provide additional selection mechanisms to isolate quantum correlations and distinct principal domain symmetries of polyspectra. We exemplify this through characterisation of the quantum spectrum $\tilde{S}^{(-)}(\omega)$ and the classical time-ordered bispectrum $\tilde{S}^{(++)}(\vec{\omega}_2)$. These examples serve as practical templates; alternative reconstruction routes can be obtained through inspection and algebraic manipulation of the formal expressions for $\mathcal{I}_{\alpha,\gamma}^{(k)}$ in \appref{app:direct_calc_3}, and various filter-function design techniques can be employed to aid estimation. Throughout, we assume truncation after the third-order cumulant ($k=3$), though the construction extends straightforwardly to arbitrary order.

\begin{figure}[!htb]
    \centering
    \tikzset{external/export next=false}
    \includegraphics[width=\columnwidth]{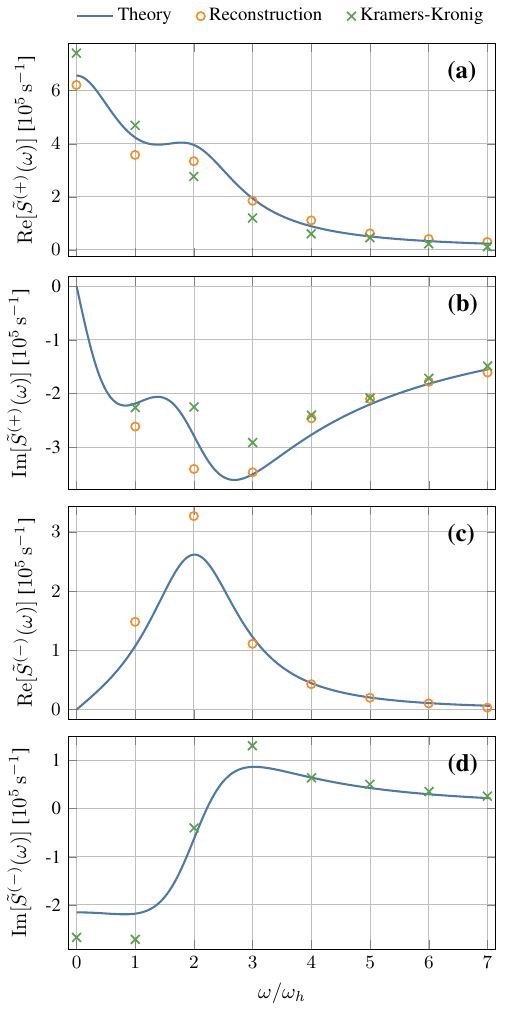}
    \caption{Comparison of theoretical and reconstructed time-ordered spectra $\tilde{S}^{(\pm)}(\omega)$ for the quantum bath operator given in Eq.~\eqref{eq:quantum_bath_operator}. 
    \textbf{(a)} Real part of the anticommutator spectrum $\mathrm{Re}[\tilde{S}^{(+)}(\omega)]$.
    \textbf{(b)} Imaginary component $\mathrm{Im}[\tilde{S}^{(+)}(\omega)]$. Crosses in (a) and (b) indicate Kramers--Kronig estimates computed by numerically integrating the complementary spectral component via Eq.~\eqref{eq:kramers_kronig}.
    \textbf{(c)} The real quantum commutator spectrum $\mathrm{Re}[\tilde{S}^{(-)}(\omega)] = S^{(-)}(\omega)/2$, directly reconstructed from the $\mathcal{I}_{y,y}$ coefficients.
    \textbf{(d)} The imaginary quantum commutator spectrum $\mathrm{Im}[\tilde{S}^{(-)}(\omega)]$: while decoupled from the observed probe response for all control configurations $\mathbf{a}\in\{x,z\}$, the Kramers--Kronig relations recover it from the directly measured $\mathrm{Re}[\tilde{S}^{(-)}(\omega)]$ in~(c).
    }
    \label{fig:second_order_plots}
\end{figure}

Extracting the noise component parallel to the measurement axis ($\gamma = \alpha$) provides an unambiguous signature of quantum bath correlations. The physical basis for this selection rule emerges from the interferometric structure of the cumulant expansion: system evolution proceeds along forward-time ($j=0$) and reverse-time ($j=1$) branches, with the toggling operator $\lambda_\alpha(\mathbf{a},\pi,\vec{j})$ controlling their interference. For classical noise, the accumulated phases of forward and reverse branches cancel upon projection onto $\gamma = \alpha$, leaving $\alpha$-populations invariant. Non-commuting bath operators, in contrast, introduce anisotropy between branches, preventing path closure; the residual net phase drives population transfer along $\alpha$. As derived in \appref{app:direct_calc_3_general} for the general control case $\mathbf{a} = \{x,y,z\}$, every bracket combination contributing to the overlap integral $\mathcal{I}^{(k)}_{\alpha,\alpha}$ contains an odd number of commutator brackets. Each contribution therefore carries at least one commutator, and $\mathcal{I}^{(k)}_{\alpha,\alpha}$ vanishes identically for classical noise.

For $\alpha = \gamma = y$, this mechanism isolates the second-order quantum spectrum $\tilde{S}^{(-)}(\omega)$. The commutator correlator $\{B(t_1), B(t_2)\}_C^-$ is anti-Hermitian and antisymmetric under time exchange, properties that together ensure the unordered spectrum $S^{(-)}(\omega)$ is purely real. Causal embedding realises the complex-valued time-ordered spectrum $\tilde{S}^{(-)}(\omega)$; however, the anti-Hermiticity of the commutator correlator reverses the parity assignments relative to the anticommutator case: $\mathrm{Re}[\tilde{S}^{(-)}(\omega)] = S^{(-)}(\omega)/2$ is odd and $\mathrm{Im}[\tilde{S}^{(-)}(\omega)]$ is even. Since the associated joint filter $(F_{xz} - F_{zx}) = 2\mathrm{i}\,\mathrm{Im}[F_{xz}]$ is purely imaginary with $\mathrm{Im}[F_{xz}]$ odd, the overlap integral samples only $\mathrm{Re}[\tilde{S}^{(-)}(\omega)]$:
\begin{align}
\mathcal{I}_{y,y}(T) &= -\frac{1}{\pi}\int_{\mathbb{R}} d\omega\,\mathrm{Im}[F_{xz}(\omega,T)] \tilde{S}^{(-)}(\omega)\nonumber\\
&=-\frac{1}{\pi}\int_{\mathbb{R}} d\omega\,\mathrm{Im}[F_{xz}(\omega,T)] \mathrm{Re}[\tilde{S}^{(-)}(\omega)]
\end{align}
The prefactor $1/\pi$ carries the factor of two produced by the fold $F_{xz}-F_{zx}=2\mathrm{i}\,\mathrm{Im}[F_{xz}]$; the unfolded phase-free form appears in the $\mathcal{I}^{(2)}_{y,y}$ row of \appref{app:direct_calc_3}, and Eq.~\eqref{eq:I_order_sum} supplies its second-order parallel-sector phase.
The imaginary component of the time-ordered quantum spectrum, $\mathrm{Im}[\tilde{S}^{(-)}(\omega)]$, is entirely decoupled from the probe response. It systematically vanishes from the coefficients $\mathcal{I}_{\alpha,\gamma}$ for any choice of component axes $(\alpha,\gamma)$ and for any control configuration (cf.\ \appref{app:direct_calc_3} and~\ref{app:direct_calc_3_general}), yet it remains a well-defined property of the bath. Since the unordered commutator spectrum $S^{(-)}(\omega)$ is purely real, $\mathrm{Im}[\tilde{S}^{(-)}(\omega)]$ arises entirely from the causal embedding and is fully determined as the Hilbert transform of the directly measured $\mathrm{Re}[\tilde{S}^{(-)}(\omega)]$. The Kramers--Kronig relations therefore furnish the sole route to this component, completing the complex-valued quantum spectrum without requiring additional measurements. In contrast to the classical case of \secref{subsec:results_gaussian_classical}, where both $\mathrm{Re}[\tilde{S}^{(+)}]$ and $\mathrm{Im}[\tilde{S}^{(+)}]$ couple independently to the probe and their Kramers--Kronig consistency serves as a diagnostic, we include it here for completeness. The classical and quantum time-ordered spectra are shown in Fig.~\ref{fig:second_order_plots}(a--d) using the control sequences of \secref{subsec:results_gaussian_classical}.

\subsubsection{Bispectrum estimation}\label{subsubsec:bispectrum_estimation}
Non-Gaussian statistics arising from $\beta^2(t)$ generate non-zero bath noise cumulants of order $k \geq 3$. These higher-order correlations introduce discrete symmetries in the interaction dynamics, which are inherited by the principal domains of the associated polyspectra. The control-centric perspective insists that the \textit{observable} system response reflects the intersection of symmetries in the joint system--environment cumulants. Characterisation of noise polyspectra becomes non-trivial, since the effective principal domain is determined by the symmetry projection imposed by the coupling filter function.

We outline an extensible characterisation protocol for the symmetry-dependent principal domains of time-ordered polyspectra using the reconstruction of the classical time-ordered bispectrum $\tilde{S}^{(++)}(\vec{\omega}_2)$ as an illustrative example. We exploit the symmetry of paired configurations $(\alpha^\prime,\gamma^\prime)\in\{(x,z),(z,x)\}$, in which the measurement and noise generator axes are interchanged. The observed signal is a superposition of single- and multi-axis filter functions that project the bispectrum onto symmetry classes $\tilde{S}_{\mathbf{a}}^{(++)}$ labelled by the control axis configuration $\mathbf{a}$:
\begin{align}
\mathcal{I}_{\alpha^\prime,\gamma^\prime}(T) &= -\frac{1}{4\pi^2} \int_{\mathbb{R}^2} d\vec{\omega}_2 \sum_{\mathbf{a}} F_{\mathbf{a}}(\vec{\omega}_2,T)\, \tilde{S}^{(++)}_{\mathbf{a}}(\vec{\omega}_2).
\end{align}
Here $\gamma^\prime\neq\alpha^\prime$, so the minus sign is the
third-order transverse phase
$(-\mathrm{i})^{3-1}=-1$ from Eq.~\eqref{eq:I_order_sum}.
For $(\alpha^\prime,\gamma^\prime) = (x,z)$, the control channels are $\mathbf{a} \in \{zzz, zxx\}$; for $(z,x)$, they are $\mathbf{a} \in \{xxx, xzz\}$. 

For a real, classical process with commuting, time-translation-invariant cumulants, the bispectrum possesses three discrete symmetries: permutation symmetry $S(\omega_1, \omega_2) = S(\omega_2, \omega_1)$, conjugation symmetry $S(\omega_1, \omega_2) = S^*(-\omega_1, -\omega_2)$, and the stationarity constraint $S(\omega_1,\omega_2)=S(-\omega_1-\omega_2,\omega_2)$~\cite{NikiasPetropulu1993HigherOrderSpectra,sungNonGaussianNoiseSpectroscopy2019a}. Together, these partition the $(\omega_1, \omega_2)$ plane into 12 equivalent regions, restricting unique spectral content to the standard principal domain $\mathcal{D}_2$, which forms a minimal wedge. The symmetry of the filter function determines which projection of the bispectrum is accessible to the probe. A filter invariant under a subgroup $G \subseteq S_3$ couples only to the $G$-symmetrised component of the bispectrum; spectral content in the kernel of the associated projection $\Pi_G$ is invisible regardless of the pulse waveform. The single-axis filter $F_{zzz}(\omega_1, \omega_2) = F_z(\omega_1)F_z(\omega_2)F_z(-\omega_1-\omega_2)$ is invariant under the full symmetric group $S_3$---all six permutations of its frequency arguments---a structural consequence of identical control axes. The effective principal domain therefore coincides with this standard principal domain, $\mathcal{D}_2^{(zzz)} = \mathcal{D}_2$. Multi-axis filters such as $F_{xzz}$ retain invariance only under the two-element subgroup $\{e, (23)\}$ generated by transposition of the two $z$-coupled arguments, extending the effective principal domain to $\mathcal{D}_2^{(xzz)} \supset \mathcal{D}_2$ (cf.\ Fig.~\ref{fig:principal_domain}).

The time-ordered bispectrum $\tilde{S}^{(++)}_{\mathbf{a}}(\vec{\omega}_2)$ decomposes into real and imaginary parts, each coupling to filter components of matching parity. As demonstrated in~\cite{paz-silvaMultiqubitSpectroscopyGaussian2017a}, these components may be isolated by engineering symmetry of the control sequence about the cycle midpoint $\tau_c/2$. Temporal mirroring ($y(\tau_c - t) = y(t)$) imposes odd parity on the toggling phase, constraining the longitudinal filter $F_z$ to be real and the transverse filter $F_x$ to be imaginary. Conversely, mirror anti-symmetry---defined by inverting the pulse amplitudes in the second half of the cycle ($y(\tau_c - t) = -y(t)$)---reverses these assignments. By selecting the appropriate symmetry, we can systematically resolve the real or imaginary components of the spectrum.

\begin{figure*}[tb]
    \tikzset{external/export next=false}
    \hspace*{-1.0cm}
    \includegraphics[]{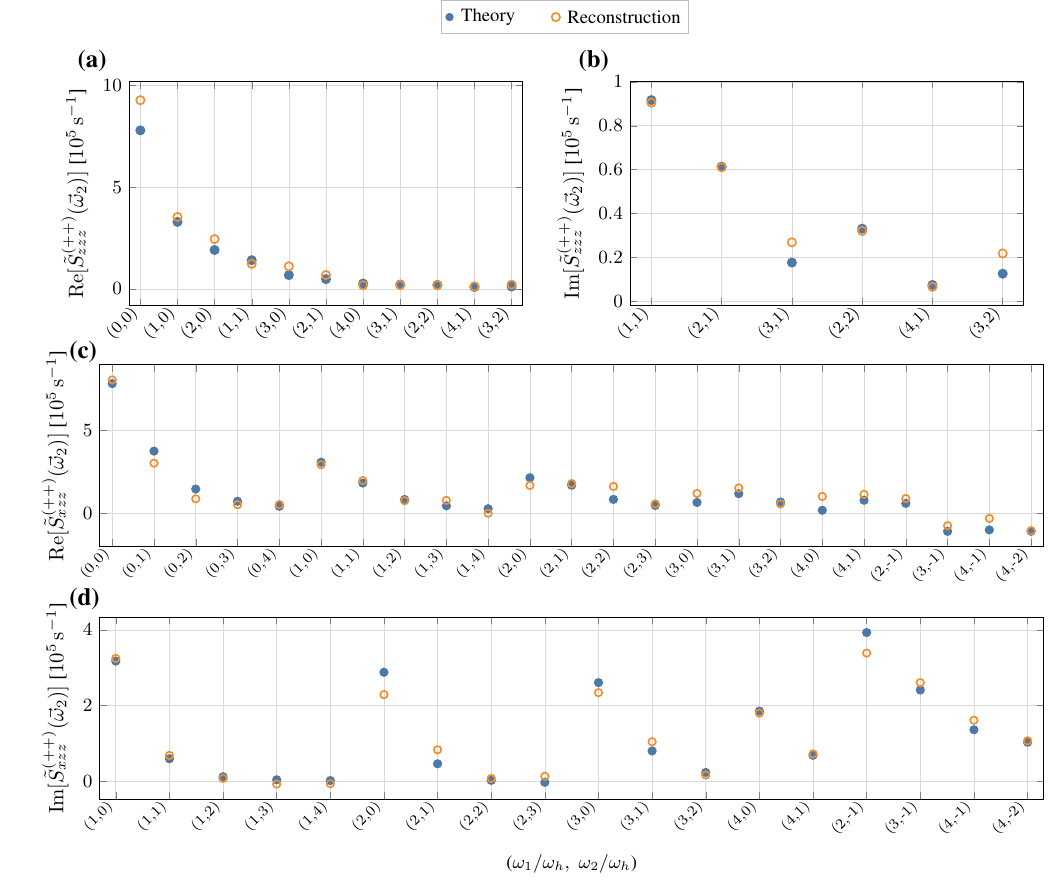}
    \caption{Symmetry-resolved reconstruction of third-order $(k=3)$ time-ordered bispectra $\tilde{S}^{(++)}(\vec{\omega}_2)$ for non-Gaussian quantum noise. Multi-axis controls induce anisotropy under interaction permutation, providing greater spectroscopic detail on the extended effective principal domain $\mathcal{D}_2^{(xzz)}\supset\mathcal{D}_2^{(zzz)}$. \textbf{(a)} Real and \textbf{(b)} imaginary components of the permutation-symmetric projection $\tilde{S}^{(++)}_{zzz}$, estimated from the linear system $\vec{\mathcal{I}}_{x,z} \approx \mathbf{G}_{zzz}\cdot\vec{\tilde{S}}^{(++)}_{zzz}$ under control configurations satisfying $|\mathbf{G}_{zzz}| \gg |\mathbf{G}_{zxx}|$. Temporally mirrored control functions filter purely real interactions, isolating $\mathrm{Re}[\tilde{S}^{(++)}_{zzz}]$ in (a); this contribution is subtracted from $\mathcal{I}_{x,z}$ to estimate $\mathrm{Im}[\tilde{S}^{(++)}_{zzz}]$ in (b). \textbf{(c)} Real and \textbf{(d)} imaginary components of the permutation-asymmetric projection $\tilde{S}^{(++)}_{xzz}$, reconstructed on the extended effective principal domain $\mathcal{D}_2^{(xzz)}$ from $\mathcal{I}_{z,x}$ where $|\mathbf{G}_{xzz}| \gg |\mathbf{G}_{xxx}|$. Temporally anti-mirrored control functions enable isolated estimation of $\mathrm{Re}[\tilde{S}^{(++)}_{xzz}]$ in (c), which is subtracted from $\mathcal{I}_{z,x}$ to recover $\mathrm{Im}[\tilde{S}^{(++)}_{xzz}]$ in (d). Filled blue circles denote theoretical values; open orange circles show numerical reconstructions.}
    \label{fig:bispectrum_combined}
\end{figure*}

The reconstruction proceeds in two phases. First, we characterise the fully symmetric sector by targeting $\tilde{S}^{(++)}_{zzz}(\vec{\omega}_2)$ via the configuration $(\alpha,\gamma) = (x,z)$. Applying temporally mirrored sequences with near-instantaneous control ($|y_x| \ll |y_z|$) suppresses the mixing term such that $|F_{zxx}| \ll |F_{zzz}|$, yielding direct access to $\mathrm{Re}[\tilde{S}^{(++)}_{zzz}(\vec{\omega}_2)]$ over $\mathcal{D}_2^{(zzz)}$. The imaginary component $\mathrm{Im}[\tilde{S}^{(++)}_{zzz}(\vec{\omega}_2)]$ follows from optimising filters satisfying $\mathrm{Re}[F_{zzz}] \ll \mathrm{Im}[F_{zzz}]$ and subtracting the previously estimated real contribution. Second, we access the symmetry-broken sector by switching to $(\alpha, \gamma) = (z, x)$. Temporal anti-mirroring combined with subtraction of the known symmetric background isolates $\mathrm{Re}[\tilde{S}^{(++)}_{xzz}(\vec{\omega}_2)]$ over the extended domain $\mathcal{D}_2^{(xzz)}$. The imaginary component $\mathrm{Im}[\tilde{S}^{(++)}_{xzz}(\vec{\omega}_2)]$ follows from subtracting all previously estimated contributions. This sequential approach exploits the domain containment $\mathcal{D}_2^{(zzz)} \subset \mathcal{D}_2^{(xzz)}$: harmonics within the symmetric wedge are determined first, then used to constrain estimates over the extended region $\mathcal{D}_2^{(xzz)} \setminus \mathcal{D}_2^{(zzz)}$.

Fig.~\ref{fig:bispectrum_combined} presents the reconstructed time-ordered bispectra on both the $S_3$-symmetric domain $\mathcal{D}_2^{(zzz)}$ (panels a--b) and the extended $\{e,(23)\}$-symmetric domain $\mathcal{D}_2^{(xzz)}$ (panels c--d). The real components capture correlations analogous to the unordered bispectrum, while the imaginary components encode causal structure through the multidimensional Hilbert transform---information absent in standard polyspectral analysis but accessible through time-ordered reconstruction.

\subsection{Reconstruction limitations}\label{subsec:reconstruction_limitations}
While the introduced control-centric framework generalises environmental characterisation across a broad range of experimental regimes, several fundamental sources of reconstruction error must be addressed to ensure high-fidelity spectral estimation. Beyond standard considerations of measurement noise and experimental overhead, three primary failure modes are particularly critical when reconstructing time-ordered polyspectra.
First, the accuracy of the spectral reconstruction is intrinsically linked to the truncation of the cumulant expansion at a finite order $k$. In environments where the noise process significantly deviates from Gaussianity--often due to long-range temporal correlations or strong system--bath coupling--unaccounted higher-order cumulants ($k > 2$) may introduce systematic biases into lower-order spectral estimates. In particular, the zero-frequency estimate of the real polyspectra, $S_{\mathbf{b}}(0,0,\ldots,0)$, requires careful treatment: any filter designed to probe this point contributes with increasing weight at higher cumulant orders, scaling as $\propto [F(0)]^k$, so that the spectral overlap becomes dominated by the highest-order term retained in the expansion.

Convergence is not universal and must be verified over the experimental time window from the decreasing size of successive connected contributions; error amplification in the resulting inversion is governed by the corresponding regularised inverse $\mathbf{G}^{+}$.

To mitigate these truncation effects, the characterisation protocol can be systematically extended to include higher-order polyspectra, allowing for the isolation of non-Gaussian signatures. When no reference spectra are available, the adequacy of a chosen truncation can be tested internally through Kramers--Kronig self-consistency, cross-channel comparison, and hold-out prediction under test controls excluded from the reconstruction system; \appref{app:internal_validation} outlines these checks.
Second, selecting an inappropriate high-frequency cut-off, $\omega_{\mathrm{max}}$, for the sampled harmonics can compromise the integrity of the discrete reconstruction. This failure mode occurs when the bath polyspectra $\tilde{S}_{\mathbf{b}}^{(k)}(\vec{\omega}_{k})$ do not exhibit sufficiently rapid decay, causing spectral contributions from frequencies outside the bounded sampling domain $\Omega_{k-1}^{(\mathbf{a},\mathbf{b})}$ to leak into the sampled harmonic coordinates. This issue is particularly pronounced when characterising the Hilbert-transformed (imaginary) components of the time-ordered polyspectra, which generally converge to zero more slowly than their unordered counterparts.
Such leakage effects may be mitigated through advanced filter design to attenuate the response at higher frequencies or by using joint spline interpolations spliced with exponential-decay fits to approximate the missing harmonic information. Finally, the frequency-comb approximation must be valid, as it determines the accuracy with which the repeated control sequences discretise the spectral overlap integrals. The fidelity of this discretisation depends on the number of sequence repetitions $M$ and the analytic properties of the underlying noise process. Specifically, sharp spectral features or rapidly varying dispersive structures can challenge the asymptotic behaviour of the multidimensional Dirac comb. This can be addressed by tuning $M$ for optimal discretisation.

\dobib

\paperExternalReferences
\paperStandaloneSection{4}{29}{4}{0}

\section{Discussion}
\label{sec:discussion}


The control-centric framework distinguishes sharply between spectral features that \emph{exist} in the environment and those that are \emph{observable} by the probe. For any fixed system--bath coupling and control protocol, the probe accesses only a restricted projection of the full spectral structure: some components contribute directly to the measured response and can therefore be reconstructed directly, whereas others do not couple to the probe under that configuration, even though they remain well-defined characteristics of the bath. Certain uncoupled components may still be inferred indirectly via analytic constraints, such as Kramers--Kronig relations, whereas others require a different coupling basis, control setting, or more powerful sensing architectures to become observable.

Accordingly, the effective principal domain $\mathcal{D}_{k-1}^{(\mathbf{a},\mathbf{b})}$ should be viewed as a boundary of probe observability rather than a boundary of spectral existence. For a real, stationary process, the full unordered polyspectrum $S^{(k)}_{\mathbf{b}}$ possesses permutation, conjugation, and stationarity symmetries~\cite{NikiasPetropulu1993HigherOrderSpectra}. However, the probe accesses only the effective principal domain determined by the permutation invariance of the multi-axis filter function (cf.\ Fig.~\ref{fig:principal_domain}). The information in this domain is sufficient to predict the system's future evolution under that specific control~\cite{chalermpusitarakFrameBasedFilterFunctionFormalism2020}. An instructive example of the second tier is the even, imaginary component $\mathrm{Im}[\tilde{S}^{(-)}(\omega)]$, which is the negative Hilbert transform of $\mathrm{Re}[\tilde{S}^{(-)}(\omega)]=S^{(-)}(\omega)/2$, the unordered commutator spectrum, in the convention of Eq.~\eqref{eq:kramers_kronig}. As established in \secref{subsec:results_quantum}, this quantity systematically vanishes from the measured coefficients $\mathcal{I}_{\alpha,\gamma}$ for any choice of component axes $(\alpha,\gamma)$ and any control configuration, yet it remains a well-defined quantity fixed by $S^{(-)}(\omega)$ through the causal embedding. Although no choice of control or measurement axis couples the probe to $\mathrm{Im}[\tilde{S}^{(-)}(\omega)]$ directly, the Kramers--Kronig relations recover it from the directly measured $\mathrm{Re}[\tilde{S}^{(-)}(\omega)]$ (Fig.~\ref{fig:second_order_plots}(c--d)). The spectral definitions themselves require only the initial product-state condition $\rho(0) = \rho_S \otimes \rho_B$; stationarity of bath correlations is invoked only at the reconstruction stage, where time-translation invariance enables discrete comb-based sampling.

At second order, the control-centric formalism places the classical and quantum contributions within a common time-ordered spectral representation. In both cases, the imaginary components of $\tilde{S}^{(\pm)}(\omega)$ emerge as Hilbert-transform partners of their real counterparts through the $\Theta(\nu)$ causal embedding~\eqref{eq:S_ordered}. This shared structure does not imply shared physical content. For classical noise, $\mathrm{Im}[\tilde{S}^{(+)}(\omega)]$ is convention-dependent: an equivalent formalism that absorbs the time ordering into the filter function would set $\mathrm{Im}[\tilde{S}^{(+)}(\omega)] = 0$ identically while leaving the power spectrum $S^{(+)}(\omega)$ unchanged. The quantum commutator spectrum $\tilde{S}^{(-)}(\omega)$, in contrast, retains its complex structure regardless of convention, because the non-commutativity of bath operators at unequal times is a physical property of the environment. The physical origins of the two imaginary parts therefore differ, as detailed below.

For classical noise, $\mathrm{Re}[\tilde{S}^{(+)}(\omega)] = S^{(+)}(\omega)/2$ governs the decay of probe coherence, shortening the Bloch vector through pure dephasing or relaxation depending on the coupling axis~\cite{szankowskiEnvironmentalNoiseSpectroscopy2017}. Its Hilbert-conjugate component, $\mathrm{Im}[\tilde{S}^{(+)}(\omega)]$, encodes a principal-value convolution of the power spectrum and induces a coherent, noise-driven rotation of the Bloch vector. This dispersive contribution acts as a non-local diagnostic: it is sensitive to spectral structure between comb harmonics that the real spectrum, sampled at discrete grid points, cannot resolve directly. Its magnitude scales with the bath correlation time $\tau_B$. For purely white noise, the power spectrum $S^{(+)}(\omega)$ is constant, and the antisymmetry of the $1/\nu$ kernel in the principal-value integral forces $\mathrm{Im}[\tilde{S}^{(+)}(\omega)]$ to vanish identically. As the noise becomes increasingly coloured, the spectral structure generates a growing imaginary component that can dominate the phase evolution. Recovering the full time-ordered spectrum, including its imaginary part, is therefore necessary to characterise open-system dynamics in the presence of non-idealised controls or structured, finite-correlation-time environments.

The quantum commutator spectrum $\tilde{S}^{(-)}(\omega)$ vanishes identically for classical noise; its presence is an unambiguous signature of $[B(t_1),B(t_2)]\neq 0$. The parallel-axis protocol $\gamma = \alpha$ isolates $\mathrm{Re}[\tilde{S}^{(-)}(\omega)] = S^{(-)}(\omega)/2$, the unordered commutator spectrum encoding the dissipative part of the bath response, as demonstrated through the $\mathcal{I}_{y,y}$ reconstruction in Fig.~\ref{fig:second_order_plots}(c). Taken together, the classical and quantum $k=2$ results reinforce the principle that \emph{filter parity selects spectrum parity}: even filters project onto the even spectral component, and odd filters isolate the odd component, irrespective of whether that component is real or imaginary.

At third order, the control-centric framework extends this principle to the time-ordered bispectrum $\tilde{S}^{(++)}(\vec{\omega}_2)$. For the fully symmetric single-axis projection, $\tilde{S}^{(++)}_{zzz}(\vec{\omega}_2)$ inherits the permutation symmetry of the control: the single-axis filter $F_{zzz}$ is structurally invariant under the full symmetric group $S_3$, regardless of the pulse waveform, so that all single-axis coefficients couple exclusively to the $S_3$-symmetrised projection of the bispectrum. Multi-axis projections such as $\tilde{S}^{(++)}_{xzz}$ break this permutation symmetry. The filter $F_{xzz}$ is invariant only under transposition of its two $z$-coupled frequency arguments, projecting onto a larger subspace and thereby extending the effective principal domain to $\mathcal{D}_2^{(xzz)} \supset \mathcal{D}_2^{(zzz)}$ (Fig.~\ref{fig:bispectrum_combined}(c--d)). This domain enlargement provides access to spectral content that arises when the toggling-frame Hamiltonian couples through non-commuting operators, establishing multi-axis controls as a physical requirement for complete bispectral characterisation.

The bath operator of Eq.~\eqref{eq:quantum_bath_operator} couples to non-commuting system axes, so the effective interaction does not commute at distinct times and the third-order coefficients couple to the complex-valued time-ordered bispectra $\tilde{S}^{(++)}_{zzz}(\vec{\omega}_2)$ and $\tilde{S}^{(++)}_{xzz}(\vec{\omega}_2)$ defined through the causal embedding~\eqref{eq:S_ordered}. Their real and imaginary components are resolved by the same filter-parity mechanism that operates at second order. The two-step comb protocol exploits this structure: temporally mirrored control sequences constrain the filter to be real-valued, isolating $\mathrm{Re}[\tilde{S}^{(++)}]$ at the comb harmonics via linear inversion; complex-valued filters from generic (non-mirrored) sequences then couple to $\mathrm{Im}[\tilde{S}^{(++)}]$, which is extracted after subtraction of the known real contribution (Figs.~\ref{fig:bispectrum_combined}(a--d)). The sequential approach exploits the domain containment $\mathcal{D}_2^{(zzz)} \subset \mathcal{D}_2^{(xzz)}$: harmonics within the fully symmetric wedge are determined first, then used to constrain estimates over the extended region $\mathcal{D}_2^{(xzz)} \setminus \mathcal{D}_2^{(zzz)}$. As at second order, the Kramers--Kronig relations~\eqref{eq:kramers_kronig} connect $\mathrm{Re}[\tilde{S}^{(++)}_{zzz}(\vec{\omega}_2)]$ and $\mathrm{Im}[\tilde{S}^{(++)}_{zzz}(\vec{\omega}_2)]$ through multidimensional Hilbert transforms. Comparing the independently reconstructed $\mathrm{Im}[\tilde{S}^{(++)}_{zzz}]$ against its Kramers--Kronig prediction from $\mathrm{Re}[\tilde{S}^{(++)}_{zzz}]$ therefore serves the same diagnostic function at $k=3$ as at $k=2$, flagging unresolved sub-harmonic structure when the two estimates diverge. Cross-channel consistency between $F_{zzz}$ and $F_{xzz}$ reconstructions further constrains the bispectral estimates.


Across all orders, the time-ordered polyspectra provide a systematic decomposition of the system--bath interaction into algebraically distinct spectral sectors. The symmetric sector \(\tilde{S}^{(+)}\) resolves into a real component \(\mathrm{Re}[\tilde{S}^{(+)}]\) and its Hilbert-conjugate imaginary component \(\mathrm{Im}[\tilde{S}^{(+)}]\), while the antisymmetric sector \(\tilde{S}^{(-)}\) isolates the commutator, irreducibly quantum part of the bath correlations. For \(k \geq 3\), these sectors multiply through additional bracket-cumulant symmetries, yielding further distinct projections of the full polyspectrum. Their physical interpretation is not fixed solely at the spectral level: whether a given component contributes to decoherence, coherent rotation, or affine drift depends on how it contracts with the control-dependent filters and on how the resulting terms populate the operator basis of the induced system map. Thus, the time-ordered polyspectra should be understood as furnishing the control-relevant spectral coordinates of the interaction, from which the observable dynamical features of the probe are constructed. Characterising the accessible components under a given control configuration, therefore, provides the information required to predict probe dynamics under future control sequences with the same frequency support~\cite{chalermpusitarakFrameBasedFilterFunctionFormalism2020}.

The symmetry-selection rules developed above translate directly into a strategy for adaptive spectral estimation applicable at any cumulant order. The reconstruction naturally begins with time-reversal-symmetric control sequences, which yield real-valued filter functions and project onto $\mathrm{Re}[\tilde{S}]$ at the comb harmonics. It is often most practical to start with the fully symmetric projection; at $k=2$ this recovers $\mathrm{Re}[\tilde{S}(\omega)] = S(\omega)/2$, the conventional power spectrum, while at $k=3$ the analogue is $\mathrm{Re}[\tilde{S}^{(++)}_{zzz}(\omega_1,\omega_2)]$. Supplementing these with antisymmetric or complex control sequences then yields estimates of $\mathrm{Im}[\tilde{S}]$, and the pair $\{\mathrm{Re}[\tilde{S}],\mathrm{Im}[\tilde{S}]\}$ provides a stringent internal consistency check via the Kramers--Kronig relations~\eqref{eq:kramers_kronig}.

The value of this consistency check becomes apparent when one recognises that comb-based QNS samples noise spectra only at discrete harmonic frequencies, leaving spectral content between comb teeth unresolved. The real spectrum is known only at discrete grid points, so any interpolation carries an inherent ambiguity regarding features that reside between them. The imaginary component provides a diagnostic for this ambiguity. Its principal-value structure $\mathrm{Im}[\tilde{S}^{(+)}(\omega)] \propto -\mathcal{P}\int S^{(+)}(\nu)/(\omega - \nu)\,d\nu$ integrates over the \emph{entire} continuous real spectrum weighted by the $1/\nu$ kernel, rendering it sensitive to features that the discrete comb grid cannot access directly. A discrepancy between the directly measured $\mathrm{Im}[\tilde{S}]$ and the Kramers--Kronig prediction computed from the sampled $\mathrm{Re}[\tilde{S}]$ therefore constitutes a quantitative error signal. At $k=2$, this signal diagnoses unresolved sub-harmonic structure in the power spectrum, as illustrated for the sharp spectral feature in Fig.~\ref{fig:gauss_ordered_spectrum}(b). The protocol can allocate resources adaptively, for instance by increasing the cycle time $\tau_c \to 2\tau_c$ in the spectral bands where the discrepancy is largest (Fig.~\ref{fig:gauss_ordered_spectrum}(a)), while carrying forward well-approximated regions to limit overhead. The net result is an adaptive protocol in which the imaginary, dispersive component serves as a diagnostic for the adequacy of the spectral sampling.

\subsection*{Outlook}

The control-centric framework suggests several directions for future investigation. First, embedding time ordering into the bath spectra affords flexibility to incorporate optimised control families, such as Slepian-based modulation sequences~\cite{freyApplicationOptimalBandlimited2017a,norrisOptimallyBandlimitedSpectroscopy2018a}, into the comb spectroscopy architecture. Such sequences offer enhanced spectral localisation and reduced leakage, which would complement the adaptive resampling strategy by suppressing finite-bandwidth artifacts at their source. Relatedly, adapting subtracted dispersion relations~\cite{mobleyKramersKronigFiniteBandwidth2000} to the multidimensional Kramers--Kronig setting could regularise the principal-value integration, replacing visual inspection of spectral convergence with a rigorous, quantitative diagnostic. The framework's compatibility with state-preparation-and-measurement-robust protocols~\cite{khanMultiaxisQuantumNoise2024} and multi-qubit extensions~\cite{vonlupkeTwoqubitSpectroscopySpatiotemporally2020} further broadens its experimental scope.

Second, the analytic structure of $\tilde{S}^{(k)}_\mathbf{b}$ in the upper half-plane, guaranteed by Titchmarsh's theorem~\cite{titchmarshIntroductionTheoryFourier1937}, constrains the pole and branch-cut structure of the spectral function. Mapping these features to finite-dimensional Markov embeddings could furnish quantitative diagnostics of non-Markovianity, identifying processes whose spectra resist approximation by rational functions of finite order.

Finally, reconstructing $\tilde{S}^{(\pm)}(\omega)$ furnishes two complementary consistency checks: a frequency-local test, in which pointwise spectral properties are evaluated directly from $\mathrm{Re}[\tilde{S}^{(\pm)}(\omega)]$, and a frequency-global test, in which the independently reconstructed $\mathrm{Im}[\tilde{S}^{(\pm)}(\omega)]$ is compared against its Kramers--Kronig prediction, probing inter-harmonic structure inaccessible to the local test. A concrete application is frequency-resolved thermometry through the Kubo--Martin--Schwinger (KMS) condition~\cite{kuboStatisticalMechanicalTheoryIrreversible1957}, which rigidly couples these spectra in thermal equilibrium. The spectral asymmetry encoded in $\tilde{S}^{(-)}(\omega)$ is directly related to the fluctuation-dissipation theorem. For a bath in thermal equilibrium at inverse temperature $\beta$, the pointwise ratio of the real parts of the quantum and classical time-ordered spectra satisfies $\mathrm{Re}[\tilde{S}^{(-)}(\omega)]/\mathrm{Re}[\tilde{S}^{(+)}(\omega)]=\tanh(\beta\hbar\omega/2)$~\cite{clerkIntroductionQuantumNoise2010}. The Kramers--Kronig relations recast this equilibrium condition as a non-local integral on $\mathrm{Im}[\tilde{S}^{(-)}]$ that probes the global spectral shape, including regions inaccessible to the standard pointwise ratio. Systematic, frequency-dependent deviations between this prediction and the directly reconstructed $\mathrm{Im}[\tilde{S}^{(-)}(\omega)]$ would flag KMS violations between the sampled harmonics or finite-bandwidth truncation artifacts. The practical significance of resolving $\tilde{S}^{(\pm)}$ is underscored by recent work showing that non-equilibrium evolution of the quantum noise spectrum during computation influences the fidelity of dynamically protected gates through circuit-history dependence~\cite{burgelmanLimitationsDynamicalError2025}. Experimental implementation on superconducting or solid-state qubit platforms is within reach, given the compatibility of the control-centric framework with realistic, finite-bandwidth control~\cite{mccourtLearningNoiseDynamical2023,wangWangGuoQingDigitalNoiseSpectroscopy2024}.

\dobib

\clearpage

\onecolumngrid
\paperExternalReferences
\paperStandaloneAppendix
\appendix
\section{Notation and derivation of the generalised overlap integrals}\label{app:notation_derivation}

This appendix derives the generalised $k^{\text{th}}$-order overlap expression $\mathcal{I}_{\alpha,\gamma}^{(k)}(T)$ for the reduced dynamics, starting from the observable expectation value and applying the cumulant expansion. Table~\ref{tab:notation} summarises the principal symbols used throughout the main text and the appendices.

\begin{table}[tb]
\caption{Principal symbols, grouped by role. Subscripts on the basis operators $\Lambda$ distinguish the roles of one orthonormal operator basis: coupling ($b$), toggling-frame expansion ($a$), measurement ($\alpha$), noise component ($\gamma$), and preparation ($\xi$).}
\label{tab:notation}
\footnotesize
\begin{tabular}{lll}
\toprule
Symbol & Meaning & Defined \\
\midrule
\multicolumn{3}{l}{\emph{Operator bases and axis labels}}\\
$\Lambda_b$, $b\in I_b$ & Bath-coupled system operators of $H_{SB}^{(\mathrm{lab})}$ & \secref{sec:theory} \\
$\Lambda_a$, $a\in I_a\supseteq I_b$ & Basis of the effective Hamiltonian; control mixes $I_b$ across $I_a$ & Eq.~\eqref{eq:H_I} \\
$\Lambda_\alpha$ & Measurement axis, defining $V_{\Lambda_\alpha}$ and $H_{\Lambda_\alpha}$ & Eq.~\eqref{eq:unexpanded V} \\
$\Lambda_\gamma$ & Noise-component axis in the exponent of $V_{\Lambda_\alpha}$ & Eq.~\eqref{eq:V_lambda} \\
$\Lambda_\xi$, $\rho_\xi$ & Preparation axis and initial states of the tomography scheme & \appref{appendix:cbar_calcs} \\
$s(\gamma,\alpha)$ & Sector selector: 1 (identity, parallel), 0 (transverse) & Eq.~\eqref{eq:I_order_sum} \\
\addlinespace
\multicolumn{3}{l}{\emph{Control}}\\
$y_{a,b}(t)$, $\mathbf{Y}(t)$ & Control functions and control matrix & below Eq.~\eqref{eq:H_I} \\
$\mathbf{Y}^{(k)}_{\mathbf{a},\mathbf{b}}$ & Product of $k$ control functions along the axis strings $\mathbf{a}$, $\mathbf{b}$ & Eq.~\eqref{eq:Y_k} \\
$\tilde{F}^{(k)}_{\mathbf{a},\mathbf{b}}$, $F^{(k)}_{\mathbf{a},\mathbf{b}}$ & Time-ordered (simplex) and unrestricted filter functions & Eq.~\eqref{eq:time_to_freq} \\
\addlinespace
\multicolumn{3}{l}{\emph{Toggling-frame selection}}\\
$\lambda_\alpha$, $\bar{\lambda}_{\alpha,\gamma}$ & Toggling operator and its scalar projection onto $\Lambda_\gamma$ & below Eq.~\eqref{eq:I_a_gamma_k} \\
$\vec{j}\in\mathcal{J}_k$, $\vec{l}\in\mathcal{L}_{\vec{j}}$ & Branch vector (forward/reverse) and admissible time orderings & Eq.~\eqref{eq:cumulant_overlap} \\
$\pi$, $\mu(\pi)$, $\varpi$ & Ordered partition, weight $(-1)^{r-1}/r$, cumulant partition index & below Eq.~\eqref{eq:I_a_gamma_k} \\
\addlinespace
\multicolumn{3}{l}{\emph{Bath statistics}}\\
$B_b(t)$ & Interaction-picture bath operators & \secref{sec:theory} \\
$C^{(k)}$, $\mathcal{C}^{(k)}_{\mathbf{b}}$, $\mathcal{B}^{(k)}$ & Hamiltonian cumulant; bath-operator cumulant; block moment product & Eqs.~\eqref{eq:cumulant_overlap}, \eqref{eq:I_a_gamma_k} \\
$S^{(k)}_{\mathbf{b}}$, $\tilde{S}^{(k)}_{\mathbf{b}}$ & Unordered and time-ordered noise polyspectra & Eq.~\eqref{eq:S_tilde_def} \\
$\tilde{S}^{(\pm)}$, $\tilde{S}^{(\pm\pm)}$ & Nested anticommutator ($+$) and commutator ($-$) bracket spectra & \secref{sec:results} \\
$\Theta(\nu)$ & Causal kernel $\pi\delta(\nu)-\mathrm{i}\mathcal{P}(1/\nu)$ & Eq.~\eqref{eq:theta_kernel} \\
\addlinespace
\multicolumn{3}{l}{\emph{Reduced dynamics}}\\
$V_{\Lambda_\alpha}(T)$ & Effective propagator for measurement axis $\alpha$ & Eq.~\eqref{eq:unexpanded V} \\
$\hat{\mathcal{I}}^{(k)}_\alpha$ & Operator-valued $k$th-order cumulant term of $\log V_{\Lambda_\alpha}$ & Eq.~\eqref{eq:cumulant_overlap} \\
$\mathcal{I}^{(k)}_{\alpha,\gamma}$, $\mathcal{I}_{\alpha,\gamma}$ & Phase-free overlap; real coefficient with phase $(-\mathrm{i})^{k-1+s(\gamma,\alpha)}$ & Eqs.~\eqref{eq:I_projection}, \eqref{eq:I_order_sum} \\
\addlinespace
\multicolumn{3}{l}{\emph{Comb sampling and reconstruction}}\\
$\tau_c$, $M$, $\omega_h$ & Cycle time, repetition count, comb frequency $\omega_h=2\pi/\tau_c$ & Eq.~\eqref{eq:comb_overlap} \\
$\Sha(\vec{\omega}_{k-1},M\tau_c)$ & Comb function: asymptotic $(k-1)$-dimensional Dirac comb & below Eq.~\eqref{eq:comb_overlap} \\
$\mathcal{D}_{k-1}$, $\mathcal{D}^{(\mathbf{a},\mathbf{b})}_{k-1}$ & Principal domain; effective principal domain under control symmetry & \secref{subsec:spectral_domain} \\
$\Omega^{(\mathbf{a},\mathbf{b})}_{k-1}$, $N_h$, $N_s$ & Sampled harmonic set; number of harmonics; number of sequences & Eq.~\eqref{eq:spectrum_comb_estimate} \\
$\mathbf{G}_{\mathbf{a},\mathbf{b}}$, $m_{\mathbf{a},\mathbf{b}}(\vec{s})$, $\kappa$ & Filter matrix $(M/\tau_{c,j}^{k-1})\,m_{\mathbf{a},\mathbf{b}}F^{(k)[j]}_{\mathbf{a},\mathbf{b}}$; multiplicity; condition number & below Eq.~\eqref{eq:spectrum_comb_estimate} \\
\bottomrule
\end{tabular}
\end{table}

\subsection{Preliminaries}
We consider an open quantum system with Hilbert space $\mathcal{H}_S$ of dimension $d_s$ coupled to a bath $\mathcal{H}_B$. The total Hamiltonian in the laboratory frame is
\begin{align*}
    H^{(\mathrm{lab})}(t) = H_S + H_B + H_{SB}^{(\mathrm{lab})}(t) + H_{\mathrm{ctrl}}^{(\mathrm{lab})}(t),
\end{align*}
where $H_S$ and $H_B$ generate the free system and bath dynamics, and $H_{\mathrm{ctrl}}^{(\mathrm{lab})}(t)$ acts non-trivially on $\mathcal{H}_S$. The system--bath interaction
\begin{align*}
    H_{SB}^{(\mathrm{lab})}(t) = \sum_{b \in I_b} \Lambda_b \otimes \beta_b(t)
\end{align*}
couples the set $\{\Lambda_b \mid b \in I_b\}$ of orthonormal, invertible system operators to bath operators $\{\beta_b(t)\}$ that may be classical, quantum or both.

\subsection{Reduced dynamics and the effective propagator}\label{app:reduced_dynamics}
Let $O \in \mathcal{B}(\mathcal{H}_S)$ be an invertible system observable. For an initially uncorrelated system--bath state $\rho(0) = \rho_S \otimes \rho_B$, the expectation value at time $T$ is
\begin{align}\label{eq:app_expect_O}
    \langle O(T) \rangle = \left\langle \Tr_{SB}\left[U^{(\mathrm{lab})}(T)\, (\rho_S \otimes \rho_B)\, U^{(\mathrm{lab})\dagger}(T)\, (O \otimes \mathbbm{1}_B)\right] \right\rangle_c,
\end{align}
where $\langle \cdot \rangle_c$ averages over classical noise realisations. To isolate the noise-induced dynamics, we transform to the interaction picture defined by the free Hamiltonians $H_S + H_B$. The free system and bath propagators are $U_S(t) = e^{-\mathrm{i}H_S t}$ and $U_B(t) \equiv \mathcal{T}_+ e^{-\mathrm{i}\int_0^t ds\, H_B(s)}$, respectively. In this frame, bath operators $B_b(t) = U_B^\dagger(t)\, \beta_b(t)\, U_B(t)$ evolve under the free bath dynamics. To accommodate controls that do not commute with the free system Hamiltonian, $[H_S, H_{\mathrm{ctrl}}^{(\mathrm{lab})}(t)] \neq 0$, the control propagator is defined in the interaction picture as $U_{\mathrm{ctrl}}(t)\equiv\mathcal{T}_+\exp\!\left(-\mathrm{i}\int_0^t ds\,U_S^\dagger(s)\, H_{\mathrm{ctrl}}^{(\mathrm{lab})}(s)\, U_S(s)\right)$.

Expanding the system operators in the complete orthonormal basis of invertible operators $\{\Lambda_a \mid a \in I_a\}$ yields the effective Hamiltonian
\begin{align}\label{eq:app_HI}
    H_I(t) = \sum_{a \in I_a} \sum_{b \in I_b} y_{a,b}(t)\, \Lambda_a \otimes B_b(t).
\end{align}
Here, the control functions are defined by the joint action of the system and control propagators,
\begin{align*}
    y_{a,b}(t) = \frac{1}{d_s}\,\Tr\left[\Lambda_a^\dagger\, U_{\mathrm{ctrl}}^\dagger(t)\, U_S^\dagger(t)\, \Lambda_b\, U_S(t)\, U_{\mathrm{ctrl}}(t)\right],
\end{align*}
encoding the redistribution of the interaction across the operator basis. Consequently, the full lab-frame propagator factorises as $U^{(\mathrm{lab})}(T) = U_S(T)\, U_{\mathrm{ctrl}}(T)\, U_B(T)\, U_I(T)$, where $U_I(T)$ is the propagator generated by the effective interaction.

The toggling-frame observable
\begin{align*}
    \bar{O}(T) = U_{\mathrm{ctrl}}^\dagger(T)\, U_S^\dagger(T)\, O\, U_S(T)\, U_{\mathrm{ctrl}}(T)
\end{align*}
is expanded in the orthonormal basis as $\bar{O}(T) = \sum_\alpha o_\alpha(T)\, \Lambda_\alpha$ with coefficients $o_\alpha(T) = \Tr[\Lambda_\alpha^\dagger\, \bar{O}(T)]/d_s$. Substituting the factorised propagator into Eq.~\eqref{eq:app_expect_O} and exploiting the cyclic property of the trace yields
\begin{align*}
    \langle O(T) \rangle 
    &= \left\langle \Tr_{SB}\left[U_I(T)\, (\rho_S \otimes \rho_B)\, U_I^\dagger(T)\, \bar{O}(T)\right] \right\rangle_c.
\end{align*}
Inserting the identity $\bar{O}(T)\, \bar{O}^{-1}(T) = \mathbbm{1}$ between $U_I^\dagger(T)$ and the remaining operators, then applying the cyclic property of the trace and the factorisation $\Tr_{SB}[\cdot] = \Tr_S[\Tr_B[\cdot]]$ for the product state, the expectation value becomes
\begin{align*}
    \langle O(T) \rangle
     &= \left\langle \Tr_S\left[\Tr_B\left[
        \bar{O}^{-1}(T)\, U_I^\dagger(T)\, \bar{O}(T)\, U_I(T)\,
        \rho_B\right] \rho_S\, \bar{O}(T)\right] \right\rangle_c\\
     &=\sum_\alpha o_\alpha(T)\, \Tr_S\left[V_{\Lambda_\alpha}(T)\,
        \rho_S\, \Lambda_\alpha\right],
\end{align*}
where all bath-mediated effects are captured by the effective propagator
\begin{align}
    V_{\Lambda_\alpha}(T) = \left\langle \Tr_B\left[\Lambda_\alpha^{-1}\, U_I^\dagger(T)\, \Lambda_\alpha\, U_I(T)\, \rho_B\right] \right\rangle_c.
\end{align}

\begin{figure}[t]
\centering
\includegraphics[width=0.65\columnwidth]{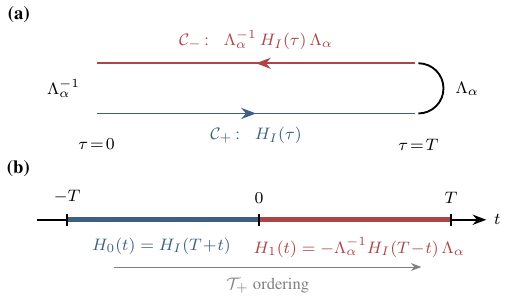}
\caption{Closed-time-path contour and its unfolded representation.
\textbf{(a)}~The Keldysh contour $\mathcal{C}$ in physical time $\tau$,
comprising a forward branch $\mathcal{C}_+$ and backward branch
$\mathcal{C}_-$, with the observable $\Lambda_\alpha$ inserted at the
turning point. The forward branch carries the interaction Hamiltonian
$H_I(\tau)$ (coefficient $-\mathrm{i}$ in the exponential), while the
backward branch carries the conjugated Hamiltonian
$\Lambda_\alpha^{-1}\, H_I(\tau)\, \Lambda_\alpha$ (coefficient
$+\mathrm{i}$).
\textbf{(b)}~Extended time parameterisation on the contour-time axis
$t \in [-T, T]$. The forward branch maps to $t \in [-T, 0)$ via
$t = \tau - T$, and the backward branch maps to $t \in (0, T]$ via
$t = T - \tau$. The sign difference between branches is absorbed into
$H_1(t)$, yielding a uniform $-\mathrm{i}$ coefficient.}
\label{fig:keldysh_contour}
\end{figure}

We now derive the representation of the operator $\Lambda_\alpha^{-1}\, U_I^\dagger(T)\, \Lambda_\alpha\, U_I(T)$ as a single time-ordered exponential on the extended interval $[-T,T]$, following the Schwinger--Keldysh closed-time-path contour construction~\cite{schwingerBrownianMotionQuantum1961,keldyshDiagramTechniqueNonequilibrium1965}. The closed-time-path contour $\mathcal{C}$ traverses the physical-time axis twice (Fig.~\ref{fig:keldysh_contour}a), first along a forward branch $\mathcal{C}_+$ from $\tau=0$ to $\tau=T$ carrying the unconjugated Hamiltonian $H_I(\tau)$, and then along a backward branch $\mathcal{C}_-$ from $\tau=T$ to $\tau=0$ carrying the observable-conjugated Hamiltonian. The observable $\Lambda_\alpha$ enters at the turning point $\tau=T$, while $\Lambda_\alpha^{-1}$ closes the contour at $\tau=0$. The derivation proceeds in three steps.
\paragraph*{Forward branch.}~The forward propagator $U_I(T) = \mathcal{T}_+ \exp\!\left(-\mathrm{i}\int_0^T H_I(\tau)\, d\tau\right)$ is reparameterised by setting $t = \tau - T$, so that $\tau = T + t$ and $t \in [-T,0]$. This gives
\begin{align*}
U_I(T) = \mathcal{T}_+ \exp\!\left(-\mathrm{i}\int_{-T}^{0} H_I(T+t)\, dt\right).
\end{align*}

\paragraph*{Conjugated reverse branch.}
Taking the adjoint reverses the time ordering and flips the sign of the exponent,
\begin{align*}
U_I^\dagger(T) = \mathcal{T}_- \exp\!\left(+\mathrm{i}\int_0^T H_I(\tau)\, d\tau\right).
\end{align*}
Since $\Lambda_\alpha$ is time-independent, it distributes through the Dyson series of $U_I^\dagger(T)$ by insertion of $\Lambda_\alpha\, \Lambda_\alpha^{-1} = \mathbbm{1}$ between each adjacent pair of Hamiltonians, yielding
\begin{align*}
\Lambda_\alpha^{-1}\, U_I^\dagger(T)\, \Lambda_\alpha
= \mathcal{T}_- \exp\!\left(+\mathrm{i}\int_0^T \Lambda_\alpha^{-1}\, H_I(\tau)\, \Lambda_\alpha\, d\tau\right).
\end{align*}
The substitution $\tau = T - t$ maps $\tau \in [0, T]$ to $t \in [0, T]$ with unit Jacobian and reverses the temporal ordering ($\tau_1 > \tau_2 \Leftrightarrow t_1 < t_2$), so that anti-time-ordering in $\tau$ becomes time ordering in $t$. At $n$th order in the Dyson series, the anti-time-ordered simplex $\tau_1 \geq \cdots \geq \tau_n$ maps onto the time-ordered simplex $t_1 \geq \cdots \geq t_n$ via $\tau_j = T - t_{n+1-j}$, with the operator product $(\Lambda_\alpha^{-1} H_I \Lambda_\alpha)(\tau_n) \cdots (\Lambda_\alpha^{-1} H_I \Lambda_\alpha)(\tau_1)$ becoming the time-ordered sequence $(\Lambda_\alpha^{-1} H_I \Lambda_\alpha)(T-t_1) \cdots (\Lambda_\alpha^{-1} H_I \Lambda_\alpha)(T-t_n)$. The sign of the exponent is preserved, giving
\begin{align}
\Lambda_\alpha^{-1}\, U_I^\dagger(T)\, \Lambda_\alpha
= \mathcal{T}_+ \exp\!\left(+\mathrm{i}\int_0^T \Lambda_\alpha^{-1}\, H_I(T-t)\, \Lambda_\alpha\, dt\right).
\end{align}
\paragraph*{Combination.}
Since every $t' \in (0, T]$ from the reverse branch exceeds every $t \in [-T, 0)$ from the forward branch, the $\mathcal{T}_+$ ordering of the product is already satisfied on $[-T, T]$, and the full operator product becomes
\begin{align*}
\Lambda_\alpha^{-1}\, U_I^\dagger(T)&\, \Lambda_\alpha\, U_I(T)= \mathcal{T}_+ \exp\!\left(
+\mathrm{i}\int_0^T \Lambda_\alpha^{-1}\, H_I(T\!-\!t)\, \Lambda_\alpha\, dt
-\mathrm{i}\int_{-T}^{0} H_I(T\!+\!t)\, dt\right).
\end{align*}
The forward branch contributes with the coefficient $-\mathrm{i}$, while the conjugated reverse branch contributes with $+\mathrm{i}$. Defining the reverse-branch Hamiltonian as $H_1(t) \equiv -\Lambda_\alpha^{-1}\, H_I(T-t)\, \Lambda_\alpha$ absorbs this sign difference, so that both branches enter the exponent with a uniform factor of $-\mathrm{i}$ (Fig.~\ref{fig:keldysh_contour}b). Taking the combined classical-quantum average $\langle \cdot \rangle_{c,q} = \langle \Tr_B[\cdot\, \rho_B] \rangle_c$ yields the representation
\begin{align}\label{eq:app_V_exp}
    V_{\Lambda_\alpha}(T) = \left\langle \mathcal{T}_+ \exp\!\left(-\mathrm{i} \int_{-T}^{T} H_{\Lambda_\alpha}(t)\, dt\right) \right\rangle_{c,q},
\end{align}
with the piecewise Hamiltonian
\begin{align}\label{eq:app_piecewise_H}
    H_{\Lambda_\alpha}(t) = \begin{cases}
        H_1(t) \equiv -\Lambda_\alpha^{-1}\, H_I(T-t)\, \Lambda_\alpha, & t \in (0, T], \\[4pt]
        H_0(t) \equiv H_I(T+t), & t \in [-T, 0).
    \end{cases}
\end{align}
The binary index $j \in \{0,1\}$ labels the branch. The forward branch $j=0$ corresponds to the negative contour portion ($t < 0$), mapping to physical time $\tau = T + t \in [0,T)$ and carrying the unconjugated interaction from $U_I(T)$. The reverse branch $j=1$ corresponds to the positive portion ($t > 0$), mapping to $\tau = T - t \in [0,T)$ and carrying the observable-conjugated interaction from $U_I^\dagger(T)$.

To keep the derivation valid for any orthonormal operator basis $\{\Lambda_a\}$, we express the piecewise Hamiltonian in terms of the branch-dependent superoperator action $\mathcal{A}^{(\alpha)}_{j}$. Substituting the interaction Hamiltonian Eq.~\eqref{eq:app_HI} into Eq.~\eqref{eq:app_piecewise_H}, the forward ($j=0$) and reverse ($j=1$) branches take the unified form
\begin{align*}
H_{j}(t) = \sum_{a,b} y_{a,b}(\tau)\, \mathcal{A}^{(\alpha)}_{j}(\Lambda_a) \otimes B_b(\tau),
\end{align*}
where $\tau$ maps the contour time to physical time ($T+t$ for $j=0$, $T-t$ for $j=1$). The transformation of system basis operators induced by the measurement axis $\Lambda_\alpha$ is encoded by the branch-dependent superoperator
\begin{align*}
\mathcal{A}^{(\alpha)}_{0}(\Lambda_a) = \Lambda_a, \quad\quad
\mathcal{A}^{(\alpha)}_{1}(\Lambda_a) = -\Lambda_\alpha^{-1}\, \Lambda_a\,
\Lambda_\alpha.
\end{align*}

\subsection{Cumulant expansion}
Applying the cumulant expansion to Eq.~\eqref{eq:app_V_exp} yields
\begin{align}\label{eq:app_V_a_I_a}
    V_{\Lambda_\alpha}(T) = \exp\!\left(\sum_{k=1}^{\infty} (-\mathrm{i})^k \hat{\mathcal{I}}_\alpha^{(k)}(T)\right),
\end{align}
where the $k^{\text{th}}$-order contribution is
\begin{align*}
    \hat{\mathcal{I}}_\alpha^{(k)}(T) = \int_{-T}^{T} d_> \vec{t}_{[k]}\, C^{(k)}\!\left(H_{\Lambda_\alpha}(t_1), \ldots, H_{\Lambda_\alpha}(t_k)\right),
\end{align*}
with $C^{(k)}$ the $k^{\text{th}}$-order cumulant and $\int d_> \vec{t}_{[k]}$ denoting integration over the simplex $t_1 > t_2 > \cdots > t_k$. We transform the integration domain from the extended interval $[-T, T]$ to the physical interval $[0, T]$. This mapping ``re-folds'' the extended contour (Fig.~\ref{fig:keldysh_contour}b), exchanging the linearised time domain for a summation over the branch indices $\vec{j}$ while preserving the original time ordering
\begin{align}\label{eq:app_Ik_simplex}
    \hat{\mathcal{I}}_\alpha^{(k)}(T) = \int_0^T d_> \vec{t}_{[k]} \sum_{\vec{j} \in \mathcal{J}_k} \sum_{\vec{l} \in \mathcal{L}_{\vec{j}}} C^{(k)}\!\left(H_{j_1}(t_{l_1}), \ldots, H_{j_k}(t_{l_k})\right),
\end{align}
where $d_> \vec{t}_{[k]} = dt_1\, dt_2 \cdots dt_k$ with $T \geq t_1 \geq t_2 \geq \cdots \geq t_k \geq 0$. The binary vector $\vec{j} = (j_1, \ldots, j_k) \in \mathcal{J}_k$ indexes the sequence of Hamiltonian branches $(H_{j_1},\dots,H_{j_k})$. Since the reverse branch ($j=1$, $t > 0$) lies at later contour times than the forward branch ($j=0$, $t < 0$), $\mathcal{T}_+$ ordering places all conjugated operators to the left. The admissible branch sequences, therefore, form two contiguous blocks
\begin{align*}
\vec{j}_n = (\underbrace{1, \ldots, 1}_{n}, \underbrace{0, \ldots, 0}_{k-n}),
\end{align*}
where $n \in \{0, \ldots, k\}$ is the count of conjugated operators.

For a fixed $n$, the set $\mathcal{L}_{n,k}$ contains all permutation vectors $\vec{l}$ that map the time-ordered simplex to the operator sequence. The ordering constraints follow directly from the branch definitions in Eq.~\eqref{eq:app_piecewise_H}: the reverse branch ($j=1$) requires descending indices ($l_1 > \cdots > l_n$) due to the time-reversal inherent to conjugation, while the forward branch ($j=0$) requires ascending indices ($l_{n+1} < \cdots < l_k$):
\begin{align*}
    \mathcal{L}_{n,k} \equiv \mathcal{L}_{\vec{j}_n} = \left\{ \vec{l} = (l_1, \ldots, l_k) \in S_k \,\big|\, (l_1 > \cdots > l_n) \land (l_{n+1} < \cdots < l_k) \right\},
\end{align*}
where $S_k$ is the symmetric group on $k$ elements. The complete set of admissible time-index permutations at order $k$ is the union
\begin{align*}
    \mathcal{L}_k = \bigcup_{n=0}^{k} \mathcal{L}_{n,k}.
\end{align*}
\paragraph*{Example ($k=3$, $n=1$).} For $\vec{j}_1 = (1, 0, 0)$, the constraint $l_2 < l_3$ yields $|\mathcal{L}_{1,3}| = 3$ permutations: $(1,2,3)$, $(2,1,3)$, and $(3,1,2)$.

\subsection{Moment expansion and projection}
We seek to disentangle the control, system, and bath contributions of the reduced dynamics in Eq.~\eqref{eq:app_Ik_simplex} for the effective interaction Eq.~\eqref{eq:app_HI} of a general orthonormal operator basis. The $k^{\text{th}}$-order cumulant $C^{(k)}$ is expanded into moments using the non-commutative logarithm associated with the ordinary operator exponential in Eq.~\eqref{eq:app_V_a_I_a}~\cite{kuboGeneralizedCumulantExpansion1962}:
\begin{align*}
    C^{(k)}(X_1, \ldots, X_k)
    = \sum_{\substack{\pi=(p_1,\ldots,p_r)\\
    \in \mathrm{OP}(\{1,\ldots,k\})}}
    \mu(\pi)\prod_{v=1}^{|\pi|}\left\langle \prod_{q \in p_v}X_q\right\rangle,
\end{align*}
where $\mathrm{OP}(\{1,\ldots,k\})$ denotes ordered set partitions (set compositions): $\pi=(p_1,\ldots,p_r)$ retains the left-to-right order of its $r=|\pi|$ blocks, and the indices within each block retain their original order. Expanding the logarithm gives
\begin{align*}
    \mu(\pi)=\frac{(-1)^{r-1}}{r}.
\end{align*}
For example, at second order this convention yields
\begin{align*}
C^{(2)}(X_1,X_2)=\langle X_1X_2\rangle
-\frac{1}{2}\bigl(\langle X_1\rangle\langle X_2\rangle
+\langle X_2\rangle\langle X_1\rangle\bigr).
\end{align*}
If the block moments commute, the $r!$ block orderings give identical products and may be collected, recovering the familiar coefficient $(-1)^{r-1}(r-1)!$ for each \emph{unordered} set partition. Assigning that factorial coefficient to every ordered partition would instead overcount the non-commuting block products. Within each partition block $p \in \pi$, the Hamiltonian product evaluates to
\begin{align}\label{eq:app_block_moment}
    \left\langle \prod_{q \in p} H_{j_q}(t_{l_q}) \right\rangle = \sum_{\mathbf{a}^{(p)}, \mathbf{b}^{(p)}} \left[ \prod_{q \in p} y_{a^{(q)},b^{(q)}}(t_{l_q}) \right] 
    \left( \prod_{q \in p} \mathcal{A}^{(\alpha)}_{j_q}(\Lambda_{a^{(q)}}) \right) \left\langle \prod_{q \in p} B_{b^{(q)}}(t_{l_q}) \right\rangle_{c,q},
\end{align}
where the product over $q \in p$ respects the ordering of elements within each block of the ordered partition. The sequences $\mathbf{a}^{(p)} = (a^{(q)})_{q \in p}$ and $\mathbf{b}^{(p)} = (b^{(q)})_{q \in p}$ index the system and bath operator axes for positions belonging to block $p$, listed in ascending order. 

Substituting Eq.~\eqref{eq:app_block_moment} into Eq.~\eqref{eq:app_Ik_simplex}, and collecting $k$-length terms denoting the index vectors $\mathbf{a} = (a^{(1)}, \ldots, a^{(k)})$ and $\mathbf{b} = (b^{(1)}, \ldots, b^{(k)})$ across all partition blocks, results in the factored product of control, system, and bath components. We define the \textit{control matrix}
\begin{align*}
    \mathbf{Y}^{(k)}_{\mathbf{a},\mathbf{b}}(t_{\vec{l}}) = \prod_{u=1}^{k} y_{a^{(u)},b^{(u)}}(t_{l_u}),
\end{align*}
where $t_{\vec{l}} \equiv (t_{l_1}, \ldots, t_{l_k})$ denotes the time vector permuted according to $\vec{l}$. The system evolution is captured by the \textit{toggling operator}, defined for a general basis as the ordered product of conjugated operators
\begin{align*}
    \lambda_\alpha(\mathbf{a}, \pi, \vec{j}\,) = \mu(\pi) \prod_{v=1}^{|\pi|} \left( \prod_{q \in p_v} \mathcal{A}^{(\alpha)}_{j_q}(\Lambda_{a^{(q)}}) \right).
\end{align*}
For a qubit system ($d_s=2$), the Pauli basis operators satisfy $\Lambda_\alpha^{-1} = \Lambda_\alpha$ and the anti-commutation relations $\{\Lambda_i, \Lambda_j\} = 2\delta_{ij}\mathbbm{1}$. The reverse-branch
conjugation evaluates to $\Lambda_\alpha^{-1}\, \Lambda_a\, \Lambda_\alpha =
(-1)^{1-s(a,\alpha)}\,\Lambda_a$, where $s(a,\alpha) = \delta_{a,\alpha} + \delta_{a,\mathbbm{1}}$ equals unity
when $\Lambda_a$ commutes with $\Lambda_\alpha$. Combining this with the overall minus sign in $\mathcal{A}^{(\alpha)}_{1}$ gives $\mathcal{A}^{(\alpha)}_{1}(\Lambda_a) = (-1)^{s(a,\alpha)}\,\Lambda_a$.
The toggling operator then simplifies to the form used in the main text
\begin{align*}
    \lambda_\alpha(\mathbf{a}, \pi, \vec{j}\,)
    \xrightarrow{\mathfrak{su}(2)} \mu(\pi) \prod_{v=1}^{|\pi|} \left(
    \prod_{q \in p_v} (-1)^{j_q\, s(a^{(q)},\alpha)}\,
    \Lambda_{a^{(q)}} \right).
\end{align*}
The \textit{bath moment product} collects the environmental correlations organised by the partition $\pi$
\begin{align*}
    \mathcal{B}^{(k)}_{\mathbf{b};\vec{l};\pi}(\vec{t}_k) = \prod_{v=1}^{|\pi|} \left\langle \prod_{q \in p_v} B_{b^{(q)}}(t_{l_q}) \right\rangle_{c,q}.
\end{align*}
Converting to cumulants via the moment-cumulant relation $\mathcal{B}^{(k)}_{\mathbf{b};\vec{l};\pi}(\vec{t}_k) = \sum_{\varpi} \mathcal{C}^{(k)}_{\mathbf{b};\vec{l};\pi;\varpi}(\vec{t}_k)$, where $\varpi$ indexes partitions in the conversion, yields
\begin{align*}
    \hat{\mathcal{I}}_\alpha^{(k)}(T) = \int_0^T d_> \vec{t}_{[k]} \sum_{\substack{\vec{j} \in \mathcal{J}_k,\\ \vec{l} \in \mathcal{L}_{\vec{j}}}} \sum_{\substack{\mathbf{a},\,\mathbf{b},\\\pi,\,\varpi}}\, \lambda_\alpha(\mathbf{a}, \pi, \vec{j}\,)\, \mathbf{Y}^{(k)}_{\mathbf{a},\mathbf{b}}(t_{\vec{l}})\, \mathcal{C}^{(k)}_{\mathbf{b};\vec{l};\pi;\varpi}(\vec{t}_k).
\end{align*}
Finally, to resolve the reduced dynamics in terms of system observables, we
project the effective propagator onto the single-qubit Pauli basis used in the
reconstruction and write
\begin{align*}
V_{\Lambda_\alpha}(T)
=\exp\!\left(
\mathcal{I}_{\alpha,\mathbbm{1}}(T)\Lambda_{\mathbbm{1}}
+\mathcal{I}_{\alpha,\alpha}(T)\Lambda_\alpha
-\mathrm{i}\!\!\sum_{\gamma\in\{x,y,z\}\setminus\{\alpha\}}
\mathcal{I}_{\alpha,\gamma}(T)\Lambda_\gamma
\right).
\end{align*}
The phase-free order-resolved overlaps are obtained by the direct projection
\(
\mathcal{I}^{(k)}_{\alpha,\gamma}(T)
={d_s}^{-1}\Tr\!\left[
\hat{\mathcal{I}}^{(k)}_\alpha(T)\Lambda_\gamma^\dagger\right]
\).
The corresponding coefficients without an order superscript include the
cumulant-order phase:
\begin{align}
\mathcal{I}_{\alpha,\gamma}(T)
&\equiv
\sum_{k=1}^{\infty}
(-\mathrm{i})^{k-1+s(\gamma,\alpha)}
\mathcal{I}^{(k)}_{\alpha,\gamma}(T),
\label{eq:app_overlap_sum}
\end{align}
where the selector $s(\gamma,\alpha)$ is the one defined above:
$s=1$ for the identity and parallel components, and $s=0$ for the transverse
components.  Thus, for a transverse component, the explicit $-\mathrm{i}$
in the exponent supplies the remaining power of $-\mathrm{i}$.  At second
order the phase is $(-\mathrm{i})^2=-1$ for the identity and parallel
components, whereas the transverse phase $(-\mathrm{i})^{2-1}=-\mathrm{i}$
cancels the factor $+\mathrm{i}$ generated by the Pauli product in
$\mathcal{I}^{(2)}_{x,y}$.  The scalar
coefficient obtained by projecting the toggling operator is
$\lambda_\alpha(\mathbf{a},\pi,\vec{j}\,)$:
\begin{align*}
\bar{\lambda}_{\alpha,\gamma}
(\mathbf{a},\pi,\vec{l}\,)
&=\frac{1}{d_s}\Tr\!\left[
\left(\sum_{\vec{j}\in\mathcal{J}'_{\vec{l}}}
\lambda_\alpha(\mathbf{a},\pi,\vec{j}\,)\right)
\Lambda_\gamma^\dagger\right],
\end{align*}
where $\mathcal{J}'_{\vec{l}}\subseteq\mathcal{J}_k$ denotes the set of
branch vectors $\vec{j}$ for which $\vec{l}\in\mathcal{L}_{\vec{j}}$.
Combining these terms yields the generalised $k^{\text{th}}$-order overlap
integral
\begin{align}\label{eq:app_final_overlap}
\mathcal{I}^{(k)}_{\alpha,\gamma}(T)
&=\int_0^T d_>\vec{t}_{[k]}
\sum_{\substack{\mathbf{a},\,\mathbf{b},\\
\pi,\,\varpi,\\ \vec{l}\in\mathcal{L}_k}}
\bar{\lambda}_{\alpha,\gamma}
(\mathbf{a},\pi,\vec{l}\,)
\mathbf{Y}^{(k)}_{\mathbf{a},\mathbf{b}}(t_{\vec{l}})
\mathcal{C}^{(k)}_{\mathbf{b};\vec{l};\pi;\varpi}(\vec{t}_k).
\end{align}

\subsection{Real structure of the overlap integrals}
\label{app:real_structure}
For Hermitian system--bath couplings and a Hermitian bath state, the evolved
observable remains Hermitian.  For an involutory Pauli observable
$\Lambda_\alpha^{-1}=\Lambda_\alpha$, this gives the pseudo-Hermiticity
condition
\begin{align}
V_{\Lambda_\alpha}^{\dagger}(T)
=\Lambda_\alpha V_{\Lambda_\alpha}(T)\Lambda_\alpha^{-1}.
\end{align}
In the perturbative regime the same relation holds for the principal
logarithm and separately at every cumulant order.  Its coefficients along
$\Lambda_{\mathbbm{1}}$ and $\Lambda_\alpha$ are therefore real, whereas
those along the remaining Pauli operators are purely imaginary.  Since the
$k^{\text{th}}$ contribution to the exponent is
$(-\mathrm{i})^k\mathcal{I}^{(k)}_{\alpha,\gamma}$,
Eq.~\eqref{eq:app_overlap_sum} accounts for this distinction and renders
every $\mathcal{I}_{\alpha,\gamma}(T)$ real-valued.

\section{Fourier transform and frequency-comb approximation of \texorpdfstring{$\mathcal{I}_{\alpha,\gamma}^{(k)}(T)$}{I(k)alpha,gamma(T)}}\label{app:ft_comb}
We seek the Fourier transform of the generalised $k^{\text{th}}$-order
overlap in Eq.~\eqref{eq:app_final_overlap} to derive the linearised
frequency-comb approximation.  This section calculates the phase-free
order-resolved overlap; the phases are included only when forming
$\mathcal{I}_{\alpha,\gamma}(T)$ through
Eq.~\eqref{eq:app_overlap_sum}.  For notational convenience, we suppress
$(\vec{l},\pi,\varpi)$ below and write
$\bar{\lambda}_{\alpha,\gamma}(\mathbf{a})\equiv
\bar{\lambda}_{\alpha,\gamma}(\mathbf{a},\pi,\vec{l}\,)$.
We first absorb the ordering of the integration simplex into the integrand through the ordered correlator $\tilde{\mathcal{C}}^{(k)}_{\mathbf{b}}(\vec{t}_k)\equiv\left[\prod_{j=1}^{k-1}\theta(t_j-t_{j+1})\right]\mathcal{C}^{(k)}_{\mathbf{b}}(\vec{t}_k)$; we then compute the multidimensional convolution via the inverse Fourier transform
\begin{align*} 
\mathcal{I}^{(k)}_{\alpha,\gamma}(T)&=\int_{0}^{T} d\vec{t}_{[k]}\sum_{\mathbf{a},\mathbf{b}}\bar{\lambda}_{\alpha,\gamma}(\mathbf{a})\,\mathbf{Y}^{(k)}_{\mathbf{a},\mathbf{b}}(\vec{t}_k)\,\tilde{\mathcal{C}}^{(k)}_{\mathbf{b}}(\vec{t}_k)\\
&=\int_{0}^{T} d\vec{t}_{[k]}\sum_{\mathbf{a},\mathbf{b}}\bar{\lambda}_{\alpha,\gamma}(\mathbf{a})\,\mathbf{Y}^{(k)}_{\mathbf{a},\mathbf{b}}(\vec{t}_k)\,\left[\frac{1}{(2\pi)^k}\int_{\mathbb{R}^k}d\vec\omega_k\,e^{\mathrm{i}\vec{\omega}_k\cdot\vec{t}_k}\tilde{S}^{(k)}_{\mathbf{b}}(\vec{\omega}_k)\right]\\
&=\frac{1}{(2\pi)^k}\int_{\mathbb{R}^k}d\vec\omega_k\sum_{\mathbf{a},\mathbf{b}}\bar{\lambda}_{\alpha,\gamma}(\mathbf{a})\,\left[\int_{0}^{T} d\vec{t}_{[k]}\mathbf{Y}^{(k)}_{\mathbf{a},\mathbf{b}}(\vec{t}_k)e^{\mathrm{i}\vec{\omega}_k\cdot\vec{t}_k}\right]\tilde{S}^{(k)}_{\mathbf{b}}(\vec{\omega}_k)\\
&=\frac{1}{(2\pi)^k}\int_{\mathbb{R}^k}d\vec\omega_k\sum_{\mathbf{a},\mathbf{b}}\bar{\lambda}_{\alpha,\gamma}(\mathbf{a})\,F^{(k)}_{\mathbf{a},\mathbf{b}}(\vec\omega,T)\,\tilde{S}^{(k)}_{\mathbf{b}}(\vec{\omega}_k).
\end{align*}
In the final line we used the dual control-filter convention stated in \secref{subsec:spectral_domain}; no frequency relabelling has been made, so the spectrum retains the argument $\vec\omega_k$.
For stationary noise environments, $\tilde{S}^{(k)}_{\mathbf{b}}(\vec{\omega}_k)=2\pi\delta(|\vec{\omega}_k|)\tilde{S}^{(k)}_{\mathbf{b}}(\vec{\omega}_{k-1})$, where $|\vec{\omega}_k| = \sum_{j=1}^k \omega_j$, resulting in
$$\mathcal{I}^{(k)}_{\alpha,\gamma}(T)=\frac{1}{(2\pi)^{k-1}}\int_{\mathbb{R}^{k-1}}d\vec\omega_{k-1}\sum_{\mathbf{a},\mathbf{b}}\bar{\lambda}_{\alpha,\gamma}(\mathbf{a})\,F^{(k)}_{\mathbf{a},\mathbf{b}}(\vec\omega,T)\,\tilde{S}^{(k)}_{\mathbf{b}}(\vec\omega_{k-1}),$$
where we have defined $F^{(k)}_{\mathbf{a},\mathbf{b}}(\vec\omega,T) = F_{a^{(k)},b^{(k)}}(-|\vec{\omega}_{k-1}|,T)\prod_{j=1}^{k-1}F_{a^{(j)},b^{(j)}}(\omega_{j},T)$ after integrating over $\omega_k$ using the delta function constraint.
\subsection{Frequency-comb approximation}
To linearise the spectral expression for $\mathcal{I}^{(k)}_{\alpha,\gamma}(T)$, we invoke the frequency-comb approximation, where we divide the time interval $T = M\tau_{c}$ into $M$ repetitions of a base sequence with period $\tau_{c}$. For a smooth fundamental filter function $F^{(k)}_{\mathbf{a},\mathbf{b}}(\vec\omega,\tau_{c})$ and $M\gg1$, the frequency-comb approximation enables the factorisation $$F^{(k)}_{\mathbf{a},\mathbf{b}}(\vec\omega,M\tau_{c}) = \Sha^{(k)}(\vec\omega, M\tau_{c})F^{(k)}_{\mathbf{a},\mathbf{b}}(\vec\omega,\tau_{c}),$$ where the comb function $$\Sha^{(k)}(\vec\omega, M\tau_{c}) = \frac{\sin(|\vec{\omega}_{k-1}|M\tau_{c}/2)}{\sin(|\vec{\omega}_{k-1}|\tau_{c}/2)}\prod_{j=1}^{k-1}\frac{\sin(\omega_jM\tau_{c}/2)}{\sin(\omega_j\tau_{c}/2)}$$ behaves as a $(k-1)$-dimensional Dirac comb with uniform peaks, $\omega\in\{{\vec\nu\cdot\omega_h\,|\,\vec\nu\in\mathbb{Z}^{k-1}}\}$, centred on the fundamental comb frequency $\omega_h=2\pi/\tau_c$. The comb function partitions the integration domain $\mathbb{R}^{k-1}$ into hypercubes,
$$Q_{\vec{\nu}} = \prod_{j=1}^{k-1}\left(\nu_j\omega_h - \frac{\omega_h}{2}, \nu_j\omega_h + \frac{\omega_h}{2}\right], \quad \vec{\nu} = (\nu_1, \ldots, \nu_{k-1}) \in \mathbb{Z}^{k-1},$$ which form a disjoint covering $\mathbb{R}^{k-1}=\bigcup_{\vec{\nu}\in\mathbb{Z}^{k-1}} Q_{\vec\nu}$ and permit the integral decomposition $$\mathcal{I}^{(k)}_{\alpha,\gamma}(M\tau_{c}) = \frac{1}{(2\pi)^{k-1}}\sum_{\vec{\nu}\in\mathbb{Z}^{k-1}}\int_{Q_{\vec{\nu}}}d\vec\omega_{k-1}\sum_{\mathbf{a},\mathbf{b}}\bar{\lambda}_{\alpha,\gamma}(\mathbf{a})\,\Sha^{(k)}(\vec\omega, M\tau_{c})\,F^{(k)}_{\mathbf{a},\mathbf{b}}(\vec\omega,\tau_{c})\,\tilde{S}^{(k)}_{\mathbf{b}}(\vec\omega_{k-1}).$$
For each hypercube $Q_{\vec{\nu}}$, we substitute $\vec{u} = \vec\omega_{k-1} - \vec{\nu}\omega_h$ so that $u_j \in \left(-\frac{\omega_h}{2}, \frac{\omega_h}{2}\right]$ for each component. Using $\nu_j\omega_h\tau_c = 2\pi\nu_j$ and $\sin(x + n\pi) = (-1)^n\sin(x)$ for integer $n$, the comb function simplifies to $$\Sha^{(k)}(\vec{u} + \vec{\nu}\omega_h, M\tau_{c}) = \frac{\sin(|\vec{u}_{k-1}|M\tau_c/2)}{\sin(|\vec{u}_{k-1}|\tau_c/2)}\prod_{j=1}^{k-1}\frac{\sin(u_jM\tau_c/2)}{\sin(u_j\tau_c/2)}.$$ Defining the fundamental interval about each peak, $\mathcal{U}_h = \left(-\frac{\omega_h}{2}, \frac{\omega_h}{2}\right]$,
$$\mathcal{I}^{(k)}_{\alpha,\gamma}(M\tau_{c}) = \frac{1}{(2\pi)^{k-1}}\sum_{\vec{\nu}\in\mathbb{Z}^{k-1}}\int_{\mathcal{U}_h^{k-1}}d\vec{u}\sum_{\mathbf{a},\mathbf{b}}\bar{\lambda}_{\alpha,\gamma}(\mathbf{a})\,\Sha^{(k)}(\vec{u}, M\tau_{c})\,F^{(k)}_{\mathbf{a},\mathbf{b}}(\vec{u}+\vec{\nu}\omega_h,\tau_{c})\,\tilde{S}^{(k)}_{\mathbf{b}}(\vec{u}+\vec{\nu}\omega_h)$$ and substituting $\vec{u} = \vec{\zeta}\frac{2}{M\tau_c}$ with $d\vec{u} = \left(\frac{2}{M\tau_c}\right)^{k-1}d\vec{\zeta}$, we have
\begin{align*}
    \mathcal{I}^{(k)}_{\alpha,\gamma}(M\tau_{c}) &=\frac{1}{\left(\pi M\tau_c\right)^{k-1}}\sum_{\vec{\nu}\in\mathbb{Z}^{k-1}}\int_{-\pi M/2}^{\pi M/2}d\vec{\zeta}\,\sum_{\mathbf{a},\mathbf{b}}\bar{\lambda}_{\alpha,\gamma}(\mathbf{a})\,\frac{\sin(|\vec{\zeta}_{k-1}|)}{\sin(|\vec{\zeta}_{k-1}|/M)}\prod_{j=1}^{k-1}\frac{\sin(\zeta_j)}{\sin(\zeta_j/M)}\\
    &\hspace{6cm}\times F^{(k)}_{\mathbf{a},\mathbf{b}}\left(\vec{\zeta}\frac{2}{M\tau_c}+\vec{\nu}\omega_h,\tau_{c}\right)\tilde{S}^{(k)}_{\mathbf{b}}\left(\vec{\zeta}\frac{2}{M\tau_c}+\vec{\nu}\omega_h\right).
\end{align*}
In the limit $M\to\infty$, the small-angle approximation $\sin(\zeta_j/M) \approx \zeta_j/M$ simplifies the integration to 
\begin{align*}
\lim_{M\to\infty}\mathcal{I}^{(k)}_{\alpha,\gamma}(M\tau_{c}) &= \frac{1}{\left(\pi M\tau_c\right)^{k-1}}\sum_{\vec{\nu}\in\mathbb{Z}^{k-1}}\sum_{\mathbf{a},\mathbf{b}}\bar{\lambda}_{\alpha,\gamma}(\mathbf{a})F^{(k)}_{\mathbf{a},\mathbf{b}}(\vec{\nu}\omega_h,\tau_{c})\tilde{S}^{(k)}_{\mathbf{b}}(\vec{\nu}\omega_h)\int_{-\infty}^{\infty}d\vec{\zeta}\,\frac{\sin(|\vec{\zeta}_{k-1}|)}{\sin(|\vec{\zeta}_{k-1}|/M)}\prod_{j=1}^{k-1}\frac{\sin(\zeta_j)}{\sin(\zeta_j/M)}\\
&\approx\frac{M}{\left(\pi \tau_c\right)^{k-1}}\sum_{\vec{\nu}\in\mathbb{Z}^{k-1}}\sum_{\mathbf{a},\mathbf{b}}\bar{\lambda}_{\alpha,\gamma}(\mathbf{a})F^{(k)}_{\mathbf{a},\mathbf{b}}(\vec{\nu}\omega_h,\tau_{c})\tilde{S}^{(k)}_{\mathbf{b}}(\vec{\nu}\omega_h)\int_{-\infty}^{\infty}d\vec{\zeta}\,\frac{\sin(|\vec{\zeta}_{k-1}|)}{|\vec{\zeta}_{k-1}|}\prod_{j=1}^{k-1}\frac{\sin(\zeta_j)}{\zeta_j}
\end{align*}
Noting that the multidimensional sinc integral evaluates to $$\int_{-\infty}^{\infty}d\vec{\zeta}\,\frac{\sin(|\vec{\zeta}_{k-1}|)}{|\vec{\zeta}_{k-1}|}\prod_{j=1}^{k-1}\frac{\sin(\zeta_j)}{\zeta_j} = \pi^{k-1},$$
we conclude by re-introducing the suppressed variables $(\vec{l},\pi,\varpi)$ to obtain the discrete frequency-comb approximation to the spectral overlap integral
\begin{align*} 
\lim_{M\to\infty}\mathcal{I}^{(k)}_{\alpha,\gamma}(M\tau_{c}) &\approx \frac{M}{\tau_c^{k-1}}\sum_{\substack{\mathbf{a},\mathbf{b},\\\pi,\varpi\\ \vec{l} \in \mathcal{L}_k}}\bar{\lambda}_{\alpha,\gamma}(\mathbf{a},\pi,\vec{l}\,)\sum_{\vec{\nu}\in\mathbb{Z}^{k-1}}F^{(k)}_{\mathbf{a},\mathbf{b}}(\vec{\nu}\omega_h,\tau_{c})\tilde{S}^{(k)}_{\mathbf{b};\vec{l};\pi;\varpi}(\vec{\nu}\omega_h).
\end{align*}
Thus the overlap integral samples the polyspectrum discretely at the harmonic frequencies $\vec{\nu}\omega_h$, with each component weighted by the corresponding filter function value.
\section{Direct calculation of \texorpdfstring{$\mathcal{I}^{(k)}_{\alpha,\gamma}(T)$}{I(k)alpha,gamma(T)} up to \texorpdfstring{$k=3$}{k=3}}\label{app:direct_calc_3}
The following rows give the phase-free order-resolved overlaps of
Eq.~\eqref{eq:app_final_overlap}.  Their contributions to the
real-valued coefficients $\mathcal{I}_{\alpha,\gamma}(T)$ in the exponent
are obtained from
Eq.~\eqref{eq:app_overlap_sum}; explicit factors of $\mathrm{i}$ in the rows
below arise from the Pauli products rather than the cumulant-order phase.

For toggling-control axes $\mathbf{a}=\{x,z\}$, the second-order dynamics
correspond to $\gamma\in\{\mathbbm{1},y\}$:
\begin{align*}
\mathcal{I}^{(2)}_{x,\mathbbm{1}}(T)
&=\frac{1}{2\pi}\int_{\mathbb{R}}d\omega\,
F_{zz}(\omega,T)\tilde{S}^{(+)}(\omega),
&
\mathcal{I}^{(2)}_{x,y}(T)
&=\frac{\mathrm{i}}{2\pi}\int_{\mathbb{R}}d\omega\,
F_{zx}(\omega,T)\tilde{S}^{(+)}(\omega),
\\
\mathcal{I}^{(2)}_{y,\mathbbm{1}}(T)
&=\frac{1}{2\pi}\int_{\mathbb{R}}d\omega\,
\left(F_{xx}(\omega,T)+F_{zz}(\omega,T)\right)
\tilde{S}^{(+)}(\omega),
&
\mathcal{I}^{(2)}_{y,y}(T)
&=-\frac{\mathrm{i}}{2\pi}\int_{\mathbb{R}}d\omega\,
\left(F_{xz}(\omega,T)-F_{zx}(\omega,T)\right)
\tilde{S}^{(-)}(\omega),
\\
\mathcal{I}^{(2)}_{z,\mathbbm{1}}(T)
&=\frac{1}{2\pi}\int_{\mathbb{R}}d\omega\,
F_{xx}(\omega,T)\tilde{S}^{(+)}(\omega),
&
\mathcal{I}^{(2)}_{z,y}(T)
&=-\frac{\mathrm{i}}{2\pi}\int_{\mathbb{R}}d\omega\,
F_{xz}(\omega,T)\tilde{S}^{(+)}(\omega).
\end{align*}

The third-order dynamics correspond to $\gamma\in\{x,z\}$.  For $\gamma=x$,
\begin{align*}
\mathcal{I}^{(3)}_{x,x}(T)
&=-\frac{1}{(2\pi)^2}
\int_{\mathbb{R}^2}d\vec\omega_2\,
\left(F_{zxz}(\vec\omega_2,T)-F_{zzx}(\vec\omega_2,T)\right)
\tilde{S}^{(+-)}(\vec\omega_2),
\\
\mathcal{I}^{(3)}_{y,x}(T)
&=\frac{1}{(2\pi)^2}
\int_{\mathbb{R}^2}d\vec\omega_2\,\Big[
\left(F_{xxx}(\vec\omega_2,T)+F_{zzx}(\vec\omega_2,T)\right)
\tilde{S}^{(++)}(\vec\omega_2)
+\left(F_{xzz}(\vec\omega_2,T)-F_{zxz}(\vec\omega_2,T)\right)
\tilde{S}^{(--)}(\vec\omega_2)\Big],
\\
\mathcal{I}^{(3)}_{z,x}(T)
&=\frac{1}{(2\pi)^2}
\int_{\mathbb{R}^2}d\vec\omega_2\,
\left(F_{xxx}(\vec\omega_2,T)+F_{xzz}(\vec\omega_2,T)\right)
\tilde{S}^{(++)}(\vec\omega_2).
\end{align*}
For $\gamma=z$,
\begin{align*}
\mathcal{I}^{(3)}_{x,z}(T)
&=\frac{1}{(2\pi)^2}
\int_{\mathbb{R}^2}d\vec\omega_2\,
\left(F_{zxx}(\vec\omega_2,T)+F_{zzz}(\vec\omega_2,T)\right)
\tilde{S}^{(++)}(\vec\omega_2),
\\
\mathcal{I}^{(3)}_{y,z}(T)
&=\frac{1}{(2\pi)^2}
\int_{\mathbb{R}^2}d\vec\omega_2\,\Big[
\left(F_{xxz}(\vec\omega_2,T)+F_{zzz}(\vec\omega_2,T)\right)
\tilde{S}^{(++)}(\vec\omega_2)
-\left(F_{xzx}(\vec\omega_2,T)-F_{zxx}(\vec\omega_2,T)\right)
\tilde{S}^{(--)}(\vec\omega_2)\Big],
\\
\mathcal{I}^{(3)}_{z,z}(T)
&=\frac{1}{(2\pi)^2}
\int_{\mathbb{R}^2}d\vec\omega_2\,
\left(F_{xxz}(\vec\omega_2,T)-F_{xzx}(\vec\omega_2,T)\right)
\tilde{S}^{(+-)}(\vec\omega_2).
\end{align*}

\subsection{Direct calculation of \texorpdfstring{$\mathcal{I}^{(k)}_{z,\gamma}(T)$}{I(k)z,gamma(T)} for \texorpdfstring{$\mathbf{a}\in\{x,y,z\}$}{a in x,y,z}}\label{app:direct_calc_3_general}
For general interaction-control axes $\mathbf{a}=\{x,y,z\}$ and
$\alpha=z$, the second-order overlaps are
\begin{align*}
\mathcal{I}^{(2)}_{z,\mathbbm{1}}(T)
&=\frac{1}{2\pi}\int_{\mathbb{R}}d\omega\,
\left(F_{xx}(\omega,T)+F_{yy}(\omega,T)\right)
\tilde{S}^{(+)}(\omega),
\\
\mathcal{I}^{(2)}_{z,x}(T)
&=\frac{\mathrm{i}}{2\pi}\int_{\mathbb{R}}d\omega\,
F_{yz}(\omega,T)\tilde{S}^{(+)}(\omega),
\\
\mathcal{I}^{(2)}_{z,y}(T)
&=-\frac{\mathrm{i}}{2\pi}\int_{\mathbb{R}}d\omega\,
F_{xz}(\omega,T)\tilde{S}^{(+)}(\omega),
\\
\mathcal{I}^{(2)}_{z,z}(T)
&=-\frac{\mathrm{i}}{2\pi}\int_{\mathbb{R}}d\omega\,
\left(F_{yx}(\omega,T)-F_{xy}(\omega,T)\right)
\tilde{S}^{(-)}(\omega).
\end{align*}
The corresponding third-order overlaps are
\begin{align*}
\mathcal{I}^{(3)}_{z,\mathbbm{1}}(T)
&=-\frac{\mathrm{i}}{(2\pi)^2}
\int_{\mathbb{R}^2}d\vec\omega_2\,\Big[
\left(F_{xzy}(\vec\omega_2,T)-F_{yzx}(\vec\omega_2,T)\right)
\tilde{S}^{(++)}(\vec\omega_2)
-\left(F_{xyz}(\vec\omega_2,T)-F_{yxz}(\vec\omega_2,T)\right)
\tilde{S}^{(--)}(\vec\omega_2)\Big],
\\
\mathcal{I}^{(3)}_{z,x}(T)
&=\frac{1}{(2\pi)^2}
\int_{\mathbb{R}^2}d\vec\omega_2\,\Big[
\left(F_{xxx}(\vec\omega_2,T)+F_{xzz}(\vec\omega_2,T)
+F_{yyx}(\vec\omega_2,T)\right)\tilde{S}^{(++)}(\vec\omega_2)
\nonumber\\[-2pt]
&\hspace{9.0cm}
+\left(F_{xyy}(\vec\omega_2,T)-F_{yxy}(\vec\omega_2,T)\right)
\tilde{S}^{(--)}(\vec\omega_2)\Big],
\\
\mathcal{I}^{(3)}_{z,y}(T)
&=\frac{1}{(2\pi)^2}
\int_{\mathbb{R}^2}d\vec\omega_2\,\Big[
\left(F_{xxy}(\vec\omega_2,T)+F_{yyy}(\vec\omega_2,T)
+F_{yzz}(\vec\omega_2,T)\right)\tilde{S}^{(++)}(\vec\omega_2)
\nonumber\\[-2pt]
&\hspace{9.0cm}
-\left(F_{xyx}(\vec\omega_2,T)-F_{yxx}(\vec\omega_2,T)\right)
\tilde{S}^{(--)}(\vec\omega_2)\Big],
\\
\mathcal{I}^{(3)}_{z,z}(T)
&=\frac{1}{(2\pi)^2}
\int_{\mathbb{R}^2}d\vec\omega_2\,
\left(F_{xxz}(\vec\omega_2,T)-F_{xzx}(\vec\omega_2,T)
+F_{yyz}(\vec\omega_2,T)-F_{yzy}(\vec\omega_2,T)\right)
\tilde{S}^{(+-)}(\vec\omega_2).
\end{align*}

\section{Generalised \texorpdfstring{$\mathcal{I}_{\alpha,\gamma}(T)$}{I alpha,gamma(T)} calculation from qubit tomography}\label{appendix:cbar_calcs}
\noindent~Using the all-real component convention of
\appref{app:real_structure}, the effective propagator is
$$
V_{\Lambda_\alpha}(T)
=\exp\left\{
\mathcal{I}_{\alpha,\mathbbm{1}}(T)\Lambda_{\mathbbm{1}}
+\mathcal{I}_{\alpha,\alpha}(T)\Lambda_\alpha
-\mathrm{i}\!\!\sum_{\gamma\in\{x,y,z\}\setminus\{\alpha\}}
\mathcal{I}_{\alpha,\gamma}(T)\Lambda_\gamma
\right\}.
$$
The expectation value of a traceless observable $O$ at time $T$ is determined by decomposing the toggling-frame observable $\bar{O}(T)$ into the Pauli basis, with coefficients $o_\alpha(T)=\Tr[\bar{O}(T)\Lambda_\alpha^\dagger]/d_s$. Linearity of the trace gives
\begin{align*}
\langle O(T) \rangle
&=\sum_{\alpha\in\{x,y,z\}}o_\alpha(T)\,
\Tr[V_{\Lambda_\alpha}(T)\rho_S\Lambda_\alpha]\nonumber\\
&=\frac{1}{d_s}\sum_{\alpha\in\{x,y,z\}}
\Tr[\bar{O}(T)\Lambda_\alpha^\dagger]\,
\Tr[V_{\Lambda_\alpha}(T)\rho_S\Lambda_\alpha].
\end{align*}
The notation is now specialised to the single-qubit case. To determine the noise-dependent action, we must calculate $\mathcal{I}_{\alpha,\gamma}(T)$ for the three traceless observables $\alpha\in\{x,y,z\}$ with respect to the orthonormal basis $\gamma\in\{\mathbbm{1},x,y,z\}$. For each fixed $\alpha$, a minimal informationally complete scheme uses the four initialisation states $\rho_\xi=(\Lambda_\xi + \mathbbm{1})/\Tr[\Lambda_\xi+\mathbbm{1}]$, with $\xi\in\{\mathbbm{1},x,y,z\}$. Thus, the generic Pauli-qubit scheme requires at least $3\times4=12$ independent preparation--measurement configurations. Further configurations may be added to form a redundant, overcomplete tomographic dataset.

Introducing
\begin{align*}
\bar{\mathcal{I}}
\equiv\left(
\sum_{\gamma\in\{x,y,z\}\setminus\{\alpha\}}
\mathcal{I}_{\alpha,\gamma}(T)^2
-\mathcal{I}_{\alpha,\alpha}(T)^2
\right)^{1/2},
\end{align*}
the traceless part of the exponent squares to
$-\bar{\mathcal{I}}^2\mathbbm{1}$, and the effective propagator expands as
\begin{align*}
\langle \Lambda_\alpha(T)\rangle_\xi 
&= \Tr[V_{\Lambda_\alpha}(T)\rho_\xi \Lambda_\alpha] \nonumber\\
&= e^{\mathcal{I}_{\alpha,\mathbbm{1}}(T)}
\Tr\!\left[
\left(
\cos(\bar{\mathcal{I}})\mathbbm{1}
+\frac{\sin(\bar{\mathcal{I}})}{\bar{\mathcal{I}}}
\left(
\mathcal{I}_{\alpha,\alpha}(T)\Lambda_\alpha
-\mathrm{i}\!\!\sum_{\gamma\in\{x,y,z\}\setminus\{\alpha\}}
\mathcal{I}_{\alpha,\gamma}(T)\Lambda_\gamma
\right)
\right)\rho_\xi\Lambda_\alpha
\right].
\end{align*}
Although $\bar{\mathcal{I}}^2$ may take either sign,
$\cos(\bar{\mathcal{I}})$ and
$\sin(\bar{\mathcal{I}})/\bar{\mathcal{I}}$ are even functions of
$\bar{\mathcal{I}}$, so the expressions below remain real.
By evaluating the trace for the different preparation states $\xi$, we obtain the system of equations
\begin{equation}
\langle\Lambda_\alpha\rangle_\xi = 
\begin{cases}
\displaystyle
e^{\mathcal{I}_{\alpha,\mathbbm{1}}(T)}
\frac{\sin(\bar{\mathcal{I}})}{\bar{\mathcal{I}}}
\mathcal{I}_{\alpha,\alpha}(T),
& \xi=\mathbbm{1},
\\[10pt]
\displaystyle
e^{\mathcal{I}_{\alpha,\mathbbm{1}}(T)}
\left[
\delta_{\alpha,\xi}\cos(\bar{\mathcal{I}})
+\frac{\sin(\bar{\mathcal{I}})}{\bar{\mathcal{I}}}
\left(
\mathcal{I}_{\alpha,\alpha}(T)
+\sum_{\gamma\in\{x,y,z\}}
\epsilon_{\gamma\xi\alpha}\mathcal{I}_{\alpha,\gamma}(T)
\right)
\right],
& \xi\in\{x,y,z\},
\end{cases}
\end{equation}
where $\epsilon_{\gamma\xi\alpha}$ is the Levi-Civita symbol. Evaluated for the relevant expectation values, these relations suffice to solve for the 12 unknown $\mathcal{I}_{\alpha,\gamma}(T)$ values. In particular, $\mathcal{I}_{\alpha,\mathbbm{1}}(T)$ follows in closed form:
$$
\mathcal{I}_{\alpha,\mathbbm{1}}(T) = \frac{1}{2}\log\left(-\langle \Lambda_\alpha \rangle_\mathbbm{1}^2 + \sum_{\xi \in \{x,y,z\}} (\langle \Lambda_\alpha \rangle_\xi -\langle \Lambda_\alpha \rangle_\mathbbm{1} )^2\right).
$$
Equally, $\bar{\mathcal{I}}$ follows from
\begin{align}
\tan^2(\bar{\mathcal{I}}) =  \frac{-\langle \Lambda_\alpha \rangle_{\mathbbm{1}}^2+\sum_{\xi \in \{x,y,z\}\setminus\{\alpha\}} (\langle \Lambda_\alpha \rangle_{\xi} - \langle \Lambda_\alpha \rangle_{\mathbbm{1}})^2}{\text{sign}(\langle \Lambda_\alpha \rangle_{\alpha} - \langle \Lambda_\alpha \rangle_{\mathbbm{1}})(\langle \Lambda_\alpha \rangle_{\alpha} - \langle \Lambda_\alpha \rangle_{\mathbbm{1}})^2},
\end{align}
where the arc-tangent recovers $\bar{\mathcal{I}}$. The signs suppressed by the squares must be tracked through this operation to select the correct branch.

\section{Construction of the effective principal domain}\label{app:effective_domain}

This appendix details the operational procedure by which the effective principal domain $\mathcal{D}^{(\mathbf{a},\mathbf{b})}_{k-1}$ is constructed from the retained symmetries and then used to set up the reconstruction of Sec.~\ref{sec:theory}. The construction is the higher-order-spectra procedure of Chandran and Elgar~\cite{chandranGeneralProcedureDerivation1994}, specialised to the time-ordered polyspectra and to the symmetry content that survives a given control. We first state the general procedure, then carry it out explicitly for the second-order bispectral domains $\mathcal{D}^{(zzz)}_2$ and $\mathcal{D}^{(xzz)}_2$ reconstructed in Sec.~\ref{subsubsec:bispectrum_estimation}.

\subsection{General procedure}\label{app:effective_domain_procedure}

The unordered polyspectrum $S^{(k)}_{\mathbf{b}}(\vec{\omega}_{k-1})$ of a real, stationary process is invariant under a group of transformations of its $(k-1)$ independent frequency arguments generated by three structural symmetries~\cite{chandranGeneralProcedureDerivation1994,NikiasPetropulu1993HigherOrderSpectra}. Conjugation, $S^{(k)}_{\mathbf{b}}(\vec{\omega}_{k-1}) = S^{(k)}_{\mathbf{b}}(-\vec{\omega}_{k-1})^{*}$, follows from the reality of the correlation and holds for any real bath. Stationarity supplies the reflection that accompanies the delta constraint $\omega_k = -|\vec{\omega}_{k-1}|$ of Eq.~\eqref{eq:stationary_cond}, exchanging any argument with the dependent coordinate $\omega_k$. Argument permutation, $S^{(k)}_{\mathbf{b}}(\ldots,\omega_i,\ldots,\omega_j,\ldots) = S^{(k)}_{\mathbf{b}}(\ldots,\omega_j,\ldots,\omega_i,\ldots)$, follows from the permutation invariance of the fully symmetrised cumulant and generates a subgroup of the symmetric group $S_k$ acting on the arguments. Together these generate the full symmetry group of the unordered spectrum, whose fundamental region is the standard principal domain $\mathcal{D}_{k-1}$.

The effective principal domain is obtained by the same construction with one substitution: the permutation subgroup is replaced by the subgroup that the control leaves intact. The projection onto a given control-axis configuration $\mathbf{a}$ couples the probe to the component $\tilde{S}^{(k)}_{\mathbf{a}}$ whose filter $F^{(k)}_{\mathbf{a}}$ inherits only those argument permutations that fix $\mathbf{a}$. Writing $\mathrm{Stab}(\mathbf{a}) \subseteq S_k$ for the stabiliser of the axis word $\mathbf{a}$ under simultaneous permutation of axis labels and frequency arguments, the retained symmetry group is
\begin{equation}
\mathrm{Sym}_{\mathbf{a},\mathbf{b}} = \langle\, \mathcal{K},\; \mathcal{S},\; \mathrm{Stab}(\mathbf{a}) \,\rangle,
\label{eq:sym_group_def}
\end{equation}
the group generated by conjugation $\mathcal{K}$, the stationarity reflection $\mathcal{S}$, and the axis-preserving permutations. Conjugation and stationarity are common to every configuration; the permutation content, and hence the size of $\mathrm{Sym}_{\mathbf{a},\mathbf{b}}$, is set by the control alone. A configuration with all axes equal retains the full permutation subgroup and the largest symmetry group, giving the smallest fundamental region; a configuration whose axes are all distinct retains no non-trivial permutation and gives the largest fundamental region.

The domain follows as a fundamental region of the action of $\mathrm{Sym}_{\mathbf{a},\mathbf{b}}$ on frequency space, obtained by the standard procedure of retaining one representative from each orbit~\cite{chandranGeneralProcedureDerivation1994}: enumerate the group elements as linear maps on $\vec{\omega}_{k-1}$, form the orbit of a generic point, and select the wedge bounded by the fixed-point hyperplanes of the generators. The smaller group $\mathrm{Sym}_{\mathbf{a},\mathbf{b}}$ (relative to the fully permutation-symmetric case) partitions frequency space into fewer, larger orbits, its fundamental region therefore contains the standard one
\begin{equation}
\mathcal{D}_{k-1} \subseteq \mathcal{D}^{(\mathbf{a},\mathbf{b})}_{k-1},
\end{equation}
with equality if and only if the control retains the full permutation subgroup. This is the operational content of the multiplicity factor $m_{\mathbf{a},\mathbf{b}}(\vec{s}_{k-1})$ in Eq.~\eqref{eq:spectrum_comb_estimate}: it counts the size of the orbit under $\mathrm{Sym}_{\mathbf{a},\mathbf{b}}$ of the lattice point $\vec{s}_{k-1}$, that is, the number of sampled harmonics that collapse onto a single representative in $\mathcal{D}^{(\mathbf{a},\mathbf{b})}_{k-1}$.

In the reconstruction, the domain plays a single concrete role: it fixes which harmonics of the sampling lattice $\mathcal{R}^{(\mathbf{a},\mathbf{b})}_{k-1}$ are independent unknowns. Once $\mathcal{D}^{(\mathbf{a},\mathbf{b})}_{k-1}$ is truncated to the bounded set $\Omega^{(\mathbf{a},\mathbf{b})}_{k-1}$ of $N_h$ harmonics below the cutoff, each row of $\mathbf{G}_{\mathbf{a},\mathbf{b}}$ is one control sequence evaluated at those $N_h$ representatives, weighted by $m_{\mathbf{a},\mathbf{b}}$, and the reconstruction solves for the polyspectrum at exactly those points. Harmonics lying outside the domain are not measured independently; they are images of interior points under $\mathrm{Sym}_{\mathbf{a},\mathbf{b}}$ and are recovered by symmetry once the representatives are known.

\subsection{Bispectral domains \texorpdfstring{$\mathcal{D}^{(zzz)}_2$ and $\mathcal{D}^{(xzz)}_2$}{D(zzz) and D(xzz)}}\label{app:effective_domain_bispectrum}

At second order in the reduced frequencies ($k=3$, two independent arguments $\vec{\omega}_2 = (\omega_1,\omega_2)$) the three structural symmetries of the classical bispectrum generate a group of order twelve, partitioning the $(\omega_1,\omega_2)$ plane into twelve equivalent sectors whose fundamental region is the minimal wedge $\mathcal{D}_2 = \{0 \le \omega_2 \le \omega_1\}$~\cite{NikiasPetropulu1993HigherOrderSpectra,sungNonGaussianNoiseSpectroscopy2019a}. The permutation content here is the full argument-swap subgroup $\{e,(12)\}$ together with the two stationarity reflections, which close into the six-element dihedral action on the plane; conjugation supplies the remaining factor of two.

For the single-axis configuration $\mathbf{a}=zzz$ the filter $F_{zzz}(\omega_1,\omega_2) = F_z(\omega_1)F_z(\omega_2)F_z(-\omega_1-\omega_2)$ is built from a single control axis and is therefore invariant under every permutation of its three arguments: $\mathrm{Stab}(zzz)=S_3$. The retained group $\mathrm{Sym}_{zzz}$ is the full order-twelve group, and the effective domain coincides with the standard wedge,
\begin{equation}
\mathcal{D}^{(zzz)}_2 = \mathcal{D}_2 = \{\, (\omega_1,\omega_2) \mid 0 \le \omega_2 \le \omega_1 \,\}.
\end{equation}

For the multi-axis configuration $\mathbf{a}=xzz$ the two $z$-coupled arguments remain interchangeable but the $x$-coupled argument does not, so the retained permutation subgroup collapses to the single transposition of the like axes, $\mathrm{Stab}(xzz)=\{e,(23)\}$. The stationarity reflection that would exchange the distinguished $x$ argument with the dependent coordinate is likewise broken, since it does not fix the axis word. The retained group $\mathrm{Sym}_{xzz}$ is correspondingly smaller, its orbits are smaller, and its fundamental region is larger: the reconstruction must resolve a region that the single-axis protocol folds away,
\begin{equation}
\mathcal{D}^{(xzz)}_2 \supset \mathcal{D}^{(zzz)}_2,
\end{equation}
comprising the minimal wedge together with the adjacent sector related to it by the broken $x$-argument permutation. This is the extended domain over which the permutation-asymmetric projection $\tilde{S}^{(++)}_{xzz}$ is reconstructed in Fig.~\ref{fig:bispectrum_combined}(c,d), and the containment $\mathcal{D}^{(zzz)}_2 \subset \mathcal{D}^{(xzz)}_2$ is what allows the symmetric sector to be characterised first and then used to constrain the estimate on the extended region.

The construction depends on the ordering of the axis word, not merely on its multiset of axes, since the stabiliser $\mathrm{Stab}(\mathbf{a})$ acts on labelled positions. The configurations $xzz$ and $zxz$ carry the same axes but place the distinguished $x$ coupling in different argument positions, so their retained transpositions differ, $\{e,(23)\}$ against $\{e,(13)\}$, and the associated fundamental regions are reflections of one another across the wedge bisector,
\begin{equation}
\mathcal{D}^{(xzz)}_2 \neq \mathcal{D}^{(zxz)}_2 .
\end{equation}
The two domains are equivalent under relabelling and carry identical spectral information; the distinction matters operationally because the control axes are fixed by the experiment, and the harmonics to be sampled must be taken from the domain matching the realised axis ordering rather than an arbitrary member of its relabelling class.

\section{Control sequence parameters}
\label{app:control_sequences}

This appendix specifies the control sequences used for the spectral reconstructions presented in the main text. The parameters for the second-order spectra reconstructions in Figs.~\ref{fig:gauss_ordered_spectrum} and~\ref{fig:second_order_plots} are provided in Table~\ref{tab:k2_sequences}. Due to the large number of pulses and sequences used for the bispectrum reconstructions in \secref{subsec:results_quantum}, the complete parameter sets are provided as supplementary data files in machine-readable format. Each file contains the pulse timings $t_j$, amplitudes $A_j$, widths $w_j$, and phase rotations $\phi_R$ for all sequences in the corresponding reconstruction. The supplementary files are named according to the convention \texttt{bispectrum\_<filter>\_<component>.csv} where \texttt{<filter>} is either \texttt{zzz} or \texttt{xzz}, and \texttt{<component>} is either \texttt{real} or \texttt{imag}.

\subsection{Pulse parametrisation}
Each control sequence comprises a train of $N_p$ pulses with a cosine envelope,
\begin{equation}
f(t) = \sum_{j=1}^{N_p} \frac{A_j}{w_j}\left[\cos\left(\pi + \frac{2\pi(t - t_j)}{w_j}\right) + 1\right] \mathbf{1}_{[t_j, t_j + w_j]}(t),
\label{eq:pulse_envelope}
\end{equation}
where $\mathbf{1}_{[a,b]}(t)$ denotes the indicator function on the interval $[a,b]$. The parameters $t_j$, $A_j \in [-\pi, \pi]$, and $w_j$ specify the start time, amplitude, and temporal width of the $j$th pulse, respectively. Following the experimental configuration of~\cite{sungNonGaussianNoiseSpectroscopy2019a}, all sequences employ a fixed cycle time $\tau_c = \SI{960}{\nano\second}$ and minimum pulse width $w_{\min} = \SI{11}{\nano\second}$. Each base sequence is repeated $M = 10$ times to establish the frequency-comb structure required for spectral sampling at harmonics $\omega_m = 2\pi m / \tau_c$. Additionally, a global rotation $R_y(\phi_R)$ with phase $\phi_R \in [0, \pi/2]$ is applied to the qubit state immediately before and after the controlled evolution.

To reconstruct $N_h$ harmonics of a given spectrum, we engineer a minimal set of $N_s = N_h$ linearly independent control sequences. In Table~\ref{tab:k2_sequences}, each column corresponds to a control sequence, with pulse parameters given as tuples $(t_j, A_j, w_j)$ where $t_j$ and $w_j$ are in nanoseconds and $A_j$ is in units of $\pi$.

\subsection{Second-order spectra}

Table~\ref{tab:k2_sequences} details the control sequences used to reconstruct the second-order time-ordered spectra $\tilde{S}^{(\pm)}(\omega)$ for the classical Gaussian (\secref{subsec:results_gaussian_classical}) and quantum non-Gaussian (\secref{subsec:results_quantum}) environments. Sequences $n = 1$--$8$ are designed to reconstruct the real component $\mathrm{Re}[\tilde{S}^{(+)}(\omega)]$ on the eight principal harmonics of the sampling domain $\Omega_1^{(zz)}$. The same set, excluding $n = 1$, is used for the spectroscopy of the dispersive components $\mathrm{Im}[\tilde{S}^{(\pm)}(\omega)]$ on the sampling domain $\Omega_1^{(zx)}$. The column labelled ``$2\tau_c$'' specifies the parameters for the sequence with a doubled cycle time $2\tau_c = \SI{1920}{\nano\second}$, used to resolve the sharp spectral feature within the interval $(3\omega_h, 4\omega_h)$ as discussed in Fig.~\ref{fig:gauss_ordered_spectrum}(a).

\begin{figure*}[htbp]
\centering
\includegraphics[width=\columnwidth]{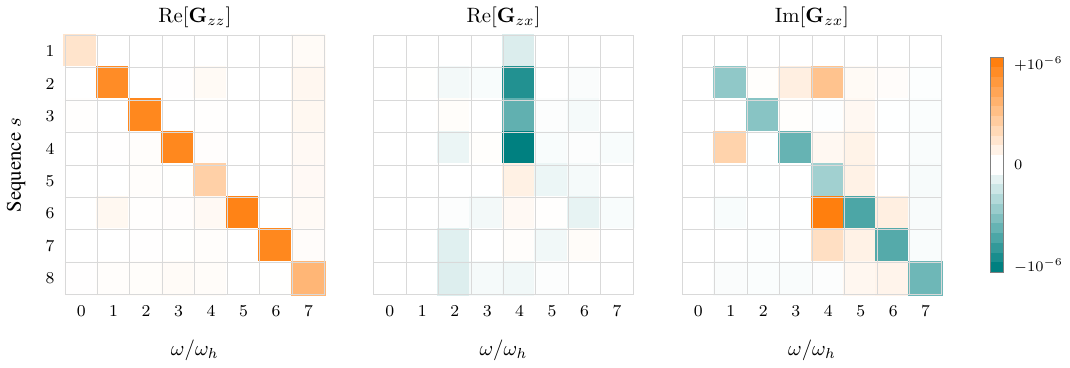}
\caption{Heat maps of the filter function matrices $\mathbf{G}$, composed of $N_s$-rows of optimised control sequence filter functions evaluated at the $N_h$ harmonic frequencies of the reconstruction domains of $\tilde{S}^{(\pm)}(\omega)$. (Left) The approximately diagonal matrix $\mathrm{Re}[\mathbf{G}_{zz}]$ supports near linearly independent row-vectors of control filter functions, enabling well-conditioned inversion and recovery of the real, classical noise spectra $\mathrm{Re}[\tilde{S}^{(+)}(\omega)]$ via $\vec{\mathcal{I}}_{x,\mathbbm{1}}$. (Centre) The real component of the transverse filter function matrix, $\mathrm{Re}[\mathbf{G}_{zx}]$, exhibits suppressed spectral power relative to (Right) the well-conditioned imaginary matrix component $\mathrm{Im}[\mathbf{G}_{zx}]$. This differential suppression, $|\mathrm{Re}[\mathbf{G}_{zx}]| \ll |\mathrm{Im}[\mathbf{G}_{zx}]|$, ensures that the linear system for $\vec{\mathcal{I}}_{x,y}$ predominantly samples the $\mathrm{Im}[\tilde{S}^{(\pm)}(\omega)]$ on $\Omega_1^{(zx)}=\{\omega_h,\dots,7\omega_h\}$.}
\label{fig:k2_filter_matrices}
\end{figure*}

The $N_s$ filter functions generated by these sequences are band-limited to the sampling domain to suppress spectral leakage. Evaluating Eq.~\eqref{eq:spectrum_comb_estimate} at the $N_h$ harmonic frequencies yields the filter function matrices $\mathrm{Re}[\mathbf{G}_{zz}]$, $\mathrm{Re}[\mathbf{G}_{zx}]$, and $\mathrm{Im}[\mathbf{G}_{zx}]$ shown in Fig.~\ref{fig:k2_filter_matrices}. The matrices $\mathrm{Re}[\mathbf{G}_{zz}]$ and $\mathrm{Im}[\mathbf{G}_{zx}]$ in the left and right panels exhibit an approximately diagonal structure and $\mathcal{O}(1)$ condition numbers, indicating well-conditioned inversion. Furthermore, the sequences for $\mathrm{Im}[\tilde{S}^{(\pm)}(\omega)]$ estimation were optimised to satisfy $|\mathrm{Re}[\mathbf{G}_{zx}]| \ll |\mathrm{Im}[\mathbf{G}_{zx}]|$, thereby minimising cross-contamination of errors from the $\mathrm{Re}[\tilde{S}^{(+)}(\omega)]$ channel.

\begin{table*}[htbp]
\centering
\caption{Control sequences for second-order spectra $\tilde{S}^{(\pm)}(\omega)$. Sequences $n = 1$--$8$ reconstruct the principal harmonics of $\mathrm{Re}[\tilde{S}^{(+)}(\omega)]$ and $\mathrm{Im}[\tilde{S}^{(\pm)}(\omega)]$ for both classical (Fig.~\ref{fig:gauss_ordered_spectrum}) and quantum (Fig.~\ref{fig:second_order_plots}) environments. The adaptive sequence employs cycle time $2\tau_c$ to sample the inter-harmonic feature in Fig.~\ref{fig:gauss_ordered_spectrum}(a).}
\label{tab:k2_sequences}
\resizebox{\textwidth}{!}{%
\begin{tabular}{cccccccccc}
\toprule
 & \multicolumn{8}{c}{Sequence index $n$} & \\
\cmidrule(lr){2-9}
$j$ & $1$ & $2$ & $3$ & $4$ & $5$ & $6$ & $7$ & $8$ & $2\tau_c$ \\
\midrule
$1$ & $(10,\,1,\,60)$ & $(146,\,1,\,150)$ & $(8,\,1,\,150)$ & $(10,\,1,\,150)$ & $(54,\,1,\,16)$ & $(4,\,1,\,150)$ & $(8,\,1,\,45)$ & $(47,\,1,\,138)$ & $(35,\,1,\,60)$ \\
$2$ & $(40,\,1,\,60)$ & $(226,\,1,\,84)$ & $(224,\,1,\,150)$ & $(183,\,1,\,150)$ & $(150,\,0.76,\,26)$ & $(101,\,1,\,150)$ & $(91,\,1,\,23)$ & $(108,\,1,\,150)$ & $(170,\,1,\,60)$ \\
$3$ & $(105,\,1,\,60)$ & $(244,\,1,\,11)$ & $(235,\,1,\,45)$ & $(185,\,1,\,58)$ & $(244,\,0.82,\,25)$ & $(198,\,1,\,150)$ & $(134,\,1,\,114)$ & $(178,\,1,\,150)$ & $(310,\,1,\,60)$ \\
$4$ & $(135,\,1,\,60)$ & $(267,\,1,\,84)$ & $(260,\,1,\,11)$ & $(221,\,1,\,11)$ & $(290,\,0.92,\,144)$ & $(293,\,1,\,150)$ & $(216,\,1,\,111)$ & $(247,\,1,\,150)$ & $(445,\,1,\,60)$ \\
$5$ & $(200,\,1,\,60)$ & $(281,\,1,\,150)$ & $(292,\,1,\,83)$ & $(322,\,1,\,150)$ & $(416,\,1,\,93)$ & $(391,\,1,\,150)$ & $(305,\,1,\,92)$ & $(313,\,1,\,150)$ & $(585,\,1,\,60)$ \\
$6$ & $(230,\,1,\,60)$ & $(283,\,1,\,11)$ & $(298,\,1,\,150)$ & $(490,\,1,\,150)$ & $(434,\,1,\,60)$ & $(485,\,1,\,150)$ & $(364,\,1,\,134)$ & $(384,\,1,\,150)$ & $(720,\,1,\,60)$ \\
$7$ & $(295,\,1,\,60)$ & $(323,\,1,\,11)$ & $(311,\,1,\,11)$ & $(618,\,1,\,150)$ & $(449,\,1,\,18)$ & $(582,\,1,\,150)$ & $(466,\,1,\,88)$ & $(420,\,1,\,11)$ & $(860,\,1,\,60)$ \\
$8$ & $(325,\,1,\,60)$ & $(626,\,1,\,150)$ & $(348,\,1,\,11)$ & $(680,\,1,\,150)$ & $(534,\,1,\,148)$ & $(677,\,1,\,150)$ & $(513,\,1,\,150)$ & $(420,\,1,\,11)$ & $(995,\,1,\,60)$ \\
$9$ & $(390,\,1,\,60)$ & $(706,\,1,\,84)$ & $(507,\,1,\,67)$ & $(689,\,1,\,73)$ & $(577,\,1,\,58)$ & $(828,\,1,\,32)$ & $(602,\,1,\,137)$ & $(452,\,1,\,150)$ & $(1135,\,1,\,60)$ \\
$10$ & $(420,\,1,\,60)$ & $(724,\,1,\,11)$ & $(512,\,1,\,150)$ & $(703,\,1,\,11)$ & $(601,\,1,\,15)$ & $(928,\,1,\,32)$ & $(695,\,1,\,94)$ & $(521,\,1,\,150)$ & $(1270,\,1,\,60)$ \\
$11$ & $(485,\,1,\,60)$ & $(747,\,1,\,84)$ & $(548,\,1,\,11)$ & $(735,\,1,\,11)$ & $(667,\,1,\,123)$ & --- & $(700,\,1,\,122)$ & $(589,\,1,\,150)$ & $(1410,\,1,\,60)$ \\
$12$ & $(515,\,1,\,60)$ & $(761,\,1,\,150)$ & $(734,\,1,\,44)$ & $(807,\,1,\,150)$ & $(894,\,-0.5,\,66)$ & --- & $(746,\,1,\,11)$ & $(657,\,1,\,150)$ & $(1545,\,1,\,60)$ \\
$13$ & $(580,\,1,\,60)$ & $(763,\,1,\,11)$ & $(735,\,1,\,150)$ & --- & --- & --- & $(771,\,1,\,119)$ & $(727,\,1,\,150)$ & $(1685,\,1,\,60)$ \\
$14$ & $(610,\,1,\,60)$ & $(803,\,1,\,11)$ & $(760,\,1,\,11)$ & --- & --- & --- & $(862,\,1,\,98)$ & $(799,\,1,\,140)$ & $(1820,\,1,\,60)$ \\
$15$ & $(675,\,1,\,60)$ & --- & --- & --- & --- & --- & --- & --- & --- \\
$16$ & $(705,\,1,\,60)$ & --- & --- & --- & --- & --- & --- & --- & --- \\
$17$ & $(770,\,1,\,60)$ & --- & --- & --- & --- & --- & --- & --- & --- \\
$18$ & $(800,\,1,\,60)$ & --- & --- & --- & --- & --- & --- & --- & --- \\
$19$ & $(865,\,1,\,60)$ & --- & --- & --- & --- & --- & --- & --- & --- \\
$20$ & $(895,\,1,\,60)$ & --- & --- & --- & --- & --- & --- & --- & --- \\
\midrule
$\phi_R/\pi$ & $0$ & $0$ & $0$ & $0$ & $0$ & $0$ & $0$ & $0$ & $0$ \\
\bottomrule
\end{tabular}%
}
\parbox{\textwidth}{\vspace{1mm}\footnotesize
Each cell shows $(t_j\,[\si{\nano\second}],\, A_j/\pi,\, w_j\,[\si{\nano\second}])$. Sequences $n = 1$--$8$ use $\tau_c = \SI{960}{\nano\second}$; the adaptive sequence uses $\tau_c = \SI{1920}{\nano\second}$. Dashes indicate that the corresponding sequence contains fewer pulses than the maximum for that set.}
\end{table*}

\FloatBarrier

\section{Formal residual and adaptive sampling algorithm}\label{app:adaptive_sampling}

This appendix converts the Kramers--Kronig consistency check of \secref{subsec:results_gaussian_classical} into an adaptive sampling rule. For separately reconstructed conjugate components $\tilde{S}'$ and $\tilde{S}''$, it defines a noise-normalised residual, uses its sign structure to identify unresolved intervals, and resamples until the discrepancy falls below tolerance. The construction extends to higher-order spectra by applying the same procedure along each frequency axis of Eq.~\eqref{eq:kramers_kronig}.

\subsection{Kramers--Kronig residual and statistical trigger}

Let $\vec{\tilde{S}}'$ and $\vec{\tilde{S}}''$ be the source- and test-channel reconstructions over the reconstruction domains $\Omega'$ and $\Omega''$, obtained by regularised inversion of filter matrices $\mathbf{G}'$ and $\mathbf{G}''$. The channels are reconstructed separately but may share measurement records. For Fig.~\ref{fig:gauss_ordered_spectrum}, $\tilde{S}'=\mathrm{Re}[\tilde{S}^{(+)}]$ and $\tilde{S}''=\mathrm{Im}[\tilde{S}^{(+)}]$, with $(\Omega',\mathbf{G}')=(\Omega_1^{(zz)},\mathbf{G}_{zz})$ and $(\Omega'',\mathbf{G}'')=(\Omega_1^{(zx)},\mathrm{Im}[\mathbf{G}_{zx}])$. The source samples define
\begin{equation}
S_{\mathrm{int}}(\omega) =
\begin{cases}
\mathrm{interp}\big[\vec{\tilde{S}}'\big](|\omega|), & |\omega| \leq \omega_{\mathrm{max}},\\[2pt]
\big[\vec{\tilde{S}}'\big]_{N_h}\, e^{-(|\omega| - \omega_{\mathrm{max}})/\omega_t}, & |\omega| > \omega_{\mathrm{max}},
\end{cases}
\label{eq:spliced_interpolant}
\end{equation}
where $\mathrm{interp}[\cdot]$ is a shape-preserving piecewise-cubic interpolant on the evenly mirrored nodes, and the exponential-tail rate $\omega_t$ is fitted to the two outermost samples. Shape preservation prevents oscillations after sharp samples are inserted. At fixed nodes and tail rate, interpolation followed by the Hilbert transform is the linear map $\mathbf{W}$, giving the test-channel residual
\begin{equation}
\Delta(s_j) \equiv \big[\vec{\tilde{S}}''\big]_j - \big[\mathbf{W}\, \vec{\tilde{S}}'\big]_j,
\qquad s_j \in \Omega''.
\label{eq:kk_residual}
\end{equation}

Propagating the tomography covariance through the regularised pseudoinverse gives $\Sigma_{\tilde{S}}=\mathbf{G}^{+}\Sigma_{\mathcal{I}}(\mathbf{G}^{+})^{\mathsf{T}}$. Shared records also produce the cross-covariance $\Sigma^{\times}_{\tilde{S}}=(\mathbf{G}')^{+}\Sigma^{\times}_{\mathcal{I}}\big[(\mathbf{G}'')^{+}\big]^{\mathsf{T}}$, so
\begin{equation}
\sigma^2(s_j) = \big[\Sigma''_{\tilde{S}} + \mathbf{W}\, \Sigma'_{\tilde{S}}\, \mathbf{W}^{\mathsf{T}} - \mathbf{W}\, \Sigma^{\times}_{\tilde{S}} - (\mathbf{W}\, \Sigma^{\times}_{\tilde{S}})^{\mathsf{T}}\big]_{jj}.
\label{eq:residual_variance}
\end{equation}
The cross term vanishes for independently acquired channels.

The normalised score
\begin{equation}
\eta_j = \frac{|\Delta(s_j)|}{\sigma(s_j)}
\label{eq:eta_trigger}
\end{equation}
flags $s_j$ when $\eta_j>\eta_{\mathrm{th}}$ (e.g.\ $\eta_{\mathrm{th}}=3$). In the deterministic limit, $\sigma(s_j)$ is replaced by $\|\vec{\tilde{S}}'\|_{\mathrm{rms}}$ and the threshold by a relative tolerance $\vartheta$.

\subsection{Adaptive resampling algorithm}\label{app:refinement_algorithm}

Unmodelled weight $A$ near $\omega_0$ contributes $\Delta(s_j)\approx-\tfrac{A}{\pi}(s_j-\omega_0)^{-1}$. A positive-to-negative flip therefore indicates under-represented weight inside the interval, while a negative-to-positive flip indicates over-represented weight. On the base comb either flip signals unresolved structure; after resampling, a reversed flip may instead signal that the interpolant renders a recovered sharp sample too broadly. A same-sign run indicates tail misfit or out-of-band leakage and calls for a splice update rather than local resampling. Sign-changing intervals are ranked by
\begin{equation}
\chi_j =
\begin{cases}
\min(\eta_j,\, \eta_{j+1}), & \Delta(s_j)\,\Delta(s_{j+1}) < 0,\\
0, & \text{otherwise},
\end{cases}
\label{eq:interval_score}
\end{equation}
and $\chi^{*}=\max_j\chi_j$ sets the stopping test and target interval. After a midpoint measurement, a persistent flip calls for further subdivision; a reversed flip indicates that the amplitude is pinned but the model width is inaccurate. The latter may be corrected without further acquisition using
\begin{equation}
S_{\varsigma}(\omega) = S_{\mathrm{int}}(\omega) + \big[\tilde{S}'(\omega_0) - S_{\mathrm{int}}(\omega_0)\big]\, e^{-(\omega - \omega_0)^2/2\varsigma^2},
\label{eq:bump_model}
\end{equation}
centred at the new comb point $\omega_0$ and evenly mirrored. Here $S_{\mathrm{int}}$ is rebuilt without the $\omega_0$ node, so the measured value fixes the correction amplitude and $\varsigma$ alone controls its width.

\begin{algorithm}[Adaptive comb resampling]\label{alg:adaptive_refinement}
\emph{Input:} reconstructions $\vec{\tilde{S}}'$ and $\vec{\tilde{S}}''$ with their covariance matrices; threshold $\eta_{\mathrm{th}}$ (or deterministic tolerance $\vartheta$); resampling budget $K$; optional line-shape model.
\begin{enumerate}
\item Construct $S_{\mathrm{int}}$ and $\mathbf{W}$; compute $\Delta(s_j)$, $\eta_j$, and $\chi_j$.
\item For a significant same-sign run without a flip, refit $\omega_t$ and repeat step 1. Record any persistent run as leakage bias.
\item Stop if $\chi^{*}\leq\eta_{\mathrm{th}}$ or $K=0$.
\item Otherwise set $j^*=\arg\max_j\chi_j$. If this interval was just resampled and its flip reversed, proceed to step 6.
\item Add a sequence whose doubled cycle time places a harmonic at the interval midpoint. Solve for that sample by regularised inversion, using $S_{\mathrm{int}}$ at the other finer-comb harmonics. Further subdivision introduces two quarter-point unknowns and requires one sequence per unknown. Insert the samples, decrement $K$, and repeat step 1.
\item \emph{Optional shape step.} Fit $\varsigma$ in Eq.~\eqref{eq:bump_model} by minimising the root-mean-square $\eta_j$. Accept only if $\chi^{*}\leq\eta_{\mathrm{th}}$ and a held-out source- or test-channel sequence is predicted within threshold; otherwise return to step 5.
\end{enumerate}
\emph{Output:} the resampled spectrum, terminal scores, and any accepted line-shape width and held-out score.
\end{algorithm}

Steps 1--5 are measurement-driven; without a line-shape model, a reversed flip also returns to further subdivision. Because step 5 substitutes interpolated finer-comb values, step 1 must re-test the result. Measuring the test channel at the new midpoint provides an additional localisation check at the cost of one sequence per iteration.

\subsection{Classical Gaussian validation}\label{app:residual_demo}

For Fig.~\ref{fig:gauss_ordered_spectrum}, the deterministic pipeline reproduces the plotted Kramers--Kronig estimates to a mean (maximum) relative deviation of $1.6\%$ ($8\%$). With $\vartheta=0.1$, the base comb flags only $(3\omega_h,4\omega_h)$ at $\chi^{*}=0.22$. The $2\tau_c$ sequence recovers $\tilde{S}^{(+)}(3.5\omega_h)=1.36\times10^{5}\,\mathrm{s}^{-1}$; the ensuing flip reversal identifies excess interpolant width. The optional shape fit gives a full width $0.20\omega_h$ against the true $0.18\omega_h$, reduces $\chi^{*}$ to $0.001$, and recovers the spectral weight within $5\%$. Table~\ref{tab:residual_demo} shows the interval scores and resulting action at each stage.

\begin{table}[htbp]
\centering
\caption{Interval scores $\chi_j$~\eqref{eq:interval_score} for the adaptive resampling of Fig.~\ref{fig:gauss_ordered_spectrum}, with deterministic tolerance $\vartheta=0.1$. Iteration 1 adds the targeted $2\tau_c$ sequence; the shape step fits Eq.~\eqref{eq:bump_model} without further acquisition.}
\label{tab:residual_demo}
\begin{tabular}{lccc}
\toprule
Interval & Iter.~0 & Iter.~1 & Shape step \\
\midrule
$(\omega_h,2\omega_h)$   & 0     & 0     & 0 \\
$(2\omega_h,3\omega_h)$  & 0.02  & 0     & 0 \\
$(3\omega_h,4\omega_h)$  & $\mathbf{0.22}$ & $\mathbf{0.46}$ & 0 \\
$(4\omega_h,5\omega_h)$  & 0     & 0     & 0 \\
$(5\omega_h,6\omega_h)$  & 0     & 0     & 0 \\
$(6\omega_h,7\omega_h)$  & 0     & 0     & 0.001 \\
\midrule
Flip in $(3\omega_h,4\omega_h)$ & $+/-$ & $\mathbf{-/+}$ & none \\
$\chi^{*}$ vs.\ $\vartheta=0.1$ & flag & flag & pass \\
Action & resample & tune shape & stop \\
\bottomrule
\end{tabular}
\end{table}

\FloatBarrier

\section{Internal validation of truncated reconstructions}\label{app:internal_validation}

The reconstructions of \secref{sec:results} truncate the cumulant expansion at a finite order, and their accuracy is verified directly against the exact reference spectra available for the simulated environments (Figs.~\ref{fig:gauss_ordered_spectrum}, \ref{fig:second_order_plots}, and~\ref{fig:bispectrum_combined}). In an experiment no such reference exists, and the adequacy of the truncation must be established from the measurement record itself. Three complementary internal checks serve this purpose, listed here in order of increasing experimental cost.

\paragraph*{Kramers--Kronig self-consistency.}
The residual construction of \appref{app:adaptive_sampling} compares the directly reconstructed dispersive component against the Hilbert transform of the sampled real component. The normalised residual of Eq.~\eqref{eq:kk_residual} bounds unresolved structure within the sampled band and requires no measurements beyond those already used for the reconstruction.

\paragraph*{Cross-channel consistency.}
The same spectral object enters several $(\alpha,\gamma)$ channels with different filter parities and different sensitivities to higher-order contamination. At $k=2$, $\mathrm{Re}[\tilde{S}^{(+)}(\omega)]$ enters both the identity channel $\mathcal{I}_{x,\mathbbm{1}}$ through $F_{zz}$ and the transverse channel $\mathcal{I}_{x,y}$ through $\mathrm{Re}[F_{zx}]$; at $k=3$, the interchanged configurations $(\alpha,\gamma)\in\{(x,z),(z,x)\}$ of \secref{subsec:results_quantum} constrain overlapping regions of the bispectrum. Neglected cumulant orders do not cancel identically across channels, so agreement between independently inverted channels supports the assumed truncation.

\paragraph*{Hold-out prediction under test controls.}
The most stringent check reserves a set of validation sequences that are excluded from the reconstruction system $\mathbf{G}_{\mathbf{a},\mathbf{b}}$ --- ideally controls outside the comb family altogether, such as non-repeated sequences or incommensurate cycle times, whose filter functions weight the spectra between and beyond the reconstruction harmonics. Inserting the reconstructed polyspectra, interpolated with the splice of Eq.~\eqref{eq:spliced_interpolant}, into the overlap of Eq.~\eqref{eq:I_a_k_spectra} predicts the coefficients $\mathcal{I}_{\alpha,\gamma}$ that the validation sequences should measure, and the relative prediction error is an operational figure of merit for the truncated model. Unmodelled cumulant orders weight the filters differently --- most sharply through the zero-frequency scaling $\propto[F(0)]^k$ noted in \secref{sec:results} --- so validation sequences with contrasting low-frequency weight expose truncation bias directly, and systematic growth of the prediction error with increasing $|F(0)|$ localises the bias to the neglected orders.

We do not deploy the hold-out protocol in this work: the simulated environments supply the exact target spectra, the direct comparisons in the figures confirm the fit at the retained orders, and the Kramers--Kronig diagnostics of \appref{app:adaptive_sampling} are exercised in situ. All three checks, however, use only the preparation, control, and measurement capabilities assumed throughout, and are therefore available in experimental settings where no reference spectra exist.

\dobib

\bibliography{gauss_bibtex}

\end{document}